\documentclass[12pt]{article}
\usepackage{amsmath,amsthm,amsfonts,amssymb,cite,graphicx}

\newlength{\xtrawidth}
\newlength{\xtraheight}
\numberwithin{equation}{section}

\def\model{skew symplectic sigma model}

\begin{document}
\begin{titlepage}
\begin{center}
\hfill BONN--TH--2015--05\\
\vskip 0.6in
{\Large\bf{Dual Pairs of Gauged Linear Sigma Models}}\\[2ex]
{\Large\bf{and Derived Equivalences of Calabi--Yau threefolds}}
\vskip 0.3in
{\large{Andreas Gerhardus and Hans Jockers}}\\
\vskip 0.1in
{\it Bethe Center for Theoretical Physics, Physikalisches Institut\\
der Universit\"at Bonn, Nussallee 12, D-53115 Bonn, Germany}
\vskip 0.1in
{\tt gerhardus@th.physik.uni-bonn.de}\\
{\tt jockers@uni-bonn.de}
\end{center}
\vskip 0.1in
\begin{center} {\bf Abstract} \end{center}
In this work we study the phase structure of \model s, which are a certain class of two-dimensional $N=(2,2)$ non-Abelian gauged linear sigma models. At low energies some of them flow to non-linear sigma models with Calabi--Yau target spaces, which emerge from non-Abelian strong coupling dynamics. The observed phase structure results in a non-trivial duality proposal among \model s and connects non-complete intersection Calabi--Yau threefolds --- that are non-birational among another --- in a common quantum K\"ahler moduli space. As a consequence we find non-trivial identifications of spectra of topological B-branes, which from a modern algebraic geometry perspective imply derived equivalences among Calabi--Yau varieties. To further support our proposals, we calculate the two sphere partition function of \model s to determine geometric invariants, which confirm the anticipated Calabi--Yau threefold phases. We show that the two sphere partition functions of a pair of dual \model s agree in a non-trivial fashion. To carry out these calculations, we develop a systematic approach to study higher-dimensional Mellin--Barnes type integrals. In particular, these techniques admit the evaluation of two sphere partition functions for gauged linear sigma models with higher rank gauge groups, but are applicable in other contexts as well.

\vfill
\noindent May, 2015

\end{titlepage}
\tableofcontents
\newpage

\section{Introduction}
In the seminal work on two-dimensional $N=(2,2)$ supersymmetric gauged linear sigma models \cite{Witten:1993yc}, Witten has introduced a powerful machinery to analyze both geometric and non-geometric phases of supersymmetric worldsheet theories. In particular the study of Abelian gauged linear sigma models --- which physically realize toric varieties in terms of $\mathbb{C}^*$~quotients in geometric invariant theory --- led to far-reaching developments in our understanding of the phase structure of string compactifications. For instance, gauged linear sigma model techniques continuously connect supersymmetric Landau--Ginzburg models of  rational conformal field theories to semi-classical non-linear sigma models with geometric Calabi--Yau target spaces \cite{Witten:1993yc}. The identification of explicit rational conformal field theories in the moduli space of gauged linear sigma models has historically been an important ingredient in finding Calabi--Yau mirror pairs \cite{Greene:1990ud}, which resulted in the remarkable combinatorial construction of mirror manifolds for complete intersection Calabi--Yau manifolds in toric varieties \cite{Batyrev:1994hm,Batyrev:1994pg,Morrison:1995yh,Hori:2000kt} and led to a mirror theorem for these classes of mirror pairs \cite{MR1408320,MR1621573}.

While the use of Abelian gauged linear sigma models in string theory has become textbook knowledge,\footnote{See for example refs.~\cite{Greene:1996cy,Cox:2000vi,Hori:2003ic} and references therein.} the role of $N=(2,2)$ supersymmetric non-Abelian gauged linear sigma models for string compactifications is less systematically explored. Such non-Abelian gauge theories are of recent interest in their own right \cite{Doroud:2012xw,Benini:2012ui,Benini:2014mia,Gerchkovitz:2014gta}, and they allow us to describe a much broader class of Calabi--Yau geometries such as certain non-complete intersections in toric varieties \cite{Witten:1993yc,Witten:1993xi,Lerche:2001vj,Hori:2006dk,Donagi:2007hi,Hori:2011pd,Jockers:2012zr,Hori:2013gga,Sharpe:2012ji}. 

Compared to their Abelian cousins, the non-Abelian gauged linear sigma models exhibit further interesting properties, such as strongly coupled phases, which sometimes can be mapped by duality to weakly coupled semi-classical phases \cite{Hori:2006dk,Hori:2011pd,Jockers:2012zr,Hori:2013gga}. A convincing argument to justify such dualities is given by identifying two sphere partition functions of such dual gauge theories \cite{Doroud:2012xw}.\footnote{For the Calabi--Yau phases examined in refs.~\cite{MR1775415,Hori:2006dk}, this approach is demonstrated in ref.~\cite{Jockers:2012dk}, where the two sphere partition function of the associated strongly coupled two-dimensional gauge theory is matched to its weakly coupled dual and semi-classical Calabi--Yau phase.} The two sphere partition function encodes the Zamolodchikov metric of the $N=(2,2)$ superconformal infrared fixed point of the renormalization group flow \cite{Doroud:2012xw,Benini:2012ui,Jockers:2012dk,Gomis:2012wy,Halverson:2013eua,Gerchkovitz:2014gta}, which is exact in the worldsheet coupling~$\alpha'$. As result the two sphere sphere partition function encodes both perturbative and non-perturbative worldsheet corrections, which in the semi-classical large volume regime are respectively identified with certain characteristic classes and genus zero Gromov--Witten invariants \cite{Jockers:2012dk,Halverson:2013qca}.

As these two-dimensional theories have four supercharges, these infrared strong-weak coupling dualities can be viewed as the two-dimensional analog of $N=1$ Seiberg dualities in four dimensions \cite{Seiberg:1994bz}, which --- due to recent progress on localizing supersymmetric gauge theories on curved spaces so as to compute  partition functions on compact backgrounds \cite{Pestun:2007rz,Festuccia:2011ws} --- have passed similarly impressive consistency checks by matching partition functions of dual four-dimensional $N=1$ gauge theories \cite{Romelsberger:2005eg,Dolan:2008qi}. 

An immediate consequence of connecting semi-classical Calabi--Yau phases through strong--weak coupling dualities is that seemingly distinct $N=(2,2)$ non-linear sigma models with Calabi--Yau threefold target spaces can emerge as two regimes in the same moduli space of the underlying common family of $N=(2,2)$ worldsheet theories. The first and prominent example of this kind furnishes the pair of the degree $14$ R\o{}dland Pfaffian Calabi--Yau threefold and a certain degree $42$ complete intersection Calabi--Yau threefold in the Grassmannian $\operatorname{Gr}(2,7)$ \cite{MR1775415},\footnote{Borisov and Libgober recently studied a generalization of this correspondence to higher dimensional Calabi--Yau varieties \cite{Borisov:2015vqa}.}  which --- as shown in the important work by Hori and Tong \cite{Hori:2006dk} --- arise respectively as a weakly and a strongly coupled phase of a non-Abelian gauge linear sigma model. An immediate consequence of this result is that the associated Calabi--Yau categories of B-branes must be equivalent,\footnote{Mathematically, the B-brane category of a geometric Calabi--Yau threefold phase is described by its derived categories of bounded complexes of coherent sheaves \cite{Sharpe:1999qz,Douglas:2000gi,Diaconescu:2001ze}.} as for the above example is mathematically shown in ref.~\cite{MR2475813,Kuznetsov:2006arxiv,Addington:2014sla}. Hosono and Takagi present another interesting example of such an equivalence \cite{MR3166392}, which also fits into the framework of non-Abelian gauged linear sigma models \cite{Jockers:2012zr}. More generally, two-dimensional dualities of gauged linear sigma models connect phases of worldsheet theories in a common moduli space, which then implies equivalences of the associated B-brane categories. It should be stressed that the categorical equivalences that arise from strong--weak coupling dualities are clearly distinct from equivalences due to birational transformations studied in ref.~\cite{MR1949787} or other typical phase transitions in Abelian gauged linear sigma models --- such as, e.g., the Landau--Ginzburg and large volume correspondence for Calabi--Yau hypersurfaces in weighted projective spaces \cite{MR2641200,Herbst:2008jq}.\footnote{For non-birational phase transitions in Abelian gauged linear sigma models see refs.~\cite{Caldararu:2007tc,Addington:2012zv}.} As proposed in ref.~\cite{Caldararu:2007tc}, including these more general identifications of Calabi--Yau geometries on the level of their B-brane categories yields derived equivalences that sometimes realize particular instances of homological projective duality introduced by Kuznetsov \cite{MR2354207}.

In this work we study the phase structure of a certain class of non-Abelian gauged linear sigma models with symplectic gauge groups, to which we refer to as the \model s. We formulate a strong-weak coupling duality among \model s together with non-trivial consistency checks in support of our proposal. In particular, we study in detail two families of Calabi--Yau threefolds with a single K\"ahler modulus arising as weakly and strongly coupled phases in the moduli space of certain \model s. Furthermore, we argue that these Calabi--Yau geometries are related by a strong-weak coupling dualities of a dual pair of \model s. Geometrically, this pair of threefolds gives rise to interesting examples for non-trivial derived equivalences. For our analysis the  two sphere partition function furnishes an important tool, which we use to study the phase structure of \model s and to confirm the proposed duality relation. In the analyzed large volume phases we extract from the two sphere partition function both perturbative and non-perturbative quantum corrections of the corresponding geometric large volume string compactifications.

As the studied gauged linear sigma models involve higher rank gauge groups, their two sphere partition functions arise from higher-dimensional Mellin--Barnes type integrals, which are technically challenging to evaluate. Therefore, we extend the technique of Zhdanov and Tsikh \cite{MR1631772} --- which transforms two-dimensional Mellin--Barnes type integral to a suitable combinatorial sum of local Grothendieck residues --- to Mellin--Barnes type integral of arbitrary dimension. While this generalization is essential to carry out the calculations in this work, we hope that these technical advances will prove useful in another context as well.

The organization of this work is as follows: In Section~\ref{sec:SSSM} we introduce the \model s --- a certain class of two-dimensional $N=(2,2)$ non-Abelian gauged linear sigma models. We study their phase structure, which generically exhibit a weakly and a strongly coupled phase. Both phases become in the infrared non-linear sigma models with interesting Calabi--Yau threefold target spaces. These threefolds are projective varieties of the non-complete intersection type, and we analyze their geometric properties. Finally, the observed phase structure together with the established geometric properties of the semi-classical infrared non-linear sigma model regimes allow us to propose a non-trivial duality relation among \model s. In Section~\ref{sec:ZS2}, we support our duality proposal by matching the two sphere partition function for a pair of dual \model s and for certain self-dual \model s. While the Mellin--Barnes type integral expression for the two sphere partition functions of the dual models look rather distinct --- e.g., even the dimensionality of the integration domains are distinct due to different ranks of the gauge group in dual \model s ---  we demonstrate that the dual two sphere partition functions are indeed identical. In Section~\ref{sec:derivedequiv} we discuss further implications of the phase structure of the \model s with geometric Calabi--Yau threefold target space regimes. We argue that the \model s give rise to an equivalence of topological B-brane spectra, which in the formulation of modern algebraic geometry conjectures a non-trivial equivalence between the derived category of bounded complexes of coherent sheaves of the studied Calabi--Yau threefolds. Then we present our conclusions and outlook in Section~\ref{sec:con}. In Appendix~\ref{sec:MB} ---building upon previous work on two-dimensional Mellin--Barnes integrals by Zhdanov and Tsikh and using multidimensional residue calculus --- we present a systematic approach to evaluate Mellin--Barnes type integrals in arbitrary dimensions. These technical results are necessary to carry out the computations described in Section~\ref{sec:ZS2}. More generally, this method can be applied to systematically evaluate two sphere partition functions of two-dimensional $N=(2,2)$ gauged linear sigma model with higher rank gauge groups.

\section{The \model} \label{sec:SSSM}
For the Calabi--Yau threefolds studied in this work, we introduce a class of two-dimensional $N=(2,2)$ supersymmetric non-Abelian gauged linear sigma models based on the compact Lie group\footnote{Note that the compact Lie group $\operatorname{USp}(2k) = U(2k)\cap \operatorname{Sp}(2k,\mathbb{C})$ is often denoted by $\operatorname{Sp}(k)$. Here we use the first notation to indicate the dimension of the defining representation.} 
\begin{equation} \label{eq:G}
  G \,=\, \frac{U(1) \times \operatorname{USp}(2k)}{\mathbb{Z}_2} \ .
\end{equation}
Here the $\mathbb{Z}_2$ quotient is generated by the element $\left( e^{i \pi} , - \mathbf{1}_{2k\times 2k} \right)$ of order two in the center of $U(1) \times \operatorname{USp}(2k)$. Furthermore, the studied gauged linear sigma model possesses a vector $U(1)_V$ R-symmetry and an axial $U(1)_A$ R-symmetry, under which the $N=(2,2)$ supercharges transform as $(Q_\pm,\bar Q_\pm) \to (e^{i\alpha} Q_\pm,e^{-i\alpha} \bar Q_\pm)$ and $(Q_\pm,\bar Q_\pm) \to (e^{\pm i\beta} Q_\pm,e^{-\mp i\beta} \bar Q_\pm)$, respectively. While the $U(1)_V$ R-symmetry is always preserved at the quantum level, the the $U(1)_A$ R-symmetry is generically anomalous.

In addition to the vector multiplets $V_{U(1)}$ and $V_{\operatorname{USp}(2k)}$ of the gauge group $G$, the chiral matter fields of the studied non-Abelian gauged linear sigma model are listed in Table~\ref{tb:spec1}. Thus the entire non-Abelian gauged linear sigma model is determined by three positive integers $(k,m,n)$. Due to the skew-symmetric multiplicity labels of the chiral field $P^{[ij]}$, we call these models the \model s or in short the $SSSM_{k,m,n}$.

A generic gauge invariant superpotential of R-charge $+2$ for the \model{} reads
\begin{equation} \label{eq:W}
  W\,=\, \operatorname{tr}\left[ P \left(A(\phi) + X^T \epsilon X \right) \right] + B(\phi)  \cdot Q^T\epsilon X  \ , \qquad
    \epsilon=\begin{pmatrix} 0 & \mathbf{1}_{k\times k} \\ -\mathbf{1}_{k\times k} & 0 \end{pmatrix} \ ,
\end{equation}
in terms of $A(\phi) = \sum_a A^a \phi_a$ constructed out of $m$ skew-symmetric matrices $A^a_{[ij]}$ of dimension $n \times n$ and the column vector $B(\phi) = \sum_a B^a \phi_a$ constructed out of the vectors $B^{ai}$ of dimension $n$. Here the trace is taken over the symplectic multiplicity indices $i,j$ of the multiplets $P^{[ij]}$ and $X_i$. While this superpotential does not represent the most general form, for generic superpotential couplings it can always be cast into this form with the help of suitable field redefinitions.

Due to the Abelian factor in the gauge group $G$, the \model{} allows for a Fayet--Iliopoulos terms, which together with the theta angle generates the two-dimensional twisted superpotential \cite{Witten:1993yc}
\begin{equation} \label{eq:Wt}
  \widetilde W \,=\, \frac{i \tilde{t}}{2\sqrt{2}} \Sigma_{U(1)} \ ,
\end{equation}
in terms of the twisted chiral field strength $\Sigma_{U(1)}=\frac{1}{\sqrt{2}}\overline{D}_+ D_- V_{U(1)}$ and the complexified Fayet--Iliopoulos coupling
\begin{equation}
  \tilde{t} \,=\, i r + \frac{\theta}{2\pi} \ .
\end{equation}
In this note, the complexified Fayet--Iliopoulos coupling $\tilde{t}$ is the key player, as we study the phase structure of the \model s as a function of this parameter $\tilde{t}$. 

Note that the constructed non-Abelian gauged linear sigma model is somewhat related to the PAX and PAXY model studied in ref.~\cite{Jockers:2012zr}. Essentially the unitary gauge group factor is replaced by the symplectic gauge group $\operatorname{USp}(2k)$, while the matter contents exhibit a similar structure.

\begin{table}[t]
\centering
\hfil\vbox{
\offinterlineskip
\tabskip=0pt
\halign{\vrule height2.6ex depth1.4ex width1pt~#~\hfil\vrule&~\hfil~#~\hfil\vrule&~\hfil~#~\hfil\vrule&\hfil~#~\hfil\vrule height2.6ex depth1.4ex width 1pt\cr
\noalign{\hrule height 1pt}
\hfil chiral multiplets & multiplicity & $G$ representation & $U(1)_V$ R-charge \cr
\noalign{\hrule height 1pt}
\ $P^{[ij]}$, $1\le i < j \le n$ & $\binom{n}2$ & $\mathbf{1}_{-2}$ & $2-2\mathfrak{q}$ \cr
\noalign{\hrule}
\ $Q$ & $1$ & $\mathbf{2k}_{-3}$ & $2-3\mathfrak{q}$ \cr
\noalign{\hrule}
\ $\phi_a$, $a=1,\ldots,m$ & $m$ & $\mathbf{1}_{+2}$ & $2\mathfrak{q}$ \cr
\noalign{\hrule}
\ $X_i$, $i=1,\ldots,n$ & $n$ & $\mathbf{2k}_{+1}$ & $\mathfrak{q}$ \cr
\noalign{\hrule height 1pt}
}}\hfil
\caption{Listed are the matter fields of the \model{} together with their multiplicities, their representations of the gauge group $G$, and their $U(1)_V$ R-charges. The gauge group representations are labeled by representations of $\operatorname{USp}(2k)$ with a subscript for the $U(1)$ gauge charge. As the $U(1)_V$ R-symmetry is ambiguous up to transformations with respect to the $U(1)$ gauge symmetry, the $U(1)_V$ charges of the multiplets exhibit the real parameter $\mathfrak{q}$.}\label{tb:spec1}
\end{table}

\subsection{Infrared limit of the \model}
For the \model{} to flow to a non-trivial infrared $N=(2,2)$ superconformal fixed point in the infrared, we follow the general strategy and require that the \model{} possesses in addition to the vector $U(1)_V$ R-symmetry a non-anomalous axial $U(1)_A$ R-symmetry. Then in the infrared limit these two R-symmetries yield the left- and right-moving $U(1)$~currents of the $N=(2,2)$ superconformal algebra. The axial anomaly $\partial_\mu j_A^\mu$ appears at one loop from the operator product of the axial $U(1)_A$ current and the gauge current. From the axial $U(1)_A$ R-charges of the fermions in the chiral multiplets we arrive at
\begin{equation} \label{eq:axialcurr}
\begin{aligned}    
     \partial_\mu j_A^\mu \,&\sim\,  \underbrace{(-2)\binom{n}2}_{P_{[ij]}}  + \underbrace{\left( - 3 \right)  (2k)}_Q  
     + \underbrace{(+2) m}_{\phi_a}  + \underbrace{\left(+1 \right) (2kn)}_{X_i} \\
     \,&=\, (n-3)(2k-n) + 2(m-n) \ .
\end{aligned}     
\end{equation}
As argued in ref.~\cite{Witten:1993yc}, we find that the axial $U(1)_A$ R-symmetry is non-anomalous, if the $U(1)$ gauge charge of the matter multiplets (weighted with the dimension of the non-Abelian $\operatorname{USp}(2k)$ representation) add up to zero, and the condition for a conserved axial current $\partial_\mu j_A^\mu = 0$ reads
\begin{equation} \label{eq:axial}
   m  \,=\, \frac12 (n-3)(n-2k) + n \ .
\end{equation}
Furthermore, if the axial $U(1)_A$ R-symmetry is non-anomalous, we can also calculate the central charge $c$ of the $N=(2,2)$ superconformal field theory at the infrared fixed point. It arises at one loop from the operator product $\Gamma_{A-V}$ of the vector $U(1)_V$ and the axial $U(1)_A$ currents
\begin{equation}
\begin{aligned}
  \Gamma_{A-V}(\mathfrak{q}) \,=\, 
  \underbrace{(2\mathfrak{q}-1) \binom{n}2}_{P_{[ij]}} &+\underbrace{(3\mathfrak{q}-1)(2k)}_Q 
     + \underbrace{(-2\mathfrak{q}+1)m}_{\phi_a} \\
     &\qquad+\underbrace{(-\mathfrak{q}+1)(2 k n)}_{X_i} 
     +\underbrace{(-1)(2k+1)k}_{V_{\operatorname{USp(2k)}}} +\underbrace{(-1)}_{V_{U(1)}} \ .
\end{aligned}  \label{eq:mixed}
\end{equation}
Inserting the axial anomaly cancellation condition~\eqref{eq:axial}, we arrive at the central charge~$c$ of the infrared $N=(2,2)$ SCFT
\begin{equation} \label{eq:central}
   \frac{c}{3} = \Gamma_{A-V}(\mathfrak{q}) = k(n-2k)-1 \ .
\end{equation}

Thus the two integers $k$ and $n$ together with the requirement of a non-anomalous axial $U(1)_A$ current determine the multiplicity label $m$ and therefore specify the associated \model{} with a controlled renormalization group flow to a $N=(2,2)$ superconformal fixed point. 

\subsection{Phases of the \model}
Our first task is to analyze the scalar potential $U$ of the \model
\begin{equation}
\begin{aligned}
   U \,=\, & \frac12 D_{U(1)}^2 + | \sigma_{U(1)} |^2 
          \left(8\!\!\!\sum_{1\le i<j\le n}\!\!\!|P^{[ij]}|^2+18\,Q^\dagger Q 
          + 8\sum_{a=1}^m|\phi_a|^2+2\sum_{i=1}^n X^\dagger_i X_i \right)\\
   &+\frac12 \operatorname{tr} D_{\operatorname{USp}(2k)}^2 
          + 2 ( \sigma_{\operatorname{USp}(2k)} Q)^\dagger  \sigma_{\operatorname{USp}(2k)} Q
          + 2 \sum_{i=1}^n ( \sigma_{\operatorname{USp}(2k)} X_i)^\dagger  \sigma_{\operatorname{USp}(2k)} X_i\\
   &+ \sum_{1\le i<j\le n} \left| F_{P^{[ij]}} \right|^2 + F_{Q}^\dagger F_Q + \sum_{a=1}^m \left|F_{\phi_a}\right|^2 
       + \sum_{i=1}^n F_{X_i}^\dagger F_{X_i}  \ ,
\end{aligned}   
\end{equation}
which is a sum of non-negative terms given in terms of the complex scalar fields $\sigma_{U(1)}$ and $\sigma_{\operatorname{USp}(2k)}$ and the auxiliary D-terms of the $N=(2,2)$ vector multiplets, and in terms of the complex scalar fields of the $N=(2,2)$ chiral multiplets and their auxiliary F-terms. For a stable supersymmetric vacuum --- i.e., for $U=0$ --- all non-negative D- and F-terms must vanish separately.  

The D-terms of the vector multiplets $V_U(1)$ and $V_{\operatorname{USp}(2k)}$ become
\begin{equation} \label{eq:DU(1)}
    D_{U(1)} \,=\, 2\sum_a |\phi_a|^2 + \sum_i X_i^\dagger X_i - 3\,Q^\dagger Q -2 \sum_{1\le i<j\le n} |P^{[ij]}|^2 - r \ ,
\end{equation}
including the Fayet--Iliopoulos parameter $r$, and 
\begin{equation} \label{eq:DUSp}
   D_{\operatorname{USp}(2k)}^A \,=\, \sum_i X_i^\dagger T^A X_i + Q^\dagger T^A Q \ , \quad A=1,\ldots, (2k+1)k \ ,
\end{equation}
where $T^A$ are the generators of the Lie algebra $\mathfrak{usp}(2k)$.\footnote{The semi-simple Lie algebra $\mathfrak{usp}(2k)=\mathfrak{u}(2k) \cap\mathfrak{sp}(2k,\mathbb{C})$ can be represented by the set of complex matrices $\left\{\begin{pmatrix} A & B \\ -B^\dagger & -A^T \end{pmatrix} \in \operatorname{Mat}(2k,\mathbb{C}) \,\middle|\,A + A^\dagger = 0 \,,\ B = B^T \right\}$.} Geometrically, the D-terms enjoy the interpretation as the moment map $\mu: V_\text{chiral} \to \mathfrak{g}^*$ with respect to the group action of the Lie group $G$ on the complex vector space $V_\text{chiral}$ of the chiral multiplets and its canonical $G$-invariant symplectic K\"ahler form.

The F-terms of the chiral multiplets are determined from the gradient of the superpotential 
\begin{equation} \label{eq:Fterms}
\begin{aligned}
  F_{P^{[ij]}} &= A_{[ij]}(\phi) + X_i^T\epsilon X_j  \ , &
  F_{Q}&= B(\phi)\cdot X \, \\ 
  F_{\phi_a}&=\operatorname{tr} \left[P A^a\right] +B^a \cdot Q^T \epsilon X\ , &
  F_{X_i}&=2 \sum_{j} P^{[ij]} X_j + B^i(\phi) Q\ .   
\end{aligned}
\end{equation}

\subsubsection{The \model{} phase $r\gg0$} \label{sec:Xphase}
Let us first analyze the \model{} in the regime $r\gg 0$ of the Fayet--Iliopoulos parameter $r$. Then setting the D-term~\eqref{eq:DU(1)} to zero enforces at least one of the scalar fields in the chiral multiplets $\phi_a$ or $X_i$ to have a non-vanishing expectation value. Next we consider the F-term $F_{\phi_a}$ and $F_{X_i}$, which  impose $m+2kn$ constraints on the $\binom{n}2+2k$ degrees of freedom arising from the fields $P^{[ij]}$ and $Q$. Assuming genericness of all these constraints, for $m\ge\frac12(n-1)(n-4k)$ we obtain mass terms for the fields $P^{[ij]}$ and $Q$, which set their scalar expectation values to zero. Note that the condition $m\ge\frac12(n-1)(n-4k)$ is automatically fulfilled for \model s with conserved axial current~\eqref{eq:axial}, which are the models of our main interest. 

With the field $Q$ set to zero, combining the constraints from the Abelian and the non-Abelian D-terms further requires that at least one scalar $\phi_a$ must be non-vanishing. As a result the F-term condition $F_{P^{[ij]}}=0$ equates the expectation values of the bilinears $\Lambda_{[ij]}:=X_i^T\epsilon X_j$ with $A_{[ij]}(\phi)$. However, not all bilinears $\Lambda_{[ij]}$ can acquire independent expectation values, because they are quadratic in the fields $X_i, i=1,\ldots,n$, which in turn are collectively represented by the $(2k)\times n$-matrix~$X$. As a result the skew-symmetric matrix $\Lambda = X^T \epsilon X$ and also $A(\phi)$ has at most rank $2k$. Conversely, together with the non-Abelian D-term constraint for $n\ge 2k$ any skew-symmetric matrix $\Lambda$ of rank $2k$ can --- up to a $\operatorname{USp}(2k)$ transformation --- uniquely be written as the bilinear form of a $(2k)\times n$-matrix $X$. Finally, multiplying the F-term $F_Q$ with $X_i$ implies the constraint $A(\phi) \cdot B(\phi)=0$. Therefore, altogether we obtain in the semi-classical $r\gg 0$ phase the target space geometry
\begin{equation} \label{eq:XSSSM}
    \mathcal{X}_{k,m,n} \,=\, \left\{\, \phi \in \mathbb{P}^{m-1} \,\middle|\,
      \operatorname{rk} A(\phi) \le 2k \ \text{and}\ A(\phi)  \cdot B(\phi) = 0 \, \right\}  \ ,
\end{equation}
in terms of the $n\times n$ skew-symmetric matrix $A(\phi)$ and the $n$-dimensional vector $B(\phi)$. We note that the target space variety $\mathcal{X}_{k,m,n}$ can alternatively be written as\footnote{We would like to thank Sergey Galkin for sharing this construction of Calabi--Yau threefolds with us, which among other places he first presented in a lecture at Tokyo University \cite{Galkin:2014Talk}. It pointed us towards the discovery of the \model s.}
\begin{equation}\label{eq:Xvariety}
   \mathcal{X}_{k,m,n} \,=\, \left\{ [\phi,\tilde P] \in \mathbb{P}( V, \Lambda^2 V^*) \, \middle| \, 
   \operatorname{rk} \tilde P \le 2k \ \text{and}\ \phi \in \ker\tilde P \right\}\,\cap\, \mathbb{P}(L) \ , \quad
   V \simeq \mathbb{C}^n \ , 
\end{equation}
with the linear subspace $L$ of dimension $m$ in $V\oplus \Lambda^2 V^*$ given by
\begin{equation} \label{eq:L}
    L \,=\, \left\{ \, \left(B(\phi), A(\phi)\right)  \in  V\oplus \Lambda^2 V^* \,\right\} \ . 
\end{equation}
This construction for projective varieties has been put forward in refs.~\cite{Galkin:2014Talk,Galkin:2015InPrep}. The dimension of the non-complete intersection variety $\mathcal{X}_{k,m,n}$ is given by
\begin{equation} \label{eq:dimX}
  \dim_\mathbb{C} \mathcal{X}_{k,m,n} \,=\, (m-1) - \frac12 (n-2k-1)(n-2k) - 2k \ .
\end{equation}
where $\frac12(n-2k-1)(n-2k)$ yields the codimension of the rank condition imposed on the skew forms in $\Lambda^2 V^*$, while $2k$ takes into account the codimension of the kernel condition imposed on the vectors in $V$.

Typically the variety $\mathcal{X}_{k,m,n}$ is singular at those points, where the rank of $\tilde P$ is reduced, i.e., $\operatorname{rk} \tilde P < 2k$. For generic choices of the projective subspace $\mathbb{P}(L)$ this occurs in codimension $\frac12 (n-2k+1)(n-2k+2) + 2k - 2$ (or bigger). Hence, for
\begin{equation} \label{eq:redrk}
  \dim_\mathbb{C} \mathcal{X}_{k,m,n}<2(n-2k)-1 \ ,
\end{equation}
the generic variety $\mathcal{X}_{k,m,n}$ has constant rank $\operatorname{rk}\tilde P = 2k$, and thus does not acquire any singularities from a reduction of the rank of $\tilde P$. Then the validity of the presented semi-classical analysis is guaranteed. 

However, if the rank condition~\eqref{eq:redrk} is not met, it does not necessarily imply that the variety $\mathcal{X}_{k,m,n}$ acquires singularities in codimension $2(n-2k)-1$ where $\operatorname{rk}\tilde P<2k$. Note that for $n=2k+2$ the rank constraint in the definition~\eqref{eq:Xvariety} becomes redundant, because the second constraint $\phi\in\ker \tilde P$ automatically implies that the even-dimensional skew symmetric matrix $\lambda$ obeys $\operatorname{rk} \tilde P \le n-2$, i.e.,
\begin{equation} \label{eq:spec}
   \mathcal{X}_{k,m,2k+2} \,=\, \left\{ [\phi,\tilde P] \in \mathbb{P}( V, \Lambda^2 V^*) \, \middle| \, \phi \in \ker \tilde P \right\}\cap \mathbb{P}(L) \ , 
   \quad V \simeq \mathbb{C}^{2k+2} \ .
\end{equation}
This kernel condition imposes $2k+2$ bilinear constraints of the form $f_i = \sum_k \lambda_{[ik]} \phi^k$ with one relation $\sum_i \phi^i f_i = 0$. Hence, we find for the dimension
\begin{equation}
    \dim_\mathbb{C} \mathcal{X}_{k,m,2k+2} \,=\, (m-1) - (2k +1 ) = m - 2k -2 \ ,
\end{equation}
which is in agreement with eq.~\eqref{eq:dimX}. For generic choice of the linear subspace $L$ in $V\oplus \Lambda^2 V^*$ the variety $\mathcal{X}_{k,m,2k+2}$ is indeed smooth.

Note that if the rank condition $\operatorname{rk}\tilde P=2k$ is saturated the non-Abelian gauge group factor $\operatorname{USp}(2k)$ is spontaneously broken at all points in the target space variety. Then the discussed phase is weakly coupled, and we do not expect any strong coupling dynamics in the infrared. If, however, there are points in the target space variety $\mathcal{X}_{k,m,n}$, where the rank condition is not saturated, the non-Abelian gauge group $\operatorname{USp}(2k)$ is not entirely broken and strong coupling dynamics in the infrared becomes important at such points \cite{Hori:2006dk,Hori:2011pd}. While this is not surprising at singular points of the target space varieties, such strong coupling effects are in principle even present for the models $SSSM_{k,m,2k+2}$, which as discussed naively give rise to the smooth target space varieties~$\mathcal{X}_{k,m,2k+2}$. For smooth target space varieties we heuristically expect that the infrared dynamics of the gauge theory --- as for instance described by its correlation functions --- varies continuously with respect to the scalar field expectation values, which realize coordinates on the target space. For this reason we suppose that, due to the smoothness of the target space, such strong coupling effects are not relevant in the infrared. We do not further examine such strong coupling effects here, but instead give a stronger indirect argument --- by evaluating the two sphere partition function of such a model in Section~\ref{sec:ZS2} --- that the models $SSSM_{k,m,2k+2}$ reduce for $r\gg 0$ to semi-classical non-linear sigma model with the smooth target space variety $\mathcal{X}_{k,m,2k+2}$, even at the loci of reduced rank.

In this work we focus on those \model{}s with an expected semi-classical non-linear sigma model phase for $r \gg 0$, which by the above argument includes the models $SSSM_{k,m,2k+2}$.

\subsubsection{The \model{} phase $r\ll0$} \label{sec:strongphase}
Let us now turn to the regime $r\ll 0$ for the Fayet--Iliopoulos parameter $r$ of the \model. For this phase non-Abelian strong coupling effects become essential \cite{Hori:2015priv}, and the analysis of this phase can be tackled analogously to the strongly coupled gauged linear sigma model phases studied by Hori and Tong~\cite{Hori:2006dk}.\footnote{We are deeply grateful to Kentaro Hori for explaining in detail to us, how to analyze this strong coupling phase.}

In this phase the abelian D-term~\eqref{eq:DU(1)} ensures a non-vanishing expectation value for the chiral multiplets $P^{[ij]}$ or $Q$. For generic choices of the parameters $B^{ai}$, the F-term $F_{\phi_a}=0$ together with the non-Abelian D-term~\eqref{eq:DUSp} even implies that at least one chiral multiplet $P^{[ij]}$ must be non-zero. Thus the Abelian gauge group factor of the gauge group is broken, and we are left with the non-Abelian $\operatorname{USp}(2k)$ gauge theory, which we view adiabatically fibered over the total space
\begin{equation}
  \mathcal{V} \,=\, \operatorname{Tot}\left(\mathcal{O}(-1)^{\oplus m}\to \mathbb{P}^{{\binom{n}2}-1}\right) \ ,
\end{equation}
where the homogeneous coordinates of the base are given by the expectation values of the $\binom{n}2$ chiral fields $P^{[ij]}$ and the fibers by the $m$ chiral fields $\phi_a$.

As presented by Hori and Tong for $SU(k)$ gauge theories fibered over a base \cite{Hori:2006dk}, the strategy is now to solve the infrared dynamics of the $\operatorname{USp}(2k)$ gauge theory at each point of the variety $\mathcal{V}$. For any point $(\phi,P)\in\mathcal{V}$ the $n+1$ fundamental flavors $(X_i,Q)$ couple according to the F-terms $F_{X_i}$ and $F_{Q}$ to the (skew-symplectic) mass matrix 
\begin{equation} \label{eq:Mmat}
  M_{\phi,P} \,=\,  \begin{pmatrix} P^{[ij]} & B^i(\phi) \\ - B^i(\phi)^T & 0 \end{pmatrix} \ ,
\end{equation}
such that the dimension of the kernel $M^2_{\phi,P}$ determines the number of massless flavors $n_f$ at low energies, i.e., 
\begin{equation} \label{eq:nf}
   n_f \,=\, \dim \ker M_{\phi,P} \,=\, n + 1 - \operatorname{rk} M_{\phi,P} \ .
\end{equation}

To this end we recall the structure of two-dimensional $N=(2,2)$ non-Abelian $\operatorname{USp}(2k)$ gauge theories with $n_f$ fundamental  flavors as established by Hori~\cite{Hori:2011pd}. The theory is regular for $n_f$ odd, which means that the theory does not possess any non-compact Coulomb branches. Imposing regularity thus implies that $n$ must be even according to eq.~\eqref{eq:nf},
\begin{equation}
   n  \in 2\mathbb{Z} \ ,
\end{equation}
which we assume to be the case in the remainder of this section. For $n_f < 2k+1$ the infrared theory does not have any normalizable supersymmetric ground state, for $n_f = 2k+1$ the theory becomes a free conformal field theory of mesons, and for $n_f \ge 2k+3$ the theory is non-trivial and can be described by a dual interacting symplectic theory with gauge group $\operatorname{USp}(\frac12 (n_f-1)-k)$.  

As a consequence the theory localizes at those points of $\mathcal{V}$ with $n_f\ge 2k+1$ \cite{Hori:2006dk,Hori:2015priv}, which translates into the constraint
\begin{equation} \label{eq:rkcon}
  \operatorname{rk} M_{\phi,P} \le n-2k \ .
\end{equation}  
First, we focus on the degeneration locus, where the above inequality is saturated, i.e., $(\phi,P)\in\mathcal{V}$ such that $\operatorname{rk} M_{\phi,P}= n-2k$. At such points we have $n_f = 2k +1$ and the low energy effective theory becomes a theory of mesons $\pi_{[AB]}$, $A,B=1,\ldots,n_f$, arising from the $n_f$ fundamentals in the kernel of $M_{\phi,P}$, together with the fields $T_\alpha$ and $N_\mu$ for tangential and normal fluctuations with respect to the point $(\phi,P)$ in $\mathcal{V}$, i.e.,
\begin{equation}
\begin{aligned} 
   &(\phi,P) \,\longrightarrow\, (\phi,P) + \sum_\alpha T_\alpha (\delta\phi^{\alpha},\delta P^\alpha) + \sum_\mu N_\mu (\delta\phi^\mu,\delta P^\mu) \ , \\
   &\alpha = 1,\ldots,(m-1)+\frac12n(n-1)-k(2k+1) \ ,\quad  \mu=1,\ldots,k(2k+1) \ .
\end{aligned}   
\end{equation}   
Here the multiplicities $k(2k+1)$ and $(m-1)+\frac12n(n-1)-k(2k+1)$ account for the normal and tangential directions to the point $(\phi,P)$ at the degeneration locus $\operatorname{rk} M_{\phi,P}=n-2k$. Then the effective superpotential of the resulting theory becomes
\begin{equation}
\begin{aligned}
  W_\text{eff}(\pi,T,N) \,=\,  N_\mu  \operatorname{tr}\left( \pi \, M_{\delta\phi^\mu,\delta P^\mu}  \right) 
    &+\left(T_\alpha \delta P^\alpha +  N_\mu \delta P^\mu\right)^{[ij]} A^a_{[ij]} \phi_a  \\
    & \qquad\qquad + P^{[ij]} A^a_{[ij]}(T_\alpha \delta \phi^\alpha_a + N_\mu \delta \phi^\mu_a)\ ,
\end{aligned}    
\end{equation}
with the effective F-terms
\begin{equation}
\begin{aligned}
   &F_{\pi_{[AB]}} \,=\, N_\mu\,M_{\delta\phi^\mu,\delta P^\mu}^{[AB]} \ , \qquad
   F_{T_\alpha} \,=\,  (\delta P^\alpha A)^a \phi_a + P^{[ij]}(A\delta\phi)_{[ij]}^\alpha \ , \\
   &F_{N_\mu} \,=\,  \operatorname{tr}\left( \pi \, M_{\delta\phi^\mu,\delta P^\mu}  \right) +  (\delta P^\mu A)^a \phi_a + P^{[ij]}(A\delta\phi)_{[ij]}^\mu \ .
\end{aligned}
\end{equation}

First, we observe that at a generic point $(\phi,P)$ of the degeneration locus the mesons $\pi_{[AB]}$ --- arising from the massless eigenvectors of the mass matrix $M_{\phi,P}$ --- do not have a definite $U(1)_V$ R-charge because they are general linear combinations of bilinears built from the fundamentals $X_i$ and $Q$ of $U(1)_V$ R-charge $\mathfrak{q}$ and $2-3\mathfrak{q}$, respectively. At such points the F-term constraint $F_{T_\alpha}=0$ is generically not fulfilled, but as there are as many constraints as dimensions of the degeneration locus, there could be discrete solutions for $(\phi,P)$ on the degeneration component. We do not expect any normalizable supersymmetric ground states from such points. For specific \model s (with $n$ even) this expectation is indirectly confirmed in Section~\ref{sec:ZS2} by extracting the Witten index $(-1)^F$  --- or strictly speaking at least that the sum of all such supersymmetric discrete ground states do not contribute to the Witten index $(-1)^F$.

Therefore, we now look at a non-generic component of the degeneration locus with mesons of definite $U(1)_V$ R-charge. Let us consider a point $(\phi,P)$ on such a component that
\begin{equation} \label{eq:Bconst}
  B(\phi)\,=\,B(\delta\phi^\alpha)\,=\,0 \ . 
\end{equation}
Then in the vicinity of $(\phi,P)$ of the degeneration locus the $2k+1$ massless mesons are given by
\begin{equation}
  \tilde\phi_i \,=\, X_i \epsilon Q  \quad \text{with} \quad  \sum_j P^{[ij]}\tilde\phi_j = 0 \ ,
\end{equation}   
with the definite $U(1)_V$ R-charge $2-2\mathfrak{q}$. Generically, the F-term $F_{T_\alpha}=0$ together with eq.~\eqref{eq:Bconst} requires that $\phi=0$ and that all tangential fluctuations $(\delta \phi^\alpha,\delta P^\alpha)$ to the degeneration locus take the non-generic form $(0,\delta P^\alpha)$. Finally, the F-terms $F_{\pi_{[AB]}}$ set $N_\nu$ to zero, whereas the F-terms $F_{N_\mu}$ yields the constraint
\begin{equation} \label{eq:Hyperii}
  0 \,=\, A^{a}_{[ij]}P^{[ij]} + B^{ai} \tilde\phi_i \ .
\end{equation}
Thus, we find that the low energy theory of this degeneration locus yields a non-linear sigma model with target space variety
\begin{equation} \label{eq:Yvariety}
\begin{aligned}
   &\mathcal{Y}_{k,m,n} \,=\, \left\{ [\tilde\phi,P] \in \mathbb{P}( V^*, \Lambda^2 V) \, \middle| \, 
   \operatorname{rk} P \le n-2 k \ \text{and}\ \tilde\phi \in \ker P \right\}\cap \mathbb{P}(L^\perp) \ , \\
   &V^* \simeq \mathbb{C}^n \ , \quad \dim_\mathbb{C} L^\perp= \frac12n(n+1) - m \ , \quad n \text{ even} \ .
\end{aligned}
\end{equation}
The linear subspace $L^\perp$ of $V^*\oplus \Lambda^2 V$ is determined by eq.~\eqref{eq:Hyperii}, and it is dual to linear subspace $L$ of $V \oplus \Lambda^2 V^*$ with $\dim_\mathbb{C} L = m$ defined in eq.~\eqref{eq:L}, which means that
\begin{equation} \label{eq:Lperp}
  L^\perp \,=\, \left\{ v \in  V^*\oplus \Lambda^2 V \, \middle| \, v(L) = 0 \, \right\} \ .
\end{equation}  

Let us briefly comment on the locus where the inequality $\operatorname{rk} P\le n-2k$ is not saturated. At these points the infrared non-Abelian gauge theory becomes non-trivial, which can be described in terms of a dual interacting symplectic theory with gauge group $\operatorname{USp}(n-2k-\operatorname{rk} P)$ \cite{Hori:2011pd}. Generically, at those points the target space variety $\mathcal{Y}_{k,m,n}$ becomes singular, which indicates geometrically that the derivation of the semi-classical non-linear sigma model target space $\mathcal{Y}_{k,m,n}$ breaks down. However, similarly as in the phase $r\gg0$, we assume that the resulting target space variety $\mathcal{Y}_{k,m,n}$ furnishes the correct semi-classical non-linear sigma model description, as long as the variety $\mathcal{Y}_{k,m,n}$ remains smooth --- even if there are points where the rank condition is not saturated. Borrowing the analysis of Section~\ref{sec:Xphase}, the latter phenomenon occurs in the phase $r\ll 0$ for the models $SSSM_{1,(k+1)(2k+3)-m,2k+2}$ with smooth target space varieties $\mathcal{Y}_{1,(k+1)(2k+3)-m,2k+2}$.

Furthermore, we remark that as the target spaces of the two phases $r\gg 0$ and $r\ll 0$ are of the same type, we arrive at the identification 
\begin{equation} \label{eq:XYdual}
   \mathcal{Y}_{k,m,n} \simeq \mathcal{X}_{\tilde k,\tilde m,n}  \ , \quad
   \tilde k = \frac{n}2-k \ , \quad \tilde m = \frac12n(n+1)-m \ , \quad n\text{ even} \ ,
\end{equation}
with the linear subspaces $L$ of $\mathcal{X}_{\tilde k,\tilde m,n}$ and $L^\perp$ of $\mathcal{Y}_{k,m,n}$ of dimension $\dim_\mathbb{C} L = \dim_{\mathbb{C}} L^\perp=\tilde m$ canonically identified. However, we should stress that the two varieties are realized rather differently in the two phases of the \model. While the target spaces~$\mathcal{X}_{k,m,n}$ arise as a weakly coupled phase in the regime $r\gg 0$ of the \model, the variety $\mathcal{Y}_{k,m,n}$ emerges due to strong coupling phenomena with mesonic degrees of freedom in the low energy regime of the two-dimensional gauge theory. Physically, the identification~\eqref{eq:XYdual} already points towards a non-trivial duality among \model s. However, we postpone this aspect to Section~\ref{sec:duality} and first examine in detail the geometry of the discovered target spaces $\mathcal{X}_{k,m,n}$ and $\mathcal{Y}_{k,m,n}$ in Section~\ref{sec:Geom}.

\subsection{Geometric properties of target space varieties} \label{sec:Geom}
Our next task is to further study the target space varieties $\mathcal{X}_{k,m,n}$ and $\mathcal{Y}_{k,m,n}$. For ease of notation we focus on $\mathcal{X}_{k,m,n}$, as all results immediately carry over to $\mathcal{Y}_{k,m,n}$ due to the equivalence~\eqref{eq:XYdual}.

Similarly as in refs.~\cite{MR1827863,Jockers:2012zr} --- in order to deduce geometric properties of the target space varieties $\mathcal{X}_{k,m,n}$ --- we introduce the incidence correspondence
\begin{equation} \label{eq:inc}
  \widehat{\mathcal{X}}_{k,m,n} \,=\, \left\{ (x,p) \in \mathbb{P}^{m-1} \times \operatorname{Gr}(2k,V) \,\middle|\, \Lambda(x,p)=0 \ \text{and}\ B(x,p)=0 \right\} \ .
\end{equation}  
Here $\Lambda(x,p)$ is a generic section of the bundle $\mathcal{R}$
\begin{equation}
     \mathcal{R} \,=\, \frac{\mathcal{O}(1) \otimes \Lambda^2 V^* }{ \mathcal{O}(1) \otimes \Lambda^2 \mathcal{S}  } \ ,
\end{equation}
in terms of the hyperplane line bundle $\mathcal{O}(1)$ of the projective space $\mathbb{P}^{m-1}$, and the universal subbundle $\mathcal{S}$ of rank $2k$ of the Grassmannian $\operatorname{Gr}(2k,V)$. Furthermore, $B(x,p)$ is a generic section of the rank $2k$ bundle $\mathcal{B}$
\begin{equation}
  \mathcal{B} \,=\, \mathcal{O}(1) \otimes \mathcal{S}^* \ .
\end{equation}  

Thus the incidence correspondence realizes the variety $\widehat{\mathcal{X}}_{k,m,n}$ as the transverse zero locus of a generic section of the bundle $\mathcal{R}\oplus\mathcal{B}$. Then the total Chern class of the varieties $\widehat{\mathcal{X}}_{k,m,n}$ is given by 
\begin{equation} \label{eq:totalChern}
  c(\widehat{\mathcal{X}}_{k,m,n}) \,=\, \frac{c(\mathbb{P}^{m-1})\,c(\operatorname{Gr}(2k,V))}{c(\mathcal{R})\,c(\mathcal{B})} \ .
\end{equation}
This formula can explicitly be evaluated as the tangent sheaf $\mathcal{T}_{\operatorname{Gr}(2k,V)}$ is canonically identified with $\operatorname{Hom}(\mathcal{S},\mathcal{Q})$ of the universal subbundle~$\mathcal{S}$ and the quotient bundle $\mathcal{Q}$, and the Chern class of the Grassmannian factor becomes $c(\operatorname{Gr}(2k,V))=c(\mathcal{S}^*\otimes \mathcal{Q})$.

The intersection numbers $\kappa_{\widehat{\mathcal{X}}_{k,m,n}}$ of the variety $\widehat{\mathcal{X}}_{k,n,m}$ for the (induced) hyperplane divisor $H$ and Schubert divisor $\sigma_1$ of $\mathbb{P}^{m-1}\times\operatorname{Gr}(2k,V)$ are defined as
\begin{equation}
  \kappa_{\widehat{\mathcal{X}}_{k,m,n}}(H^r,\sigma_1^s) \,=\, \int_{\widehat{\mathcal{X}}_{k,m,n}} H^r  \wedge \sigma_1^s \ , \quad r+s = \dim_\mathbb{C} \widehat{\mathcal{X}}_{k,m,n} \ .
\end{equation}
They can explicitly be computed from the top Chern classes of the bundles $\mathcal{R}$ and $\mathcal{B}$ according to
\begin{equation} \label{eq:inter}
   \kappa_{\widehat{\mathcal{X}}_{k,m,n}}(H^r,\sigma_1^s) \,=\, \int_{\mathbb{P}^{m-1} \times \operatorname{Gr}(2k,V)} H^r\wedge \sigma_1^s 
   \wedge c_\text{top}(\mathcal{R})\wedge c_\text{top}(\mathcal{B}) \ .
\end{equation}   
These formulas for the incidence variety $\widehat{\mathcal{X}}_{k,m,n}$ form our basic tools to study topological properties of the varieties $\mathcal{X}_{k,m,n}$ of interest, as detailed in the following.

For varieties $\mathcal{X}_{k,m,n}$ that fulfill the constant rank condition~\eqref{eq:redrk}, it is straight forward to demonstrate the isomorphism
\begin{equation} \label{eq:iso}
   \mathcal{X}_{k,m,n} \,\simeq \,\widehat{\mathcal{X}}_{k,m,n} \quad \text{for} \quad k(2k-1)+\frac{n}2(n-4k+3)>m \ , 
\end{equation}
because the zeros of $\Lambda(x,p)$ yield the rank condition $\operatorname{rk} \tilde P = 2k$ in $\mathbb{P}^{m-1}$ in eq.~\eqref{eq:Xvariety} and assign for each point in $\mathbb{P}^{m-1}$ a unique $2k$ plane in the Grassmannian $\operatorname{Gr}(2k,V)$. The zeros of $B(x,p)$ give rise to the kernel condition $\phi\in \ker\tilde P$ in eq.~\eqref{eq:Xvariety}. Thus, we can readily use this isomorphism to determine the total Chern class for this class of varieties $\mathcal{X}_{k,m,n}$. In particular from eq.~\eqref{eq:totalChern} we find for the first Chern class
\begin{equation} \label{eq:c1}
\begin{aligned}
    c_1(\mathcal{X}_{k,m,n}) 
     &= c_1(\mathbb{P}^{m-1}) + c_1(\operatorname{Gr}(2k,V)) - c_1(\mathcal{R}) - c_1(\mathcal{B}) \\
     &= m H  + n \sigma_1 - \left( \binom{n}2H - \binom{2k}2H + (2k-1)\sigma_1 \right) -  \left( 2kH + \sigma_1 \right) \\
     &= \frac12 (n-3)(2k-n)H+(m-n)H \ .
\end{aligned}    
\end{equation}
In the last step we have used the relation $k H - \sigma_1=0$, which holds on the level of the cohomology for the pull-backs of $H$ and $\sigma_1$ to the variety $\mathcal{X}_{k,m,n}$ and can be argued for by similar techniques presented in refs.~\cite{MR974411,MR2560663}. Thus, we find that the axial current~\eqref{eq:axialcurr} is proportional to the first Chern class of the target space Calabi--Yau manifold. This is consistent as the first Chern class is the obstruction to a conserved axial current in a semi-classical non-linear sigma model description \cite{Witten:1991zz}.

Moreover, the degree of the variety is deduced from the intersection formula \eqref{eq:inter} according to
\begin{equation}
   \deg(\mathcal{X}_{k,m,n}) \,=\, \int_{\mathbb{P}^{m-1} \times \operatorname{Gr}(2k,V)} H^{\dim_\mathbb{C} \mathcal{X}_{k,m,n}} 
   \wedge c_\text{top}(\mathcal{R})\wedge c_\text{top}(\mathcal{B}) \ .
\end{equation}   

The isomorphism~\eqref{eq:iso} does not hold for the special class of varieties $\mathcal{X}_{k,m,2k+2}$ as the constant rank condition~\eqref{eq:redrk} is not fulfilled. The projection $\tilde\pi: \mathbb{P}^{m-1} \times \operatorname{Gr}(2k,V)\to \mathbb{P}^{m-1}$ induces the map $\pi: \widehat{\mathcal{X}}_{k,m,2k+2} \to \mathcal{X}_{k,m,2k+2}$, which restricts to an isomorphism on the Zariski open set $\operatorname{rk}\lambda=2k$. The preimage $\pi^{-1}(\mathcal{I})$ of the subvariety~$\mathcal{I}$ of reduced rank $\operatorname{rk}\lambda<2k$ yields the exceptional divisor $E_\mathcal{I}$, which at each point $p\in\mathcal{I}$ maps to a Schubert cycle in $\operatorname{Gr}(2k,V)$ of dimension greater than zero. Thus the map $\pi: \widehat{\mathcal{X}}_{k,m,2k+2} \to \mathcal{X}_{k,m,2k+2}$ realizes a resolution of $\mathcal{X}_{k,m,2k+2}$ along the ideal $\mathcal{I}$ with the exceptional locus $\pi^{-1}(\mathcal{I})$ in $\widehat{\mathcal{X}}_{k,m,2k+2}$, and the isomorphism~\eqref{eq:iso} generalizes to
\begin{equation} \label{eq:iso2}
   \widehat{\mathcal{X}}_{k,m,2k+2} \simeq \operatorname{Bl}_\mathcal{I}\mathcal{X}_{k,m,2k+2} \ , \quad \mathcal{I} = \left\{ \operatorname{rk} \lambda<2k \right\} \ ,   
\end{equation} 
with $\operatorname{Bl}_\mathcal{I}\mathcal{X}_{k,m,2k+2}$ the resolution along the ideal $\mathcal{I}$. Thus the varieties $\widehat{\mathcal{X}}_{k,m,2k+2}$ are not isomorphic but instead birational to $\mathcal{X}_{k,m,2k+2}$.

\subsubsection{Calabi--Yau threefolds $\mathcal{X}_{1,12,6}$ and $\mathcal{X}_{2,9,6}$} \label{sec:CYs}
The focus in this work is mainly on the conformal \model s with a smooth semi-classical Calabi--Yau threefold target space phase. Thus imposing $\dim_\mathbb{C} \mathcal{X}_{k,m,n}=3$ in eq.~\eqref{eq:dimX} and requiring $c_1(\mathcal{X}_{k,m,n})=0$, we find the two smooth Calabi--Yau threefold $\mathcal{X}_{1,12,6}$ and $\mathcal{X}_{2,9,6}$.

The first variety $\mathcal{X}_{1,12,6}$ fulfills the rank condition \eqref{eq:redrk} and yields a Calabi--Yau threefold with vanishing first Chern class according to eq.~\eqref{eq:c1}. Using the isomorphism~\eqref{eq:iso}, it is then straight forward to calculate its topological data\footnote{Here $h^{1,1}(\mathcal{X}_{1,12,6})=1$ follows from arguments along the line of refs.~\cite{MR974411,MR2560663}, which then allows us to infer $h^{2,1}(\mathcal{X}_{1,12,6})$ from the Euler characteristic $\chi(\mathcal{X}_{1,12,6})$.}
\begin{equation} \label{eq:DataCY1}
\begin{aligned}
  &\,h^{1,1}(\mathcal{X}_{1,12,6}) \,=\, 1 \ , &
  &h^{2,1}(\mathcal{X}_{1,12,6}) \,=\, 52 \ , &
  &\chi(\mathcal{X}_{1,12,6}) \,=\, -102 \ , \\
  &\deg(\mathcal{X}_{1,12,6}) \,=\, 33 \ , &
  &c_2(\mathcal{X}_{1,12,6}) \cdot H \,=\, 78 \ .
\end{aligned} 
\end{equation}

The second variety $\mathcal{X}_{2,9,6}$ is of the special type $\mathcal{X}_{k,m,2k+2}$. Calculating the Euler characteristic $\chi(\mathcal{X}_{1,9,6})=33$ of the reduced rank locus $\mathcal{X}_{1,9,6} \subset \mathcal{X}_{2,9,6}$, we find  $33$ (smooth) points $p_1, \ldots, p_{33}$ in $\mathcal{X}_{2,9,6}$ with reduced rank. Thus we find the isomorphism
\begin{equation}
    \widehat{\mathcal{X}}_{2,9,6} \simeq \operatorname{Bl}_{\{p_1,\ldots,p_{33}\}}\mathcal{X}_{2,9,6} \ .
\end{equation}    
The exceptional divisor $E = \bigcup_k \pi^{-1}(p_k)$ consists of $33$ exceptional $\mathbb{P}^2$. The divisor $E$ is linear equivalent to $2 H-\sigma_1$, because the intersection numbers are $(2H-\sigma_1)^3 = 33$ and $(2H-\sigma_1)^2 H = (2H-\sigma_1) H^2 =0$. Furthermore, we find from eq.~\eqref{eq:totalChern} for the first Chern class
\begin{equation}
  c_1(\widehat{\mathcal{X}}_{2,9,6}) = -4 H+ 2 \sigma_1 = -2 E \ .
\end{equation}  
Due to the general relation $c_1(\widehat{\mathcal{X}}_{2,9,6}) = \pi^*c_1(\mathcal{X}_{2,9,6}) - 2 E$ for threefold varieties blown up at smooth points, we infer that $c_1(\mathcal{X}_{2,9,6})=0$ and $\mathcal{X}_{2,9,6}$ is a Calabi--Yau threefold. Finally, with $\chi(\widehat{\mathcal{X}}_{2,9,6}) = \chi(\mathcal{X}_{2,9,6}) + 33 (1 -\chi(\mathbb{P}^2))$ and $\deg(\mathcal{X}_{2,9,6})= \kappa_{\widehat{\mathcal{X}}_{2,9,6}}(H^3)$ and applying formulas~\eqref{eq:totalChern} and \eqref{eq:inter}, we arrive at the topological data for the Calabi--Yau threefold $\mathcal{X}_{2,9,6}$\footnote{Again, $h^{1,1}(\mathcal{X}_{2,9,6})=1$ determines $h^{2,1}(\mathcal{X}_{2,9,6})$ from $\chi(\mathcal{X}_{2,9,6})$.}
\begin{equation}
\begin{aligned}
  &\,h^{1,1}(\mathcal{X}_{2,9,6}) \,=\, 1 \ , &
  &h^{2,1}(\mathcal{X}_{2,9,6}) \,=\, 52 \ , &
  &\chi(\mathcal{X}_{2,9,6}) \,=\, -102 \ , \\
  &\deg(\mathcal{X}_{2,9,6}) \,=\, 21 \ , &
  &c_2(\mathcal{X}_{2,9,6}) \cdot H \,=\, 66 \ .
\end{aligned} \label{eq:DataCY2}
\end{equation}

It is intriguing to observe that the Hodge numbers of $\mathcal{X}_{1,12,6}$ and $\mathcal{X}_{2,9,6}$ (and thus their Euler characteristic) agree while their degrees are distinct. This gives a first hint that these two Calabi--Yau manifolds  are related by a non-trivial derived equivalence. Using the R\o{}dland's argument \cite{MR1775415}, we observe that the two Calabi--Yau threefolds $\mathcal{X}_{1,12,6}$ and $\mathcal{X}_{2,9,6}$ cannot be birationally equivalent. If they were birational their hyperplane classes $H$ would be related by a rational factor. But since the ratio of the degrees are $\frac{7}{11}$ --- arising from the ratio of third powers of the two hyperplane classes --- is not a third power of a rational number, the two hyperplane classes are not related by a rational multiple.

From a different geometric construction, Miura analyzes the Calabi--Yau threefold $\mathcal{X}_{1,12,6}$ in ref.~\cite{Miura:2013arxiv}. Furthermore, by means of mirror symmetry he conjectures that in the quantum K\"ahler moduli space of $\mathcal{X}_{1,12,6}$ there emerges another Calabi--Yau threefold, whose invariants match with $\mathcal{X}_{2,9,6}$.\footnote{In refs.~\cite{MR2282973,Almkvist:2005arxiv,vanStraten:2012db,Hofmann:2013PhDThesis} certain Picard--Fuchs operators for one parameter Calabi--Yau threefolds are classified, and their geometric invariants are calculated at points of maximal unipotent monodromy~\cite{vanStraten:2012db,Hofmann:2013PhDThesis} In this way --- without the knowledge of the underlying Calabi--Yau variety $\mathcal{X}_{1,12,6}$ --- the invariants~\eqref{eq:DataCY1} are determined in refs.~\cite{vanStraten:2012db,Hofmann:2013PhDThesis}.} The connection between Miura's construction and the geometric realization of the two varieties $\mathcal{X}_{1,12,6}$ and $\mathcal{X}_{2,9,6}$ of the underlying \model s is mathematically demonstrated in refs.~\cite{Galkin:2014Talk,Galkin:2015InPrep}. In the remainder of this paper we give strong evidence that the pair of Calabi--Yau threefolds $\mathcal{X}_{1,12,6}$ and $\mathcal{X}_{2,9,6}$ are actually related by duality. This provides for a physics argument that the Calabi--Yau manifolds $\mathcal{X}_{1,12,6}$ and $\mathcal{X}_{2,9,6}$ are derived equivalent.

\subsection{Dualities among \model s} \label{sec:duality}
The observed geometric correspondence~\eqref{eq:XYdual} among low energy effective target space varieties $\mathcal{X}_{k,m,n}$ and $\mathcal{Y}_{k,m,n}$ suggests a duality among the associated \model s, i.e.,
\begin{equation} \label{eq:SSSMdual}
     SSSM_{k,m,n} \,\simeq\, SSSM_{\tilde k,\tilde m,n} \ \text{for }n \text{ even}\,,\
      \tilde k=\frac{n}2 - k, \ \tilde m= \frac12n(n+1)-m \ .
\end{equation}
Thus we propose that such pairs of \model s --- which clearly have distinct degrees of freedom as ultraviolet $N=(2,2)$ gauged linear sigma models --- become two equivalent two-dimensional non-linear $N=(2,2)$ effective field theories at low energies.

While the match of effective target space varieties~\eqref{eq:XYdual} is already a strong indication, let us now collect further evidence for our proposal. Similarly, as for  four-dimensional Seiberg dual gauge theories with four supercharges \cite{Seiberg:1994bz}, we check the 't~Hooft anomaly matching conditions for global symmetry currents of the \model s. We first compare the anomaly~\eqref{eq:axialcurr} of the $U(1)_A$ axial current. Inserting the identification \eqref{eq:SSSMdual} for the labels of the dual \model s, we find with eq.~\eqref{eq:axialcurr}
\begin{equation}
   \partial_\mu j^\mu_{A, SSSM_{k,m,n} } \,=\, - \partial_\mu j^\mu_{A, SSSM_{\tilde k, \tilde m,n} } \ .
\end{equation}
Thus, the 't~Hooft anomaly matching condition for the $U(1)_A$ axial current is satisfied up to a relative sign in the $U(1)$ gauge charges between dual \model s. An overall flip in sign of the $U(1)$ charges simultaneously reverses the sign of the associated Fayet--Iliopoulos parameter $r$. This change in sign is already implicit in the geometric correspondence~\eqref{eq:XYdual}, as it relates \model{} phases at $r\gg 0$ and $r\ll 0$, respectively.

Furthermore, due to eq.~\eqref{eq:mixed} the mixed axial--vector anomaly~$\Gamma_{A-V}$ obeys
\begin{equation}
\begin{aligned}
   \Gamma_{A-V, SSSM_{k,m,n}}(\mathfrak{q}) \,=\, \Gamma_{A-V, SSSM_{\tilde k,\tilde m,n}}(\tilde{\mathfrak{q}}) \ , \qquad \tilde{\mathfrak{q}}\,=\,1-\mathfrak{q} \ . 
\end{aligned}   
\end{equation}
Since this mixed anomaly depends in general on an overall ambiguity~$\mathfrak{q}$ in the assignment of $U(1)_V$ R-charges in the spectrum of Table~\ref{tb:spec1}, the anomaly matching is not really a consistency check but instead fixes the relative assignment of $U(1)_V$ charges among dual \model s. If, however, the axial anomaly vanishes, the mixed axial--vector anomaly becomes independent of $\mathfrak{q}$ and calculates the central charge of the infrared $N=(2,2)$ superconformal field theory according to \eqref{eq:central}. In this case we find 
\begin{equation}
  c_{SSSM_{k,m,n}} = c_{SSSM_{\tilde k,\tilde m,n}} \quad \text{for} \quad \partial_\mu j^\mu_A=0 \ .
\end{equation}
For this class of \model s with vanishing $U(1)_A$ axial anomaly a further non-trivial 't~Hooft anomaly matching condition is fulfilled.

In summary, the analyzed 't~Hooft anomaly matching conditions allow us now to refine our duality proposal for \model s as follows
\begin{equation} \label{eq:dualSSSM}
\begin{aligned}
   &SSSM_{k,m,n}(r,\mathfrak{q}) \, \simeq \, SSSM_{\tilde k,\tilde m,n}(\tilde r,\tilde{\mathfrak{q}})\quad \text{for}\quad n\text{ even} \ , \\
   &\tilde k=\frac{n}2 - k, \quad \tilde m= \frac12n(n+1)-m, \quad \tilde r=-r, \quad \tilde{\mathfrak{q}}\,=1-\mathfrak{q} \ .
\end{aligned}
\end{equation}   
So far this is really a duality relation between families of \model s. While the duality map for the Fayet--Iliopoulos parameter~$r$ is spelled out explicitly, the \model s also depend on the coupling constants in the superpotential~\eqref{eq:W}. In order to match the coupling constants in the superpotential as well, the linear subspace $L_{k,m,n}\equiv L$ for the model $SSSM_{k,m,n}$ of eq.~\eqref{eq:L} must be identified with the linear subspace $L^\perp_{\tilde k,\tilde m,\tilde n}\equiv L^\perp$ for the model $SSSM_{\tilde k,\tilde m,\tilde n}$ as specified in eq.~\eqref{eq:Lperp}. As both $L$ and $L^\perp$ are entirely determined by the couplings $A_{[ij]}^a$ and $B^{ia}$ in the superpotential. The duality proposal is therefore further refined by identifying these couplings according to
\begin{equation} \label{eq:csdual}
    V \oplus \Lambda^2V^* \,\supset\, L_{k,m,n}(A,B) \,\simeq \, L^\perp_{\tilde k,\tilde m,\tilde n}(\tilde A,\tilde B) \,\subset\, \tilde V^* \oplus \Lambda^2\tilde V \ , \quad
    V \oplus \Lambda^2V^* \,\simeq\, \tilde V^* \oplus \Lambda^2\tilde V \ .
\end{equation}
Note that $\dim_\mathbb{C} L_{k,m,n}=\dim_\mathbb{C} L^\perp_{\tilde k,\tilde m,\tilde n}$, and the identification follows from comparing the phase structure for $r\gg0$ and $r\ll 0$ of the \model s together with the geometric identification~\eqref{eq:XYdual}. This relationship does not uniquely fix the correspondence between the couplings $A_{[ij]}^a$ and $B^{ia}$ of $SSSM_{k,m,n}$ and $\tilde A_{[ij]}^a$ and $\tilde B^{ia}$ of $SSSM_{\tilde k,\tilde m,\tilde n}$. However, such ambiguities are not physically relevant as they can be absorbed into field redefinitions of the chiral multiplets. 

Up to here the discussed duality is based upon the identification of target spaces. However, the proposed duality falls into the class of two-dimensional Hori dualities \cite{Hori:2011pd}, which can be argued for in the following way \cite{Hori:2015priv}.\footnote{We would like to thank Kentaro Hori for sharing his duality argument with us.} The model $SSSM_{k,n,m}$ comes with $n_f=n+1$ fundamentals of $\operatorname{USp}(2k)$, which for odd $n_f$ and $n_f \ge 2k+3$ --- that is to say for even $n$ with $n\ge 2k+2$ --- is dual to $\operatorname{USp}(2\tilde k)$ for $\tilde k=\frac{n}2 - k$ and with $\binom{n_f}2$ singlets and $n_f$ fundamentals \cite{Hori:2011pd}. The singlets are the mesons of the original gauge theory 
\begin{equation}
  \tilde P_{[ij]} \,=\, X_i \epsilon X_j \ , \qquad \tilde\phi_i \,=\, Q \epsilon X_i \ .
\end{equation}  
They couple to the fundamentals $\tilde X^i$ and $\tilde Q$ of $\operatorname{USp}(2\tilde k)$ through the superpotential $W=\operatorname{tr}\left( \tilde P \cdot \tilde X^T\tilde\epsilon\tilde X\right) + \tilde\phi \cdot \tilde Q^T\tilde\epsilon \tilde X$, where  $\tilde\epsilon$ is the epsilon tensor of the dual gauge group $\operatorname{USp}(2\tilde k)$ and the trace is taken over flavor indices of the fundamentals $\tilde X^i$. Thus in summary, we arrive at the dual theory with the chiral fields $P^{[ij]}$, $\phi_a$, $\tilde P_{[ij]}$, $\tilde \phi_i$, $\tilde X^i$ and $\tilde Q$ transforming respectively in the representations $\mathbf{1}_{-2}$, $\mathbf{1}_{+2}$, $\mathbf{1}_{+2}$, $\mathbf{1}_{-2}$, $\mathbf{2\tilde k}_{-1}$ and $\mathbf{2\tilde k}_{+3}$ of the dual gauge group $\frac{U(1)\times\operatorname{USp}(2\tilde k)}{\mathbb{Z}_2}$. The dual superpotential reads
\begin{equation}
  W \,=\, \operatorname{tr}\left[ P \left(A(\phi) + \tilde P \right) \right] + B(\phi)  \cdot \tilde\phi  
  +\operatorname{tr}\left( \tilde P \cdot \tilde X^T\tilde\epsilon\tilde X\right) + \tilde\phi  \cdot \tilde Q^T\tilde\epsilon \tilde X \ .
\end{equation}  
Following ref.~\cite{Hori:2013gga} this theory can be further simplified by integrating out the chiral multiplets $\phi_a$ of multiplicity $m$, which yields the F-term constraints
\begin{equation} \label{eq:intout}
  F_{\phi_a} \,=\, \operatorname{tr} \left( P \cdot A^a \right) + B^a \cdot \tilde \phi \,=\, 0 \ .
\end{equation}  
Generically, there are $\tilde m = \binom{n+1}2-m$ solutions to these conditions, which we parametrize by the chiral fields $\tilde\phi^a$, $a=1,\ldots,\tilde m$, as
\begin{equation}
  P^{[ij]}=\tilde A^{[ij]}_a \tilde\phi^a \ , \qquad \tilde\phi_i=\tilde B_{ia}\tilde\phi^a \ .
\end{equation} 
such that eq.~\eqref{eq:intout} is fulfilled. Altogether, we obtain the simplified dual theory with gauge group $\frac{U(1)\times\operatorname{USp}(2\tilde k)}{\mathbb{Z}_2}$ and the chiral fields $\tilde P_{[ij]}$, $\tilde \phi^a$, $\tilde X^i$ and $\tilde Q$, which interact through the superpotential 
\begin{equation}
    W\,=\, \operatorname{tr}\left[ \tilde P \left(\tilde A(\tilde \phi) + \tilde X^T \tilde\epsilon \tilde X \right) \right] + \tilde B(\tilde\phi)  \cdot \tilde Q^T\tilde\epsilon\tilde X  \ .
\end{equation}
Up to a change of sign of the $U(1)$ gauge charges, this is just the \model{} with the chiral spectrum as in Table~\ref{tb:spec1} and the superpotential~\eqref{eq:W} for the integers $(\tilde k,\tilde m,n)$. This agrees with the proposed duality \eqref{eq:dualSSSM} for even~$n$.

Next we turn to explicit \model s that give rise to $N=(2,2)$ superconformal field theories in the infrared. In this note we mainly focus on $N=(2,2)$ superconformal field theories in the context of compactifications of type~II string theories. The \model s with central charges three, six, nine and twelve are of particular interest, as such theories describe the internal worldsheet theories of type~II string compactifications to eight, six, four and two space-time dimensions, respectively. The possible models in this range of central charges --- which possess a semi-classical geometric non-linear sigma model phase --- are summarized in Table~\ref{tb:models}.
\begin{table}[t]
\centering
\hfil\vbox{
\offinterlineskip
\tabskip=0pt
\halign{\vrule height2.2ex depth1ex width1pt~#~\hfil\vrule&\hfil~#~\hfil\vrule&~#~\hfil\vrule height2.2ex depth1ex width 1pt\cr
\noalign{\hrule height 1pt}
\hfil sigma model & IR central charge & smooth $r\gg 0$ target space phase\cr
\noalign{\hrule height 1pt}
$SSSM_{1,5,4}$ & $\frac{c}3=1$ & degree $5$ $T^2$ curve\cr
\noalign{\hrule}
$SSSM_{1,8,5}$ & $\frac{c}3=2$ & degree $12$ K3 surface\cr
\noalign{\hrule}
$SSSM_{1,12,6}$ & $\frac{c}3=3$ & degree $33$ CY 3-fold ($\chi=-102$)\cr
$SSSM_{2,9,6}$ & $\frac{c}3=3$ & degree $21$ CY 3-fold ($\chi=-102$) \cr
\noalign{\hrule}
$SSSM_{1,17,7}$ & $\frac{c}3=4$ & degree $98$ CY 4-fold ($\chi=672$)\cr
\noalign{\hrule height 1pt}
}}\hfil
\caption{Listed are the \model s ($SSSM_{k,m,n}$) with vanishing axial $U(1)_A$ anomaly, infrared central charges $\frac{c}{3}=1,2,3,4$, and a smooth target space phase. They are associated to supersymmetric type~II string compactifications to $\left(10-2\cdot\frac{c}{3}\right)$-spacetime dimensions. The last column lists the geometric target space in the semi-classical non-linear sigma model regime $r\gg 0$.}\label{tb:models}
\end{table}

In particular the models $SSSM_{1,5,4}$, $SSSM_{1,12,6}$ and $SSSM_{2,9,6}$ with even~$n$ possess dual \model{} descriptions according to our proposal~\eqref{eq:dualSSSM}. The model $SSSM_{1,5,4}$ is self-dual with two equivalent geometric $T^2$ phases for $r\gg 0$ and $r\ll 0$, namely we find the self-duality relationship
\begin{equation} \label{eq:dualT2s}
   SSSM_{1,5,4}(r,\mathfrak{q}) \,\simeq\, SSSM_{1,5,4}(-r,1-\mathfrak{q}) \ .
\end{equation}
The other two models, $SSSM_{1,12,6}$ and $SSSM_{2,9,6}$, give rise to dual families of $N=(2,2)$ superconformal field theories with central charge nine, that is to say
\begin{equation} \label{eq:dualpair}
  SSSM_{1,12,6}(r,\mathfrak{q}) \,\simeq\, SSSM_{2,9,6}(-r,1-\mathfrak{q}) \ .
\end{equation}
Moreover, these models possess two geometric non-linear sigma model phases with Calabi--Yau threefold target spaces $\mathcal{X}_{1,12,6}$ and $\mathcal{Y}_{1,12,6}$, and $\mathcal{Y}_{2,9,6}$ and $\mathcal{X}_{2,9,6}$ for $r\gg 0$ and $r\ll 0$, respectively, as discussed in Section~\ref{sec:CYs}. They obey the geometric equivalences $\mathcal{X}_{1,12,6}\simeq\mathcal{Y}_{2,9,6}$ and $\mathcal{X}_{2,9,6}\simeq\mathcal{Y}_{1,12,6}$ according to eq.~\eqref{eq:XYdual}.

Finally, let us turn to the model $SSSM_{1,8,5}$ in Table~\ref{tb:models}. While a general and detailed analysis of the non-linear sigma model phase $r \ll 0$ for \model s with $n$ odd is beyond the scope of this work, we arrive at a duality proposal for the specific model $SSSM_{1,8,5}$. The degree twelve K3~surface  $\mathcal{X}_{1,8,5}$ of geometric phase for $r\gg0$ is actually well-known in the mathematical literature \cite{MR1714828,MR2047679,Hosono:2014ty}.\footnote{We would like to thank Shinobu Hosono for pointing out and explaining to us the geometric properties and the relevance of the degree twelve K3~surface as discussed in the following.} There is a degree twelve K3 surfaces $\mathcal{X}_{\mathbb{S}_5}$, which is a linear section of codimension eight of either one of the two isomorphic connected components $\operatorname{OGr}^\pm(5,10)$  of the ten-dimensional orthogonal Grassmannian $\operatorname{OGr}(5,10)=O(10,\mathbb{C})/(O(5,\mathbb{C})\times O(5,\mathbb{C}))$ \cite{MR1714828,Hosono:2014ty}. The two isomorphic components $\operatorname{OGr}^\pm(5,10)$ are known as the ten-dimensional spinor varieties~$\mathbb{S}_5$ embedded in $\mathbb{P}(V)$ with $V\simeq \mathbb{C}^{16}$. The resulting degree twelve K3 surface is therefore given by
\begin{equation} \label{eq:K31}
  \mathcal{X}_{\mathbb{S}_5} \,=\, \mathbb{S}_5 \cap \mathbb{P}(L) \,\subset\, \mathbb{P}(V) \ , \quad V \,\simeq\, \mathbb{C}^{16} \ , \quad
   L \simeq\mathbb{C}^8 \ .
\end{equation}
In addition, there is a dual degree twelve K3 surface $\mathcal{Y}_{\mathbb{S}_5^*}$ based on the projective dual variety $\mathbb{S}_5^*\subset \mathbb{P}(V^*)$ {}\footnote{For details on projective dual varieties we refer the reader to the interesting review~\cite{MR2027446}.}
\begin{equation} \label{eq:K32}
  \mathcal{Y}_{\mathbb{S}_5^*} \,=\,  \mathbb{S}_5^* \cap \mathbb{P}(L^\perp) \,\subset\, \mathbb{P}(V^*) \ , 
\end{equation}   
with the eight-dimensional linear section $L^\perp \subset V^*$ orthogonal to $L \subset V$. Since $\mathbb{S}_5^*$ is isomorphic to $\mathbb{S}_5$, the K3 surfaces $\mathcal{X}_{\mathbb{S}_5}$ and $\mathcal{Y}_{\mathbb{S}_5^*}$ are members of the same family of degree twelve polarized K3 surfaces. By embedding the spinor variety $\mathbb{S}_{5}$ into the projective space $\mathbb{P}^{15}$ with the Pl\"ucker embedding, an isomorphism between the polarized K3 surfaces $\mathcal{X}_{1,8,5}$ and $\mathcal{X}_{\mathbb{S}_5}$ is demonstrated for instance in ref.~\cite{MR2071808}. For us the important result is that both polarized K3 surfaces $\mathcal{X}_{\mathbb{S}_5}$ and $\mathcal{Y}_{\mathbb{S}_5^*}$ appear as phases in the same quantum K\"ahler moduli space \cite{MR2047679}. Therefore, it is natural to expect that the second varieties $\mathcal{Y}_{\mathbb{S}_5^*}$ is identified with a geometric target space $\mathcal{Y}_{1,8,5}$ of $SSSM_{1,8,5}$ in the phase $r\ll 0$. With both varieties $\mathcal{X}_{1,8,5}$ and $\mathcal{Y}_{1,8,5}$ as members of the same family of degree twelve polarized K3 surfaces, we are led to the self-duality proposal
\begin{equation} \label{eq:K3sd}
     SSSM_{1,8,5}(r,\mathfrak{q}) \,\simeq\, SSSM_{1,8,5}(-r,1-\mathfrak{q}) \ .
\end{equation}
Again a suitable identification of superpotential couplings is assumed to accommodate for the relationship between the complex structure moduli of the two polarized K3~surfaces~\eqref{eq:K31} and \eqref{eq:K32}.

In the remainder of this work the dual pair $SSSM_{1,12,6}$ and $SSSM_{2,9,6}$ and the two self-dual models $SSSM_{1,5,4}$ and $SSSM_{1,8,5}$ of the \model s listed in Table~\ref{tb:models} are our key player. In the following section we present additional evidence in support of our duality proposals.

\section{The two sphere partition function}\label{sec:ZS2}
In the previous section we have proposed a remarkable duality for the infrared theories of \model s. As discussed for the models $SSSM_{1,12,6}$ and $SSSM_{2,9,6}$, the duality implies a correspondence between Calabi--Yau threefold target spaces spelled out in eq.~\eqref{eq:XYdual}. While our duality proposal has passed some non-trivial checks --- such as the agreement of target space geometries of dual pairs in a semi-classical low energy analysis and the realization of 't~Hooft anomaly matching conditions, the aim of this section is to study the two sphere partition functions $Z_{S^2}$ of the \model s in Table~\ref{tb:models}. For four-dimensional $N=1$ gauge theories, the comparison of partition functions provides an impressive consistency check for Seiberg duality at the quantum level~\cite{Romelsberger:2005eg,Dolan:2008qi}. Analogously, we here use the two sphere partition function as a similarly strong argument in support of our duality proposal.

Using novel localization techniques for partition functions of supersymmetric gauge theories on curved spaces \cite{Pestun:2007rz,Festuccia:2011ws}, the two sphere partition function $Z_{S^2}$ of two-dimensional $N=(2,2)$ gauge theories is explicitly calculated in refs.~\cite{Doroud:2012xw,Benini:2012ui}\footnote{Note that the partition function stated in refs.~\cite{Doroud:2012xw,Benini:2012ui} is more general, as it includes twisted mass parameters for twisted chiral multiplets. Such twisted masses, however, do not play a role here.}
\begin{equation} \label{eq:ZS2}
  Z_{S^2}(\boldsymbol{r},\boldsymbol{\theta}) 
  \,=\, \frac{1}{(2\pi)^{\dim\mathfrak{h}}|\mathcal{W}|} \sum_{\mathfrak{m} \in\Lambda_\mathfrak{m}} \int_{\mathfrak{h}}
  \!\!d^{\dim\mathfrak{h}}\boldsymbol{\sigma}\,
  Z_G(\mathfrak{m},\boldsymbol{\sigma}) 
  Z_\text{matter}(\mathfrak{m},\boldsymbol{\sigma}) 
  Z_\text{cl}(\mathfrak{m},\boldsymbol{\sigma},\boldsymbol{r},\boldsymbol{\theta}) \ .
\end{equation}
Here $\mathfrak{h}$ is the Cartan subalgebra of the Lie algebra~$\mathfrak{g}$ --- which decomposes into the Abelian Lie algebra~$\mathfrak{u}(1)^\ell$ and the simple Lie algebra~$\mathfrak{g}_s$ according to $\mathfrak{g} =\mathfrak{u}(1)^\ell \oplus \mathfrak{g}_s$ --- and $|\mathcal{W}|$ is the cardinality of the Weyl group~$\mathcal{W}$ of the gauge group $G$. The sum of $\mathfrak{m}$ is taken over the magnetic charge lattice $\Lambda_\mathfrak{m}\subset\mathfrak{h}$ of the gauge group~$G$ \cite{Goddard:1976qe}, which is the cocharacter lattice of the Cartan torus of the gauge group $G$,\footnote{The coweight lattice of the Lie algebra $\mathfrak{g}$ is a sublattice of the cocharacter lattice~$\Lambda_\mathfrak{m}$ of the Lie group $G$. The latter is sensitive to the global structure of $G$. For instance, the coweight lattice of $\mathfrak{so}(n)$ is an index two sublattice of the cocharacter lattice of $\operatorname{SO}(n)$, but coincides with the cocharacter lattice of its double cover $\operatorname{Spin}(n)$. Similarly, the coweight lattice the gauge group $G$ of the \model{} is an index two sublattice of its cocharacter lattice due to the $\mathbb{Z}_2$ quotient in $G$.}while the integral of $\boldsymbol{\sigma}$ is performed over the Cartan subalgebra  $\mathfrak{h}$. The vector-valued parameters $\boldsymbol{r}$ and $\boldsymbol{\theta}$ of dimension $\ell$ are the Fayet--Iliopoulos terms and the theta angles of the Abelian gauge group factors, formally residing in the annihilator of the Lie bracket, i.e., $\boldsymbol{r},\boldsymbol{\theta} \in \operatorname{Ann}([\mathfrak{g},\mathfrak{g}])\simeq\mathfrak{u}(1)^{*\,\ell} \subset\mathfrak{g}^*$. Finally, with the canonical pairing~$\langle\,\cdot\,,\,\cdot\,\rangle$ between the weight and the coroot lattice the factors in the integrand are given by\footnote{Compared to refs.~\cite{Doroud:2012xw,Benini:2012ui}, the additional factor $(-1)^{\langle \alpha,\mathfrak{m} \rangle}$ in $Z_G$ has been established in ref.~\cite{Hori:2013ika}.}
\begin{equation}
\begin{aligned}
  Z_G(\mathfrak{m},\boldsymbol{\sigma})\,&=\, 
    \prod_{\alpha \in\Delta^+} (-1)^{\langle \alpha,\mathfrak{m} \rangle}\left( \frac14 \langle \alpha,\mathfrak{m} \rangle^2+ \langle \alpha,\boldsymbol{\sigma} \rangle^2\right) \ , \\
  Z_\text{matter}(\mathfrak{m},\boldsymbol{\sigma})\,&=\,
    \prod_{\rho_j\in\operatorname{Irrep}(\rho)} \prod_{\beta\in w(\rho_j)} 
    \frac{\Gamma(\tfrac12 \mathfrak{q}_j-i \langle\beta,\boldsymbol{\sigma}\rangle-\tfrac12\langle\beta,\mathfrak{m}\rangle)}
    {\Gamma(1-\tfrac12 \mathfrak{q}_j+i \langle\beta,\boldsymbol{\sigma}\rangle-\tfrac12\langle\beta,\mathfrak{m}\rangle)} \ , \\[1ex]
   Z_\text{cl}(\mathfrak{m},\boldsymbol{\sigma},\boldsymbol{r},\boldsymbol{\theta})\,&=\, 
     e^{-4\pi i \langle \boldsymbol{r},\boldsymbol{\sigma}\rangle - i \langle \boldsymbol{\theta},\mathfrak{m}\rangle} \ .
\end{aligned} \label{eq:ZS2comp}
\end{equation}
The gauge group contribution~$Z_G$ arises as a finite product over the positive roots $\Delta^+$ of the Lie algebra~$\mathfrak{g}$. The matter term $Z_\text{matter}$ is a nested finite product, where the outer product is taken over all irreducible representations $\operatorname{Irrep}(\rho)$ of the chiral matter spectrum and the inner product is over the weights~$w(\rho_j)$ of a given irreducible representation $\rho_j$ with $U(1)_V$ R-charge $\mathfrak{q}_j$.\footnote{For the chosen contour $\mathfrak{h}$ of the integral of the two sphere partition function, it is necessary to choose the $U(1)_V$ R-charges of all chiral multiplets to be bigger than zero, i.e., $\mathfrak{q}_j>0$ for all $j$.}  Finally, the classical term $Z_\text{cl}$ depends on the Fayet--Iliopoulos parameters $\boldsymbol{r}$ and the theta angles $\boldsymbol{\theta}$ of the two-dimensional $N=(2,2)$ gauge theory.  

Note that $Z_{S^2}$ is a function of the Fayet--Iliopoulos parameters and the theta angles of the $N=(2,2)$ gauge theory, but not of the superpotential couplings. In refs.~\cite{Jockers:2012dk,Gomis:2012wy,Halverson:2013eua,Gerchkovitz:2014gta} it is argued that $Z_{S^2}$ calculates the sign-reversed exponentiated K\"ahler potential of the Zamolodchikov metric for the marginal operators in the $(c,a)$ ring of the infrared $N=(2,2)$ superconformal field theory. As a consequence, in phases with a non-linear sigma model interpretation it determines the exact sign-reversed exponentiated K\"ahler potential of the quantum K\"ahler moduli space of the Calabi--Yau target space \cite{Jockers:2012dk}, which for a Calabi--Yau threefold $\mathcal{X}$ with a single K\"ahler modulus reads\footnote{As $h^{1,1}(\mathcal{X}_{1,12,6})=h^{1,1}(\mathcal{X}_{2,9,6})=1$, we only spell out the quantum K\"ahler potential for Calabi--Yau threefolds with a single K\"ahler modulus. For the general case we refer the reader to ref.~\cite{Jockers:2012dk}.}
\begin{equation}
\begin{aligned}
     e^{-K(t)} \,=\, -\frac{i}{6}& \operatorname{deg}(\mathcal{X}) (t-\bar t)^3 + \frac{\zeta(3)}{4\pi^3}\chi(\mathcal{X}) \\
      &- \sum_{d=1}^{+\infty} N_d(\mathcal{X}) \left( \frac1{4\pi^3} \operatorname{Li}_3(e^{2\pi i t d}) - \frac{i}{4\pi^2} (t-\bar t) \operatorname{Li}_2(e^{2\pi i t d}) 
      + \text{c.c.} \right) \ .
\end{aligned} \label{eq:expK}
\end{equation}
Here $\operatorname{Li}_k(x)=\sum_{n=1}^{+\infty}\frac{x^n}{n^k}$ is the polylogarithm, $t$ is the complexified K\"ahler coordinate, and $\operatorname{deg}(\mathcal{X})$, $\chi(\mathcal{X})$, and $N_d(\mathcal{X})$ are the degree, the Euler characteristic, and the degree $d$ integral genus zero Gromov--Witten invariants of the Calabi--Yau threefold $\mathcal{X}$ respectively. The real function~$K(t)$ is the exact K\"ahler potential of the Weil--Petersson metric of the quantum K\"ahler moduli space, which is expressed in terms of the flat quantum K\"ahler coordinates $t$ of the Calabi--Yau threefold $\mathcal{X}$ that is calculated by the partition function correspondence and the IR--UV map \cite{Jockers:2012dk}
\begin{equation}\label{eq:ZtoK}
   Z_{S^2}(r,\theta) \,=\, e^{-K(t)}  \ , \quad q \,=\, e^{- r + i \theta} \ , \quad q\,=\,q(t) \ .
\end{equation}   
That is to say, in matching the two sphere partition function $Z_{S^2}$ to the form~\eqref{eq:expK} both the IR--UV map $q(t)$ and the K\"ahler potential~$K(t)$ of a geometric phase are unambiguously determined, as explained in detail in ref.~\cite{Jockers:2012dk}.

In the following we calculate the two sphere partition function~$Z_{S^2}$ for both \model s $SSSM_{1,12,6}$ and $SSSM_{2,9,6}$. In particular, we determine explicitly the quantum K\"ahler potential~\eqref{eq:expK} in the geometric phases of both \model s, so as to support the $N=(2,2)$ gauge theory duality~\eqref{eq:dualpair} and the associated geometric correspondences~$\mathcal{X}_{1,12,6}\simeq \mathcal{Y}_{2,9,6}$ and $\mathcal{X}_{2,9,6}\simeq \mathcal{Y}_{1,12,6}$ arising from eq.~\eqref{eq:XYdual}. 

From a computational point of view the two sphere partition function~$Z_{S^2}$ is expressed in terms of higher-dimensional Mellin--Barnes type integrals. While the one-dimensional case can be treated by standard residue calculus in complex analysis with one variable, Zhdanov and Tsikh show that the two-dimensional Mellin--Barnes type integrals are determined by combinatorial sums of two-dimensional Grothendieck residues \cite{MR1631772}. While this technique is sufficient to evaluate the two-dimensional integrals in $Z_{S^2}$ for the model~$SSSM_{1,12,6}$, we need to evaluate three-dimensional integrals in $Z_{S^2}$ for the model~$SSSM_{2,6,9}$. Therefore, in Appendix~\ref{sec:MB} we revisit the result of Zhdanov and Tsikh and generalize it to arbitrary higher dimensions. The calculations below closely follow the general procedure described there.

\subsection{The model  $SSSM_{1,12,6}$}\label{sec:M1General}
The \model{} $SSSM_{1,12,6}$ is determined by the gauge group
\begin{equation}
   G \,=\, \frac{U(1) \times \operatorname{USp}(2)}{\mathbb{Z}_2} \,\simeq\, \frac{U(1) \times SU(2)}{\mathbb{Z}_2} \ ,
\end{equation}
and the chiral matter multiplets listed in Table \ref{tb:spec1}, which are the $SU(2)$~scalar multiplets $P_{[ij]}$ and $\phi_a$ of $U(1)$~charge $-2$ and $+2$, and the $SU(2)$ spin-$\frac12$ multiplets $Q$ and $X_i$ of $U(1)$~charge $-3$ and $+1$. These multiplets  --- together with their multiplicities as specified by the range of their indices --- form the irreducible representations~$\operatorname{Irrep}(\rho)$ that appear in the expressions~\eqref{eq:ZS2} and \eqref{eq:ZS2comp} for the two sphere partition function $Z_{S^2}$. 

\begin{figure}[tbp]
\centering
\includegraphics[width=0.65\textwidth]{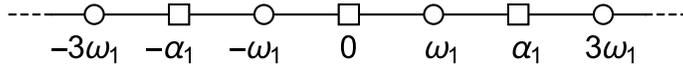}
\caption{\label{fig:SU2Lattice}Weight and root lattice of $\mathfrak{su}(2)$, roots are represented as squares.}
\end{figure}

We first discuss the relevant representation theory of the Lie algebra $\mathfrak{su}(2)$ of the non-Abelian part of the gauge group $G$. Its Cartan matrix reads
\begin{equation}
   A_{\mathfrak{su}(2)} \,=\, \begin{pmatrix}  2 \end{pmatrix} \ .
\end{equation}   
Hence, there is one fundamental weight $\omega_1$ generating the weight lattice $\Lambda_w$, and the single simple root is $\alpha_1 = 2 \omega_1$. The weight and root lattice are illustrated in Figure~\ref{fig:SU2Lattice}.
The simple root~$\alpha_1$ is the only positive root $\left\{\alpha_1\right\}=\Delta^+$, and $\left\{\omega_1, -\omega_1 \right\}$ are the weights of the  spin-$\frac12$ representations~$\mathbf{2}_s$. Linearity defines the scalar product of weights by
\begin{equation}
\left\langle \omega_1, \omega_1 \right\rangle \,=\, \frac{1}{2} \ ,
\end{equation}
which is given by the symmetric quadratic form matrix of $\mathfrak{su}(2)$. 

In order to evaluate the two sphere partition function $Z_{S^2}$, we need to carefully take into account the $\mathbb{Z}_2$ quotient in the definition of the gauge group~$G$. An irreducible representation $\rho$ of the double cover $U(1)\times SU(2)$ induces a projective representation $\hat\rho$ of $G$. However, $\hat\rho$ is an ordinary representation of $G$, if $\rho$ is invariant with respect to the $\mathbb{Z}_2$ generator $(e^{i\pi},-\mathbf{1}_{2\times 2})$ in the center of $U(1)\times SU(2)$. Such representations are represented by those highest weight vectors $\lambda \in \Lambda_w$ of $\mathfrak{u}(1)\times\mathfrak{su}(2)$ with
\begin{equation} \label{eq:U1SU2w}
  \lambda\,=\,\lambda_a q +\lambda_n \omega_1 \quad \text{for} \quad  \lambda_a+\lambda_n \in 2\mathbb{Z} \ .
\end{equation}
Here the Abelian charge $q$ is canonically normalized as
\begin{equation}
  \left\langle q, q \right\rangle \,=\, 1 \ .
\end{equation}
The restriction to the representations~\eqref{eq:U1SU2w} has important consequences for the evaluation of the partition function~$Z_{S^2}$. The localized BPS solutions contributing to $Z_{S^2}$ are classified by the magnetic quantum numbers $\mathfrak{m}$ of gauge flux configurations \cite{Doroud:2012xw,Benini:2012ui}. They obey the generalized Dirac quantization condition, which says that $\left\langle \lambda, \mathfrak{m} \right\rangle$ must be integral for all weights $\lambda$ in the representations of the gauge group~$G$ \cite{Goddard:1976qe}. This explicitly implies here
\begin{equation} \label{eq:M1P1SumRestrict}
   \mathfrak{m} \,=\, \frac12 m_a q + m_n \omega_1  \ , \quad m_a+m_n \in 2\mathbb{Z} \ ,
\end{equation}
with a conventional factor $\frac12$ to avoid half-integral magnetic quantum numbers $m_a$. Analogously, we set
\begin{equation}
   \boldsymbol{\sigma} \,=\, \frac12 \sigma_a q + \sigma_n \omega_1 \ , \qquad
   \boldsymbol{r} = 2 r q \ , \qquad 
   \boldsymbol{\theta}=2 \theta q \quad \text{with} \quad  \theta \sim \theta + 2 \pi \ .
\end{equation}
Note that due to the $\mathbb{Z}_2$ quotient in the definition of the gauge group, the electron of the $U(1)$ factor in the gauge group~$G$ --- which is a singlet of $SU(2)$ --- has charge two. Thus the Abelian background electric field contributing to the vacuum energy can be reduced by electron-positron pairs of charge $\pm 2$ \cite{Witten:1993yc}, and the periodicity of the theta angle $2\theta$ is $2\theta\sim2\theta+4\pi$, which implies the stated periodicity for $\theta$.

With the general expression for the two sphere partition function~\eqref{eq:ZS2} and taking into account the discussed magnetic charge lattice
\begin{equation}
   \Lambda_\mathfrak{m} \,=\, \left\{ \, (m_a,m_n)\in\mathbb{Z}^2 \,\middle|\, m_a+m_n \in 2\mathbb{Z} \, \right\} \ ,
\end{equation}   
we arrive at the two sphere partition function
\begin{equation} \label{eq:ZM1P1Raw}
  Z_{S^2}^{1,12,6}(r,\theta) \,=\, \frac{1}{8\pi^2}\!\!\sum_{(m_a,m_n)\in\Lambda_\mathfrak{m}}\!\!
  \int_{\mathbb{R}^2}\!\! d^2\boldsymbol{\sigma}\,
    Z_G(m_n,\sigma_n)
    Z_\text{matter}(\mathfrak{m},\boldsymbol{\sigma})
    Z_{\text{cl}}(m_a,\sigma_a,r,\theta) \ ,
\end{equation}
with $d^2\boldsymbol{\sigma} = \frac12 d\sigma_a d\sigma_n$ and where
\begin{equation}
\begin{aligned}
    &Z_G(m_n,\sigma_n)\, =\, (-1)^{m_n} \left(\frac{m_{n}^2}{4}+\sigma_{n}^2\right)\ , \qquad
    Z_{\text{cl}}(m_a,\sigma_a,r,\theta) \,=\, e^{-4 \pi i  r \,\sigma_{a}-i\theta \, m_{a}} \ , \\
    &Z_\text{matter}(\mathfrak{m},\boldsymbol{\sigma}) \,=\,
      Z_P(m_a,\sigma_a)^{15}\,Z_\phi(m_a,\sigma_a)^{12}\,Z_Q(\mathfrak{m},\boldsymbol{\sigma})\,
      Z_X(\mathfrak{m},\boldsymbol{\sigma})^6 \ .
\end{aligned}    
\end{equation}
The multiplicities of the irreducible representations of the matter multiplets are encoded by their respective powers in $Z_\text{matter}$, and they are individually given by 
\begin{align*}
  Z_P  &= \frac{\Gamma \left(\frac{m_{a}}{2}-\mathfrak{q}+i \sigma_{a}+1\right)}{\Gamma
      \left(\frac{m_{a}}{2}+\mathfrak{q}-i \sigma_{a}\right)} \ , \hspace{27mm} 
  Z_\phi = \frac{\Gamma \left(-\frac{m_{a}}{2}+\mathfrak{q}-i \sigma_{a}\right)}{\Gamma
      \left(-\frac{m_{a}}{2}-\mathfrak{q}+i \sigma_{a}+1\right)} \ ,\\
  Z_Q &= 
     \frac{\Gamma \left(\frac{3 m_{a}}{4}-\frac{m_{n}}{4}-\frac{3 \mathfrak{q}}{2}+\frac{3 i
      \sigma_{a}}{2}-\frac{i \sigma_{n}}{2}+1\right)}{\Gamma \left(\frac{3
      m_{a}}{4}-\frac{m_{n}}{4}+\frac{3 \mathfrak{q}}{2}-\frac{3 i \sigma_a}{2}+\frac{i \sigma_{n}}{2}\right)}
      \cdot\frac{\Gamma \left(\frac{3
     m_{a}}{4}+\frac{m_{n}}{4}-\frac{3 \mathfrak{q}}{2}+\frac{3 i \sigma_a}{2}+\frac{i \sigma_{n}}{2}+1\right)}{\Gamma \left(\frac{3
     m_{a}}{4}+\frac{m_{n}}{4}+\frac{3 \mathfrak{q}}{2}-\frac{3 i \sigma_a}{2}-\frac{i \sigma_{n}}{2}\right)}
      \ , \\
  Z_X &=
    \frac{\Gamma \left(-\frac{m_{a}}{4}-\frac{m_{n}}{4}+\frac{\mathfrak{q}}{2}-\frac{i
     \sigma_{a}}{2}-\frac{i \sigma_{n}}{2}\right)}{\Gamma
     \left(-\frac{m_{a}}{4}-\frac{m_{n}}{4}-\frac{\mathfrak{q}}{2}+\frac{i \sigma_a}{2}+\frac{i \sigma_{n}}{2}+1\right)}
     \cdot\frac{\Gamma
     \left(-\frac{m_{a}}{4}+\frac{m_{n}}{4}+\frac{\mathfrak{q}}{2}-\frac{i \sigma_a}{2}+\frac{i \sigma_{n}}{2}\right)}{\Gamma
     \left(-\frac{m_{a}}{4}+\frac{m_{n}}{4}-\frac{\mathfrak{q}}{2}+\frac{i \sigma_a}{2}-\frac{i \sigma_{n}}{2}+1\right)}
      \ .
\end{align*}
By the general structure of the two sphere partition function, the Weyl group $\mathcal{W}_{\mathfrak{su}(2)}\simeq \mathbb{Z}_2$ of $SU(2)$ acts as a symmetry on the integrand summed over the weight lattice $\Lambda_\mathfrak{m}$ through the action generated by
\begin{equation} \label{eq:Wsu(2)}
  \mathfrak{s}: \left(m_n,\sigma_n\right) \longmapsto \left(-m_n,-\sigma_n\right) \ .
\end{equation}
Keeping track of this symmetry transformation proves useful in evaluating $Z_{S^2}^{1,12,6}$ in the following.

For consistency let us first confirm that the two sphere partition function~\eqref{eq:ZM1P1Raw} is actually real. Using the reflection formula
\begin{equation} \label{eq:GammaReflect}
  \Gamma(x) \Gamma(1-x) = \frac{\pi}{\operatorname{sin}(\pi x)} \ , 
\end{equation}
and the generalized Dirac quantization condition~\eqref{eq:M1P1SumRestrict}, we establish with a few steps of algebra the conjugation formulas
\begin{equation}
\begin{aligned}
   &Z_P(m_a,\sigma_a)^*=(-1)^{-m_a} Z_P(-m_a,-\sigma_a) \ , 
   &&Z_\phi(m_a,\sigma_a)^*=(-1)^{m_a} Z_\phi(-m_a,-\sigma_a) \ , \\
   &Z_Q(\mathfrak{m},\boldsymbol{\sigma})^*=(-1)^{-3m_a} Z_Q(-\mathfrak{m},-\boldsymbol{\sigma}) \ , 
   &&Z_X(\mathfrak{m},\boldsymbol{\sigma})^*=(-1)^{m_a} Z_X(-\mathfrak{m},-\boldsymbol{\sigma}) \ .
\end{aligned}   
\end{equation}
Taking into account the multiplicities of the individual chiral multiplets, we find $Z_\text{matter}(\mathfrak{m},\boldsymbol{\sigma})^*=Z_\text{matter}(-\mathfrak{m},-\boldsymbol{\sigma})$, which readily shows that after substituting the summation and integration variables $(\mathfrak{m},\boldsymbol{\sigma})$ to  $(-\mathfrak{m},-\boldsymbol{\sigma})$ the two sphere partition function $Z^{1,12,6}_{S^2}$ is real.

For the detailed analysis of $Z^{1,12,6}_{S^2}$, it is convenient to work with integration variables that diagonalize the arguments of two Gamma functions and simultaneously remove the parameter $\mathfrak{q}$ of the $U(1)_V$ R-symmetry from all Gamma functions. One particular choice of such a substitution reads
\begin{equation}
  \sigma_a = -i\, (\mathfrak{q} + x_1 + x_2 ) \ , \qquad \sigma_n =- i\, (x_1-x_2) \ ,
\end{equation}
which results with $\boldsymbol{x}=(x_1,x_2)$ in
\begin{equation}
 Z_{S^2}^{1,12,6}(r,\theta) \,=\, \frac{e^{-4\pi r \mathfrak{q}}}{8\pi^2}
 \sum_{(m_a,m_n)\in\Lambda_\mathfrak{m}}
 \int_{\gamma+i\mathbb{R}^2} \omega(\mathfrak{m},\boldsymbol{x}) \,dx_1 \wedge dx_2 \ , \quad
 \gamma = -\left(\tfrac{\mathfrak{q}}2,\tfrac{\mathfrak{q}}2\right) \ .\label{eq:ZM1Integral}
\end{equation}
The integrand is given by
\begin{equation} \label{eq:ZModel1}
  \omega(\mathfrak{m},\boldsymbol{x}) = Z'_G(m_n,\boldsymbol{x})
  Z'_P(m_a,\boldsymbol{x})^{15}Z'_\phi(m_a,\boldsymbol{x})^{12}
  Z'_Q(\mathfrak{m},\boldsymbol{x})Z'_X(\mathfrak{m},\boldsymbol{x})^6
  Z'_\text{cl}(r,\theta,m_a,\boldsymbol{x}) \ ,
\end{equation}
where
\begin{align*}
  Z'_{\text{cl}} &= e^{-4 \pi r \left(x_1+x_2\right) - i\,\theta \, m_a} \ ,
  \hspace{30mm} Z'_G = (-1)^{m_n} \left(\frac{m_{n}^2}4-(x_1-x_2)^2\right) \ , \\
  Z'_P &= \frac{\Gamma \left(1+ \frac{m_a}{2} +x_1+x_2 \right)}{\Gamma \left( \frac{m_a}{2} -x_1-x_2 \right)} \ , 
  \hspace{21.5mm} Z'_\phi = \frac{\Gamma \left(-\frac{m_a}{2}-x_1-x_2\right)}{\Gamma \left(1-\frac{m_a}{2}+x_1+x_2\right)} \ , \\
  Z'_Q &= \underbrace{\frac{\Gamma \left(1+ \frac{3m_a-m_n}{4}+x_1+2x_2 \right)}{\Gamma\left(\frac{3m_a-m_n}{4}-x_1-2x_2\right)}}_{Z'_{Q_1}} \cdot
 \underbrace{\frac{\Gamma\left(1+\frac{3m_a+m_n}{4}+2x_1+x_2 \right)}{\Gamma \left(\frac{3m_a+m_n}{4}-2x_1-x_2 \right)}}_{Z'_{Q_2}} \ ,\\
  Z'_X &= \underbrace{\frac{\Gamma \left(-\frac{m_a+m_n}{4}-x_1\right)}{\Gamma \left(1-\frac{m_a+m_n}{4}+x_1\right)}}_{Z'_{X_1}} \cdot
  \underbrace{\frac{\Gamma \left(-\frac{m_a-m_n}{4}  -x_2 \right)}{\Gamma \left(1-\frac{m_a-m_n}{4} +x_2 \right)}}_{Z'_{X_2}}\ .
\end{align*}
Since the integral~\eqref{eq:ZM1Integral} is now of the same form as the Mellin--Barnes type integral~\eqref{eq:MBInt}, we proceed as in Appendix~\ref{sec:MB} and rewrite the integral as a sum of local Grothendieck residues. We also record that in terms of the new variables $\boldsymbol{x}$ the generator \eqref{eq:Wsu(2)} of the Weyl group $\mathcal{W}_{\mathfrak{su}(2)}$ acting on the integrand becomes
\begin{equation} \label{eq:Wsu(2)new}
  \mathfrak{s}: \left(m_n,x_1,x_2\right) \longmapsto \left(-m_n,x_2,x_1\right) \ ,
\end{equation}
which induces the action on the signed volume form
\begin{equation} \label{eq:Wsu(2)vol}
  \mathfrak{s}: dx_1 \wedge dx_2\longmapsto -dx_1 \wedge dx_2 \ .
\end{equation}

Firstly, we have to determine the divisors for the poles of the integrand $\omega(\mathfrak{m},\boldsymbol{x})$ in $\mathbb{R}^2 \subset\mathbb{C}^2$. Such poles arise from non-positive integral arguments of Gamma functions in the numerator. Taking into account cancellation between the poles and zeros --- arising from non-positive integral arguments of Gamma functions in the denominator --- we find the following divisors of poles in terms of the (constraint) integers $n_P, n_{Q_1},n_{Q_2},n_{X_1},n_{X_2}$
\begin{equation}\label{eq:DivisModel1}
\begin{aligned}
    D^{n_P}_P\,&=\, x_1+x_2 + n_P + \tfrac{m_a}{2} +1  &&\text{ for } n_P\geq \operatorname{Max}\left[0,-m_a\right] \ ,\\
    D^{n_{Q_1}}_{Q_1}\,&=\,x_1+2x_2 + n_{Q_1}+\tfrac{3m_a-m_n}{4}+1 &&\text{ for } n_{Q_1}\geq \operatorname{Max}\left[0,-\tfrac{3m_a-m_n}{2}\right] \ ,\\
    D^{n_{Q_2}}_{Q_2}\,&=\,2x_1+x_2 + n_{Q_2}+\tfrac{3m_a+m_n}{4}+1 &&\text{ for } n_{Q_2}\geq \operatorname{Max}\left[0,-\tfrac{3m_a+m_n}{2}\right] \ ,\\
    D^{n_{X_1}}_{X_1}\,&=\,x_1 -  n_{X_1} +\tfrac{m_a+m_n}{4} &&\text{ for } n_{X_1}\geq \operatorname{Max}\left[0,\tfrac{m_a+m_n}{2}\right] \ , \\
    D^{n_{X_2}}_{X_2}\,&=\,x_2 -  n_{X_2} +\tfrac{m_a-m_n}{4} &&\text{ for } n_{X_2}\geq \operatorname{Max}\left[0,\tfrac{m_a-m_n}{2}\right]. 
\end{aligned}
\end{equation}
Note that $Z'_\phi$ does not yield a contribution, since all its poles are canceled by the denominator of $Z'_P$.

Secondly, to determine the critical line $\partial H$ introduced in eq.~\eqref{eq:MBHn2}, we identify $\boldsymbol{p}$ in eq.~\eqref{eq:MBw} as $\boldsymbol{p} =4 \pi r (1,1)\in\mathbb{R}^2$, and therefore find
\begin{equation} \label{eq:UntiltedLine}
  \partial H = \left\{ \,(x_1,x_2) \in \mathbb{R}^2 \, \middle| \, x_1+x_2 = -\mathfrak{q}\, \right\} \ . 
\end{equation}
Since this line is parallel to the divisors $D^{n_P}_P$, we apply the method of Appendix~\ref{sec:MBParallel} and introduce an additional exponential factor such that
\begin{equation} \label{eq:OmegaPrime}
  \omega(\mathfrak{m},\boldsymbol{x}) \, \longrightarrow \,
  \omega'(\mathfrak{m},\boldsymbol{x}) = \omega(\mathfrak{m},\boldsymbol{x}) \cdot e^{-4\pi r \varepsilon\,\boldsymbol{\delta}\cdot\boldsymbol{x}} \ ,
  \qquad \boldsymbol{\delta}=(\delta_1,\delta_2) \ ,
\end{equation}
where we eventually take the limit $\varepsilon \to 0^+$ after carrying out the integral. Then the critical line $\partial H$ is modified to
\begin{equation}
\partial H' = \left\{ \, \boldsymbol{x} \in \mathbb{R}^2 \, \middle| \, \sum_{i=1}^2 \left(1+\varepsilon\,\delta_i\right)x_i = -\left(1+\varepsilon\,\sum_{i=1}^2\frac{\delta_i}{2}\right)\,\mathfrak{q} \, \right\} \ ,
\end{equation}
which for $\varepsilon$ small and $\delta_1 \ne \delta_2$ indeed removes parallelness among $\partial H'$ and the divisors $D^{n_P}_P$.\footnote{Note that the perturbation $\boldsymbol{\delta}$ with $\delta_1 \ne \delta_2$ breaks the Weyl symmetry $\mathcal{W}_{\mathfrak{su}(2)}$.} The point $\gamma = -\left(\frac{\mathfrak{q}}2,\frac{\mathfrak{q}}2\right) \in \partial H'$ splits the critical line into the two rays
\begin{equation} \label{eq:M1TwoRays}
  \partial H'_1 =\left\{ (x_1,x_2) \in \partial H' \, \middle| \, x_1>-\tfrac{\mathfrak{q}}{2} \right\} \ , 
  \quad \partial H'_2 =\left\{ (x_1,x_2) \in \partial H' \, \middle| \, x_1<-\tfrac{\mathfrak{q}}{2} \right\}\ ,
\end{equation}
which according to the conventions in Appendix~\ref{sec:2dMB} correspond in the respective regimes to the rays
\begin{equation}
  \partial H_+ \,=\,\begin{cases} \partial H'_1 & \text{for }r\gg0 \\ \partial H'_2 & \text{for }r\ll0 \end{cases} \ , \qquad
  \partial H_- \,=\,\begin{cases} \partial H'_2 & \text{for }r\gg0 \\ \partial H'_1 & \text{for }r\ll0 \end{cases} \ .
\end{equation}  

Thirdly, we calculate the intersection points of the divisors with $\partial H'$. Up to leading order as $\varepsilon \to 0^+$ they are given by
\begin{equation}\label{eq:M1qinwhichRay}
\begin{aligned}
 q^{n_P}_P\,&=\,\left(\tfrac{1+\frac{m_a}{2}+n_P-\mathfrak{q}}{\varepsilon(\delta_{1}-\delta_{2})}, -\tfrac{1+\frac{m_a}{2}+n_P-\mathfrak{q}}{\varepsilon(\delta_{1}-\delta_{2})} \right)\quad \in\, \begin{cases} \partial H'_1 & \delta_1>\delta_2\\ \partial H'_2 & \delta_2>\delta_1 \end{cases} \ ,  \\
 q^{n_{Q_1}}_{Q_1}\,&=\,\left(1+\tfrac{3 m_a-m_n}{4}+n_{Q_1}-2\mathfrak{q},-1-\tfrac{3 m_a-m_n}{4}-n_{Q_1}+\mathfrak{q}\right)\quad \in\, \partial H'_1 \ , \\
 q^{n_{Q_2}}_{Q_2}\,&=\,\left(-1-\tfrac{3 m_a+m_n}{4}-n_{Q_2}+\mathfrak{q},1+\tfrac{3 m_a+m_n}{4}+n_{Q_2}-2\mathfrak{q}\right)\quad \in\, \partial H'_2 \ , \\
 q^{n_{X_1}}_{X_1}\,&=\,\left(-\tfrac{m_a+m_n}{4}+n_{X_1},\tfrac{m_a+m_n}{4}-n_{X_1}-\mathfrak{q}\right)\quad \in\, \partial H'_1 \ ,\\
 q^{n_{X_2}}_{X_2}\,&=\,\left(\tfrac{m_a-m_n}{4}-n_{X_2}-\mathfrak{q},-\tfrac{m_a-m_n}{4}+n_{X_2}\right)\quad \in\, \partial H'_2 \ ,
\end{aligned}
\end{equation}
where the respective constraints on the integers $n_P, n_{Q_1},n_{Q_2},n_{X_1},n_{X_2}$ and the condition~$0<\mathfrak{q}<\frac23$ determine the associated rays $\partial H'_1$ and $\partial H'_2$.

Now we are ready to determine those pairs of divisors associated to poles that appear in the sets $\Pi_\pm$, which are defined in eq.~\eqref{eq:P2pm} via the index set \eqref{eq:I2pm}. To this end we need to consider the two disjoint half-spaces $H'_1$ and $H'_2$ bounded by the critical line
\begin{equation}
\begin{aligned}
  H'_1 &= \left\{\,\boldsymbol{x} \in \mathbb{R}^2 \, \middle| \, \sum_{i=1}^2 \left(1+\varepsilon\,\delta_{i}\right)x_i > -\left(1+\varepsilon\,\sum_{i=1}^2\frac{\delta_{i}}{2}\right)\,\mathfrak{q} \right\} \ ,\\
  H'_2 &= \left\{\,\boldsymbol{x} \in \mathbb{R}^2 \, \middle| \, \sum_{i=1}^2 \left(1+\varepsilon\,\delta_{i}\right)x_i < -\left(1+\varepsilon\,\sum_{i=1}^2\frac{\delta_{i}}{2}\right)\,\mathfrak{q} \right\} \ ,
\end{aligned}
\end{equation}
such that the relevant half-space $H$ for the respective gauged linear sigma model phases is
\begin{equation}
  H \,=\,\begin{cases} H'_1 & \text{for }r\gg0 \\ H'_2 & \text{for }r\ll0 \end{cases} \ .
\end{equation}  
Then choosing the orientation of the intersections conveniently, we find
\begin{equation} \label{eq:PipmM1P1}
  \Pi_+ \,=\,\begin{cases}
     \left\{\, p^{\vec n_8}_{\vec\jmath_8} \,\right\} & \text{ for } r\gg 0 \ , \\[1.5ex]
     \left\{\, p^{\vec n_i}_{\vec\jmath_i}\,\middle|\, i=1,2,3,4,5 \,\right\} & \text{ for } r\ll 0,\ \delta_1>\delta_2 \ , \\[1.5ex]
     \left\{\, p^{\vec n_i}_{\vec\jmath_i}\,\middle|\, i=3,4,5,6,7 \,\right\} & \text{ for } r\ll 0,\ \delta_2>\delta_1 \ ,
  \end{cases} \qquad
  \Pi_- = \emptyset \ ,
\end{equation}     
in terms of the (oriented) divisor intersections with the labels
\begin{equation}\label{eq:M1jLables}
\begin{aligned}
    &(\vec\jmath_1,\vec n_1)\,=\, (Q_2,P; n_{Q_2},n_P) \ , &&(\vec\jmath_2,\vec n_2)\,=\, (X_2,P; n_{X_2},n_P) \ , \\
    &(\vec\jmath_3,\vec n_3)\,=\, (Q_2,Q_1; n_{Q_2},n_{Q_1}) \ , &&(\vec\jmath_4,\vec n_4)\,=\, (X_2,Q_1; n_{X_2},n_{Q_1}) \ , \\
    &(\vec\jmath_5,\vec n_5)\,=\, (Q_2,X_1; n_{Q_2},n_{X_1}) \ , &&(\vec\jmath_6,\vec n_6)\,=\, (P,Q_1; n_{P},n_{Q_1}) \ , \\
    &(\vec\jmath_7,\vec n_7)\,=\, (P,X_1; n_{P},n_{X_1}) \ , &&(\vec\jmath_8,\vec n_8)\,=\, (X_1,X_2; n_{X_1},n_{X_2}) \ , \\
\end{aligned}    
\end{equation}
and the pole loci
\begin{equation}
\begin{aligned}
     p^{\vec n_1}_{\vec\jmath_1} \,&=\, \left(n_P - n_{Q_2} -\tfrac{m_a+m_n}{4} , -1+n_{Q_2}-2n_P -\tfrac{m_a-m_n}{4}\right) \ , \\
     p^{\vec n_2}_{\vec\jmath_2}  \,&=\,\left(-1-n_{X_2}-n_P -\tfrac{m_a+m_n}{4} , n_{X_2} -\tfrac{m_a-m_n}{4}\right) \ , \\
     p^{\vec n_3}_{\vec\jmath_1} \,&=\, \left(\tfrac{-1+n_{Q_1}-2n_{Q_2}}{3}-\tfrac{m_a+m_n}{4},\tfrac{-1+n_{Q_2}-2n_{Q_1}}{3}-\tfrac{m_a-m_n}{4}\right) \ , \\
     p^{\vec n_4}_{\vec\jmath_2}  \,&=\,\left(-1-n_{Q_1}-2n_{X_2} -\tfrac{m_a+m_n}{4},n_{X_2} -\tfrac{m_a-m_n}{4}\right) \ , \\
     p^{\vec n_5}_{\vec\jmath_1} \,&=\, \left(n_{X_1} -\tfrac{m_a+m_n}{4},-1-n_{Q_2}-2n_{X_1} -\tfrac{m_a-m_n}{4}\right) \ , \\
     p^{\vec n_6}_{\vec\jmath_2}  \,&=\, \left(-1+n_{Q_1}-2n_P-\tfrac{m_a+m_n}{4},n_P - n_{Q_1} -\tfrac{m_a-m_n}{4}\right)\ , \\
     p^{\vec n_7}_{\vec\jmath_1} \,&=\, \left( n_{X_1} -\tfrac{m_a+m_n}{4},-1-n_P-n_{X_1}-\tfrac{m_a-m_n}{2}\right) \ , \\
     p^{\vec n_8}_{\vec\jmath_2}  \,&=\,\left( n_{X_1} -\tfrac{m_a+m_n}{4}, n_{X_2} -\tfrac{m_a-m_n}{4}\right) \ .
\end{aligned}\label{eq:PolesM1P1}
\end{equation}
This allows us to express the two sphere partition function according to eq.~\eqref{eq:MB2D} as 
\begin{equation}
   Z_{S^2}^{1,12,6}(r,\theta) \,=\,- \frac{e^{-4\pi r \mathfrak{q}}}{2} \,\lim_{\varepsilon\to0^+}\,\sum_{(m_a,m_n)\in\Lambda_\mathfrak{m}}
   \, \sum_{\boldsymbol{x}\in\Pi_+}\operatorname{Res}_{\boldsymbol{x}} \omega'(\mathfrak{m},\boldsymbol{x}) \ ,\label{eq:ZM1}
\end{equation}
where $\Pi_+$ depends both on the sign of the parameter $r$ and the regulator $\boldsymbol{\delta}$. 

For consistency, in the limit $\varepsilon \to 0^+$  the two sphere partition function~$Z_{S^2}^{1,12,6}$ must not depend on the perturbation $\boldsymbol{\delta}$. According to eq.~\eqref{eq:PipmM1P1}, this is obviously true in the phase $r\gg0$, whereas it is not manifest in the regime $r\ll0$. In the latter phase, it is a consequence of the Weyl transformation~\eqref{eq:Wsu(2)} generated by $\mathfrak{s}$, which acts upon the divisor intersections~\eqref{eq:PipmM1P1} as
\begin{equation}
\begin{aligned}
  &(\vec\jmath_1,\vec n_1) \stackrel{\mathfrak{s}}{\longmapsto} (\vec\jmath_6,\vec n_6) \ ,
  &&(\vec\jmath_2,\vec n_2) \stackrel{\mathfrak{s}}{\longmapsto} (\vec\jmath_7,\vec n_7) \ ,
  &&(\vec\jmath_3,\vec n_3) \stackrel{\mathfrak{s}}{\longmapsto} (\vec\jmath_3,\vec n_3) \ , \\
  &(\vec\jmath_4,\vec n_4) \stackrel{\mathfrak{s}}{\longmapsto} (\vec\jmath_5,\vec n_5) \ ,
  &&(\vec\jmath_5,\vec n_5) \stackrel{\mathfrak{s}}{\longmapsto} (\vec\jmath_4,\vec n_4) \ , 
  &&(\vec\jmath_6,\vec n_6) \stackrel{\mathfrak{s}}{\longmapsto} (\vec\jmath_1,\vec n_1) \ ,\\
  &(\vec\jmath_7,\vec n_7) \stackrel{\mathfrak{s}}{\longmapsto} (\vec\jmath_2,\vec n_2) \ , 
  &&(\vec\jmath_8,\vec n_8) \stackrel{\mathfrak{s}}{\longmapsto} (\vec\jmath_8,\vec n_8) \ .
\end{aligned}  
\end{equation}
Here we have taken into account that the minus signs from the orientation-reversal of the ordered pair of intersecting divisors are compensated by the sign-reversal of the volume form in eq.~\eqref{eq:Wsu(2)vol}. These transformations together with eq.~\eqref{eq:PipmM1P1} demonstrate that the end result is indeed independent of the perturbation $\boldsymbol{\delta}$.

\subsubsection{The $SSSM_{1,12,6}$ phase $r\gg 0$}
In this subsection we specialize to the phase $r\gg0$. From eq.~\eqref{eq:PipmM1P1} the relevant set of poles $\Pi_+$ is seen to be independent of the regulator $\boldsymbol{\delta}$. We thus expect to manifestly recover the antisymmetry under $\mathfrak{s}$ in the limit $\varepsilon \to 0^+$. In fact, this limit will turn out to be trivial in the present case. In order to illustrate the mechanism, we shall however keep $\varepsilon >0$ and choose $\delta_{1} = 1/(2\pi r) >\delta_{2}=0$ for definiteness.

For a given $(m_a,m_n)\in\Lambda_{\mathfrak{m}}$ we have
\begin{equation}
\begin{aligned}
\Pi_+ &= \{\,p^{\vec{n}_8}_{\vec{j}_8}\,|n_{X_1}\geq \operatorname{Max}\left[0,\tfrac{m_a+m_n}{2}\right], n_{X_2}\geq \operatorname{Max}\left[0,\tfrac{m_a-m_n}{2}\right]\}\ ,\\
p^{\vec{n}_8}_{\vec{j}_8} &= \left( n_{X_1} -\tfrac{m_a+m_n}{4}, n_{X_2} -\tfrac{m_a-m_n}{4}\right)\ .
\end{aligned} 
\end{equation}
Upon changing variables to
\begin{equation}
  a = n_{X_1}+n_{X_2}- m_a\ ,\quad b=n_{X_1}-\tfrac{m_a+m_n}{2}\ , c = n_{X_1}+n_{X_2}\ , \quad d=n_{X_1} \ ,
\end{equation}
the sums in eq.~\eqref{eq:ZM1} symbolically simplify to
\begin{equation}
\sum_{(m_a,m_n)\in\Lambda_\mathfrak{m}} \, \sum_{\boldsymbol{x}\in\Pi_+} \quad \longrightarrow \quad \sum_{a=0}^\infty\, \sum_{c=0}^\infty \,\sum_{b=0}^a \,\sum_{d=0}^c \ .
\end{equation}
Introducing coordinates around the large volume point
\begin{equation}
z = e^{- 2\pi r + i \theta}, \quad \overline{z} =e^{- 2\pi r - i \theta}\ , \label{eq:zvar}
\end{equation}
the partition function takes the form
\begin{equation}
Z_{S^2,r\gg0}^{1,12,6}(r,\theta)=-\frac{\left(z \overline{z}\right)^\mathfrak{q}}{2} \lim_{\varepsilon \to 0^+} \operatorname{Res}_{\boldsymbol{x}=0} \left( e^{-2x_1\varepsilon} Z_{\text{sing}} \left|z^{x_1+x_2}\sum_{a=0}^\infty (-z)^a \sum_{b=0}^a Z_{\text{reg}} e^{-b\varepsilon}\right|^2\right), \label{eq:ZModel1Final}
\end{equation}
with
\begin{equation}\label{eq:ZModel1FinalZusatz}
\begin{aligned}
Z_{\text{sing}} &=\pi^7 \, \frac{\text{sin}\left[\pi (x_1+x_2)\right]^3\text{sin}\left[\pi (2x_1+x_2)\right]\text{sin}\left[\pi (x_1+2x_2)\right]}{\text{sin}\left(\pi x_1\right)^6\text{sin}\left(\pi x_2\right)^6}\ ,\\
Z_{\text{reg}} &=(a-2b-x_1+x_2) \frac{\Gamma (1+a+x_1+x_2)^3}{\Gamma(1+b+x_1)^6 \Gamma(1+a-b+x_2)^6}\\
&\, \qquad \quad\cdot \Gamma(1+a+b+2x_1+x_2)\Gamma(1+2a-b+x_1+2x_2)\ ,
\end{aligned}
\end{equation}
Here the residue is being evaluated with respect to the oriented pair of divisors $(x_1,x_2)$ and complex conjugation is defined to not act on $x_1$ and $x_2$. Since for any given order in $z$ and $\overline{z}$ there are only finitely many terms, the limit $\varepsilon \to 0^+$ can safely be taken before summation. An infinite sum would have been automatically regularized by $\varepsilon>0$. Note that the transformation $\mathfrak{s}$ acts on eq.~\eqref{eq:ZModel1Final} by
\begin{equation}
  \mathfrak{s}:\, (x_1,x_2,b,d) \longmapsto (x_2,x_1,a-b,c-d) \ ,
\end{equation}
under which $Z_{S^2,r\gg0}^{1,12,6}$ for $\varepsilon = 0$ indeed displays the Weyl symmetry. 

From eq.~\eqref{eq:ZModel1Final} the partition function can now be evaluated to any fixed order in $z$ and $\overline{z}$. By exploiting the relation~\eqref{eq:ZtoK}, we want to read off geometric data of the associated infrared target Calabi-Yau threefold $\mathcal{X}_{1,12,6}$ from the partition function~$Z_{S^2,r\gg0}^{1,12,6}$. For this purpose it is sufficient to keep terms of order $0$ in $\overline{z}$ only, i.e., we can set $c = d = 0$. Keeping $a$ and $b$ still arbitrary, the fundamental period $\omega^{1,12,6}_{0,r\gg0}(z)$ is then found as the coefficient function of $\left(\operatorname{log}\, z\right)^3$. Fixing its normalization such that the expansion starts with $1$, we find
\begin{equation} \label{eq:M1P1FundPeriod}
\begin{aligned}
\omega^{1,12,6}_{0,r\gg0}(z) 
& \,=\, \sum_{a=0}^\infty \sum_{b=0}^a \frac{a!^3(2a-b)!(a+b)!}{(a-b)!^6 b!^6}\bigg[ 1 + (2b-a) \left(H_{a+b}-6 H_{b}\right) \bigg] (-z)^a \\
& \,=\, 1+7 z + 199 z^2 +8\,359 z^3 +423\,751 z^4+23\,973\,757 z^5 + \ldots \, ,
\end{aligned}
\end{equation}
where $H_n$ is the $n$-th harmonic number. We then use the Euler characteristic $\chi(\mathcal{X}_{1,12,6})=-102$ calculated in eq.~\eqref{eq:DataCY1} and by a K\"ahler transformation\footnote{Here, this K\"ahler transformation is divison by $8\pi^3(z\overline{z})^{\mathfrak{q}} \left|\omega^{1,12,6}_{0,r\gg0}(z)\right|^2$ with $\omega^{1,12,6}_{0,r\gg0}(z)$ as in eq.~\eqref{eq:M1P1FundPeriod}.}  fix the overall normalization of $Z_{S^2,r\gg0}^{1,12,6}$ to match with the canonical large volume form~\eqref{eq:expK}. Subsequently, we can read off the degree and the degree $d$ integral genus zero Gromov--Witten invariants
\begin{equation} \label{eq:GW1}
\begin{aligned}
\text{deg}(\mathcal{X}_{1,12,6})& = 33 \ , \\
N_d(\mathcal{X}_{1,12,6}) &= 
\begin{cases} 
252, \quad  1\,854, \quad 27\,156, \quad 567\,063, \quad 14\,514\,039, \\
424\,256\,409, \quad 13\,599\,543\,618 \quad 466\,563\,312\,360, \\
16\,861\,067\,232\,735, \quad 634\,912\,711\,612\,848, \quad \ldots
\end{cases}
\end{aligned}
\end{equation}
These results are agreement with eq. \eqref{eq:DataCY1} and the Gromov--Witten invariants determined in ref.~\cite{Miura:2013arxiv}, and they coincide with the calculated data obtained in the classification program of one parameter Picard--Fuchs operators \cite{vanStraten:2012db,Hofmann:2013PhDThesis}.

\subsubsection{The $SSSM_{1,12,6}$ phase $r\ll0$}\label{sec:Model1Strong}
We turn to the discussion of the phase $r\ll 0$. According to eq.~\eqref{eq:PipmM1P1}, the set $\Pi_+$ now depends on the sign of $\delta_{1}-\delta_{2}$. Therefore we cannot expect to obtain a result that in the limit $\varepsilon \to 0^+$ is manifestly symmetric with respect to the Weyl symmetry generated by $\mathfrak{s}$. An equivalent argument is that --- irrespective of which case is chosen --- the relevant pairs of divisors do not separate into full orbits under $\mathfrak{s}$, as has been the case for $r\gg0$. We could restore manifest antisymmetry by adding the two terms obtained for the two different choices. As was shown at the end of Section~\ref{sec:M1General}, these two terms are, however, equal anyway. Adding them would thus be mere cosmetics. We proceed by choosing $\delta_{1} = -\frac1{2\pi r} >\delta_{2}=0$.

As seen from eq.~\eqref{eq:PipmM1P1} there are now five sets of poles contributing to $\Pi_+$. Since these five sets might have non-trivial intersections, independently summing over them would wrongly count certain poles several times. In general this can be accounted for by dividing the contribution from each pole by the number of pairs it belongs to. We, however, continue along a different way, which allows us to write the partition function in a more compact form. As we shall demonstrate 
\begin{equation} \label{eq:M1P2TwoSets}
\Pi_+ = \left\{\, p^{\vec n_i}_{\vec\jmath_i}\,\middle|\, i=1,2,3,4,5 \,\right\} = \Pi^1_+ \cup \Pi^2_+ \ , 
\end{equation}
in terms of the disjoint sets
\begin{equation}
\begin{aligned}
\Pi^1_+ &=  \left\{\, p^{\vec n_1}_{\vec\jmath_1}\,\middle|\,n_{Q_2}\geq \operatorname{Max}\left[0,-\frac{3m_a+m_n}{2}\right],n_{P}\geq \operatorname{Max}\left[0,-m_a\right] \,\right\},\\
\Pi^2_+ &=  \left\{\, p^{\vec n_3}_{\vec\jmath_3}\,\middle|\,n_{Q_2/Q_1}\geq \operatorname{Max}\left[0,-\frac{3m_a\pm m_n}{2}\right],n_{Q_1}+n_{Q_2} \notin 3 \mathbb{N}_0+1 \,\right\}.
\end{aligned}
\end{equation}
In order to show an inclusion of the type $\{p^{\vec n}_{\vec \jmath}\} \subset \{p^{\vec m}_{\vec k}\}$ we proceed in two steps:
\begin{enumerate}
\item For given $m_a$ and $m_n$, equate $p^{\vec n}_{\vec\jmath}=p^{\vec m}_{\vec k}$ and solve for $n_{k_1}$ and $n_{k_2}$.
\item Assume that $n_{\jmath_1}$ and $n_{\jmath_2}$ fulfill their constraints given in eq.~\eqref{eq:DivisModel1}. If $n_{k_1}$ and $n_{k_2}$ --- as given by the previously obtained equations --- fulfill their constraints as well, the desired results have been established.
\end{enumerate}
Let us illustrate this procedure by showing $\{p^{\vec{n}_2}_{\vec \jmath_2}\} \subset \{p^{\vec n_1}_{\vec \jmath_1}\}$ explicitly. Denoting $\vec{n}_2 = \left(\overline{n_{X_2}},\overline{n_P}\right)$ to distinguish between common divisors, we find
\begin{equation}
\begin{aligned}
\text{(i)}& \quad \left(\begin{array}{l}n_P - n_{Q_2} -\frac{m_a+m_n}{4} \\ -1+n_{Q_2}-2n_P -\frac{m_a-m_n}{4}\end{array}\right) = \left(\begin{array}{l} -1-\overline{n_{X_2}}-\overline{n_P} -\frac{m_a+m_n}{4} \\ \overline{n_{X_2}} -\frac{m_a-m_n}{4} \end{array}\right), \\[0.3em]
\Rightarrow \text{(ii)}& \quad \begin{array}{ll} n_P = \overline{n_P} \geq \operatorname{Max}\left[0,-m_a\right]\\[0.1em] n_{Q_2} = 1 + 2 \overline{n_P} + \overline{n_{X_2}} \geq 1 +\operatorname{Max}\left[0,-\frac{3m_a+m_n}{2}\right].
\end{array}
\end{aligned}\label{eq:inclusion}
\end{equation}
Similar calculations show further inclusions and with these eq.~\eqref{eq:M1P2TwoSets} can be established. With eq.~\eqref{eq:ZM1} we then find
\begin{equation}\label{eq:ZM1P2}
\begin{aligned}
   Z_{S^2,r\ll0}^{1,12,6}(r,\theta) \,=\,&- \frac{e^{-4\pi r \mathfrak{q}}}{2} \, \lim_{\varepsilon\to0^+}\,\sum_{(m_a,m_n)\in\Lambda_\mathfrak{m}}
   \, \sum_{\boldsymbol{x}\in\Pi^1_+}\operatorname{Res}_{\boldsymbol{x}} \omega'(\mathfrak{m},\boldsymbol{x})\\
&- \frac{e^{-4\pi r \mathfrak{q}}}{2} \, \underbrace{\lim_{\varepsilon\to0^+}\,\sum_{(m_a,m_n)\in\Lambda_\mathfrak{m}}
   \, \sum_{\boldsymbol{x}\in\Pi^2_+}\operatorname{Res}_{\boldsymbol{x}} \omega'(\mathfrak{m},\boldsymbol{x})}_{A}.
\end{aligned}
\end{equation}

Next we show that $A$ is equal to zero. It will prove useful to introduce the coordinates
\begin{equation}
 w =z^{-1} = e^{2\pi r - i \theta}, \quad \overline{w} =(\overline{z})^{-1} = e^{2\pi r + i \theta}
\end{equation}
at this point already. In terms of the new variables
\begin{equation}\label{eq:M1P2QQVars1}
a =n_{Q_1}+n_{Q_2}+3m_a \ , \quad b=n_{Q_1}+\frac{3m_a-m_n}{2} \ , \quad c = n_{Q_1}+n_{Q_2} \ , \quad d = n_{Q_1} \ ,
\end{equation}
the summation --- including powers of $w$ and $\overline{w}$ --- simplifies according to
\begin{align}
\sum_{(m_a,m_n)\in\Lambda_\mathfrak{m}} \, \sum_{\boldsymbol{x}\in\Pi^2_+}\longrightarrow (w \overline{w})\sum_{a=0}^{\infty} w^{\frac{a-1}{3}}\,\sum_{c=0}^{\infty} \left(\overline{w}\right)^{\frac{c-1}{3}}\,\sum_{b=0}^{a}\,\sum_{d=0}^{b}\,_{\big| \substack{c-a \,\in\, 3\mathbb{N}_0\\ c\,\notin\,3\mathbb{N}_0 +1}}. \label{eq:ZPowQQ}
\end{align}
As for eq.~\eqref{eq:ZModel1Final}, the limit $\varepsilon \to 0^+$ in $A$ can be taken before summation, since there are only finitely many terms for a given order of $w$ and $\overline{w}$. From the right-hand side of eq.~\eqref{eq:ZPowQQ} we see that the sums produce fractional powers of $w$ and $\overline{w}$ only. Indeed there are three cases
\begin{equation}\label{eq:ZPowQQ3Cases}
\begin{tabular}{lllll}
(1) & $a \in 3\mathbb{N}_0$ & $\Rightarrow $ & $c \in 3\mathbb{N}_0$, & powers of $w$ and $\overline{w}$ in $\mathbb{N}_0-\frac{1}{3}$ \ ,\\
(2) & $a \in 3\mathbb{N}_0+2$ & $\Rightarrow $ & $c \in 3\mathbb{N}_0+2$, & powers of $w$ and $\overline{w}$ in $\mathbb{N}_0+\frac{1}{3}$\ , \\
(3) & $a \in 3\mathbb{N}_0+1$ & $\Rightarrow $ & $c \in 3\mathbb{N}_0+1$, & powers of $w$ and $\overline{w}$ in $\mathbb{N}_0$ \ ,
\end{tabular}
\end{equation}
where the last case is excluded in eq.~\eqref{eq:ZPowQQ}. Due to the residual K\"ahler transformation that still needs to be fixed, it is not obvious from the present discussion which of these three cases corresponds to integer powers in the partition function. We will later on see that the powers within the first term in eq.~\eqref{eq:ZM1P2} are spaced by integers. All of them are --- as the terms in $A$ are as well --- multiplied by $\left(w\overline{w}\right)^{1-q}$. This common prefactor will be removed by the final K\"ahler transformation, such that the first two cases in eq.~\eqref{eq:ZPowQQ3Cases} correspond two fractional powers in the partition function. Precisely these two cases are realized in $A$, hence they independently have to sum to zero. Now, we show this cancellation by an explicit calculation.

In order to treat both cases simultaneously, we replace $a$ and $c$ introduced in eq.~\eqref{eq:M1P2QQVars1} with
\begin{equation} \label{eq:M1P2QQVars2}
n_a = \frac{a-2m}{3} \ ,\quad n_c = \frac{c-2m}{3} \quad \text{for }m=1,2 \ ,
\end{equation}
such that $m=0$ corresponds to the first and $m=1$ to the second case in eq.~\eqref{eq:ZPowQQ3Cases}. With this we have
\begin{align}
A &= -\sum_{m=0}^1 \left(w\overline{w}\right)^{\frac{2+2m}{3}}\operatorname{Res}_{\boldsymbol{x}=0}\left(Z_{\text{sing}} \left| \left(w\overline{w}\right)^{-\left(x_1+x_2\right)}\sum_{n_a=0}^\infty(-w)^{n_a} \sum_{b=0}^{3n_a+2m} Z_{\text{reg}}   \right|^2 \right), \label{eq:M1P2A}
\end{align}
where
\small
$$
\begin{aligned}
Z_{\text{sing}} &= \frac{1}{\operatorname{sin}[\pi(2x_1+x_2)]\operatorname{sin}[\pi(x_1+2x_2)]}\cdot \frac{\operatorname{cos}\left[\pi\,\frac{1-8m+6x_1}{6}\right]^6\operatorname{cos}\left[\pi\,\frac{1+4m+6x_2}{6}\right]^6}{\pi^7 \operatorname{cos}\left[\pi\,\frac{1+4m-x_1-x_2}{6}\right]^3} \ ,\\
Z_{\text{reg}} &= \frac{(2b-2m-3n_a+x_1-x_2)\,\Gamma\left(\frac{1+4m}{3}-b+2n_a-x_1\right)^6\Gamma\left(\frac{1-2m}{3}+b-n_a-x_2\right)^6}{\Gamma\left(\frac{2+2m}{3}+n_a-x_1-x_2\right)^3\Gamma\left(1+b-x_1-2x_2\right)\Gamma\left(1+2m-b+3n_a-2x_1-x_2\right)} \ .
\end{aligned}
$$
\normalsize
Here, complex conjugation does not act on $x_1$ and $x_2$ and the residue is with respect to the oriented pair of divisors $(2x_1+x_2,x_1+2x_2)$. An explicit evaluation then gives
\begin{equation}
\begin{aligned}
A &= \frac{27\sqrt{3}}{512\pi^9}\sum_{m=0}^1 (-1)^{m+1}\Bigg|\sum_{n_a=0}^{\infty}\frac{(-1)^{n_a} w^{n_a+\frac{2+2m}{3}}}{\Gamma\left(\frac{2+2m}{3}+n_a\right)^3}\\
&\,\,\,\, \underbrace{\sum_{b=0}^{3n_a+2m}\frac{(2b-2m-3n_a)\Gamma\left(\frac{1+4m}{3}-b+2n_a\right)^6\Gamma\left(\frac{1-2m}{3}+b-n_a \right)^6}{\Gamma\left(1+b\right)\Gamma\left(1+2m-b+3n_a\right)}}_{=0}\Bigg|^2=0 \ .
\end{aligned}
\end{equation}
The inner sum vanishes due to antisymmetry of its summand under the transformation $b \longmapsto 3n_a+2m-b = a-b$, which is indeed nothing but $\mathfrak{s}$ expressed in terms of the variables introduced in eqs.~\eqref{eq:M1P2QQVars1} and \eqref{eq:M1P2QQVars2}. By this we have established
\begin{equation}\label{eq:ZM1P2_2}
Z_{S^2,r\ll0}^{1,12,6}(r,\theta) \,=\,-\frac{e^{-4\pi r \mathfrak{q}}}{2} \, \lim_{\varepsilon\to0^+}\,\sum_{(m_a,m_n)\in\Lambda_\mathfrak{m}}
   \, \sum_{\boldsymbol{x}\in\Pi^1_+}\operatorname{Res}_{\boldsymbol{x}} \omega'(\mathfrak{m},\boldsymbol{x}) \ . 
\end{equation}

Similar to the previous calculations we introduce new variables
\begin{equation}\label{eq:M1P2SumVars}
a =n_{P}+m_a \ , \quad c =n_{P}\ ,\quad k= n_{Q_2}+\frac{3m_a+m_n}{2} \ ,\quad l = n_{Q_2}\ , 
\end{equation}
in terms of which the sums in eq.~\eqref{eq:ZM1P2_2} simplify, then the partition function takes the form
\begin{equation}
\begin{aligned}
Z_{S^2,r\ll0}^{1,12,6}(r,\theta) =&-\frac{\left(w \overline{w}\right)^{1-\mathfrak{q}}}{2} \lim_{\varepsilon \to 0^+}\\
& \operatorname{Res}_{\boldsymbol{x}=0} \left( e^{2x_1 \varepsilon} Z_{\text{sing}} \left|w^{-(x_1+x_2)}\sum_{a=0}^\infty w^a e^{\varepsilon a} \sum_{k=0}^\infty (-1)^k Z_{2} e^{-\varepsilon k}\right|^2\right)\text{,}
\end{aligned} \label{eq:ZM1P2Final}
\end{equation}
where
\begin{equation}
\begin{aligned}
Z_{\text{sing}} &= \frac{\operatorname{sin}(\pi x_1)^6\operatorname{sin}(\pi x_2)^6\operatorname{sin}\left[\pi \left(x_1+2x_2\right)\right]}{\pi^9 \operatorname{sin}\left[\pi (x_1+x_2)\right]^3\operatorname{sin}\left[\pi(2x_1+x_2)\right]}\\
Z_2 &= \left(1+3a-2k+x_1-x_2\right) \cdot \\
&\,\qquad \frac{\Gamma (-a+k-x_1)^6\Gamma (1+2a-k-x_2)^6\Gamma(-1-3a+k+x_1+2x_2)}{\Gamma(1+k-2x_1-x_2)\Gamma(1+a-x_1-x_2)^3}.
\end{aligned}\label{eq:ZM1P2FinalTerms}
\end{equation}   
The residue is taken with respect to the ordering of divisors specified in eq.~\eqref{eq:M1jLables} and complex conjugation does not act on $x_1$ and $x_2$. At this point a few comments are in order:
\begin{enumerate}
\item For given powers of $w$ and $\overline{w}$ there are --- as opposed to eq.~\eqref{eq:ZModel1Final} --- infinitely many terms in eq.~\eqref{eq:ZM1P2Final}. At the technical level the appearance of this infinite sum can be understood from the following consideration: In eq.~\eqref{eq:ZM1P2_2} the partition function is given by summing over the residues at all poles stemming from the divisors $(D^{n_{Q_2}}_{Q_2},D^{n_P}_P)$. For fixed $n_P$ there are still infinitely many poles on the divisor $D^{n_P}_{P}$, enumerated by $n_{Q_2}$. Now recall that the line $\partial H$ --- given in eq.~\eqref{eq:UntiltedLine} --- by construction depends on the same linear combination of $x_1$ and $x_2$ that appears in the classical term $Z'_{\text{cl}}$, given after eq.~\eqref{eq:ZModel1}. Since this term gives the powers of $w$ and $\overline{w}$ once they have been introduced, their exponents are constant along lines parallel to $\partial H$, which is the case for $D^{n_P}_{P}$. Having tilted the critical line to $\partial H'$, the exponents of $w$ and $\overline{w}$ are no longer constant along $D^{n_P}_{P}$ but vary at order $\varepsilon$. This results in the factor $\operatorname{exp}(-\varepsilon k)$, which automatically regularizes the possibly divergent sum. Note that the factor $\operatorname{exp}(\varepsilon a)$ is unproblematic, as we do not require the sum over $a$ to converge. 
\item The factor $Z_2$ in eq.~\eqref{eq:ZM1P2Final} is not for all values of $a$ and $k$ regular at the origin. By construction every pole in the sum is located on the intersection of some $(D^{n_{Q_2}}_{Q_2},D^{n_P}_P)$. Depending on $a$ and $k$, the pole can, however, lie on additional divisors. This is the case if at least one of the Gamma functions in the numerator of $Z_2$ diverges at the origin. Intuitively this was to be expected due to inclusions of the type  $\{p^{\vec n}_{\vec \jmath}\} \subset \{p^{\vec m}_{\vec k}\}$. Poles located on more than two divisors are treated as described in Appendix \ref{sec:MBDegenerate}.
\item As has already been argued at the beginning of this section, antisymmetry under $\mathfrak{s}$ is not manifest in eq.~\eqref{eq:ZM1P2Final}. In fact, the variables of summation introduced in eq.~\eqref{eq:M1P2SumVars} do not even transform amongst themselves. The variable $l = n_{Q_2}$, for example, is mapped to $n_{Q_1}$, which does not appear in eq.~\eqref{eq:M1P2SumVars} at all.
\end{enumerate}

From eq.~\eqref{eq:ZM1P2Final} the partition function can in principle be evaluated up to any fixed order in $w$ and $\overline{w}$, and using eq.~\eqref{eq:ZtoK} we can read off geometric data of the Calabi-Yau threefold~$\mathcal{Y}_{1,12,6}$. For this purpose it is sufficient to keep terms of order $0$ in $\overline{w}$, we thus set $c=0$. For a given value of $a$, and by this a given power of $w$, it then depends on the values of $k$ and $l$ how many and which divisors intersect. Hence a distinction of cases is needed, which is, however, not a conceptual obstacle. For those cases in which $k$ and $l$ span an infinite range it is essential to calculate the residue without fixing their value, such that the infinitely many contributions can be summed up afterwards. This turns out to be computationally expensive. The sums would in fact be divergent without the regularization by $\operatorname{exp}(-\varepsilon k)$ and $\operatorname{exp}(-\varepsilon l)$. We observe that only finitely many terms contribute to the $(\operatorname{log}z)^3$ term at a given order, hence we are able to determine the exact fundamental period
\begin{equation} \label{eq:M1P2FundPeriod}
\begin{aligned}
\omega^{1,12,6}_{0,r\ll0}(w) &= \sum_{a=0}^\infty  \sum_{n=0}^a (1+3a-2n)\cdot\frac{(2a-n)!^6}{n!\,(1+3a-n)!\,a!^3\,(a-n)!^6}\,(-w)^a \\
&= 1 - 11w +559 w^2 -42\,923 w^3+3\,996\,751 w^4 -416\,148\,761 w^5 +\ldots\ . 
\end{aligned}
\end{equation}
Using $\chi\left(\mathcal{Y}_{1,12,6}\right)=-102$ we fix the overall normalization of the partition function such that it agrees with eq.~\eqref{eq:expK},\footnote{The required K\"ahler transformation is divison by $8\pi^3(w\overline{w})^{1-\mathfrak{q}} \left|\omega^{1,12,6}_{0,r\ll0}(w)\right|^2$ with $\omega^{1,12,6}_{0,r\ll0}(w)$ as in eq.~\eqref{eq:M1P2FundPeriod}.} then we read off the degree as well as the leading order degree $d$ integral genus zero Gromov-Witten invariants
\begin{equation} \label{eq:X1126Inv}
  \operatorname{deg}(\mathcal{Y}_{1,12,6}) \,=\, 21 \ , \qquad N_d(\mathcal{Y}_{1,12,6}) \,=\,  387, \ 4\,671, \quad \ldots \ .
\end{equation}
This is in agreement with eq.~\eqref{eq:DataCY2} and with the results in ref.~\cite{Miura:2013arxiv}, in which these Gromov--Witten invariants are determined by mirror symmetry for a conjectured second geometric Calabi--Yau threefold phase.

Note that this agreement also confirms that the performed strong coupling analysis in Section~\ref{sec:strongphase} is legitimate. In particular, we see that the Euler characteristic of the variety $\mathcal{Y}_{1,12,6}$ --- that is to say the Witten index $(-1)^F$ of the $SSSM_{1,12,6}$ --- is in accord with the two sphere partition function. Thus we justify in retrospect that --- at least for the specific \model s discussed in this section --- the analyzed non-generic component \eqref{eq:Yvariety} of the degeneration locus is the only relevant contribution in the strong coupling phase $r\ll 0$. 

\subsection{The model $SSSM_{2,9,6}$}
As the analysis of the \model{} $SSSM_{2,9,6}$ parallels in many aspects the discussion of the model $SSSM_{1,12,6}$ in Section~\ref{sec:M1General}, our presentation is less detailed here. In the following we focus on new features of the model $SSSM_{2,9,6}$. 

The model $SSSM_{2,9,6}$ comes with the gauge group
\begin{equation}
    G \,=\, \frac{U(1) \times \operatorname{USp}(4)}{\mathbb{Z}_2} \ ,
 \end{equation}   
and according to Table~\ref{tb:spec1} its irreducible representations are the $\operatorname{USp}(4)$ singlets $P_{[ij]}$ and $\phi_a$ with $U(1)$ charge $-2$ and $+2$ and the multiplets $Q$ and $X_i$ in the defining representation $\mathbf{4}$ of $\operatorname{USp}(4)$ and with $U(1)$ charge $-3$ and $+1$.\footnote{The Lie group $\operatorname{USp}(4)$ is isomorphic to $\operatorname{Spin}(5)$, and the defining representation~$\mathbf{4}$ of $\operatorname{USp}(4)$ is the spinor representation $\mathbf{4}_s$ of $\operatorname{Spin}(5)$.} Compared to the $SSSM_{1,12,6}$, the non-Abelian gauge group factor and the multiplicity of the singlet multiplets~$\phi_a$ are changed.

\begin{figure}[t]
\centering
\includegraphics[width=0.5\textwidth]{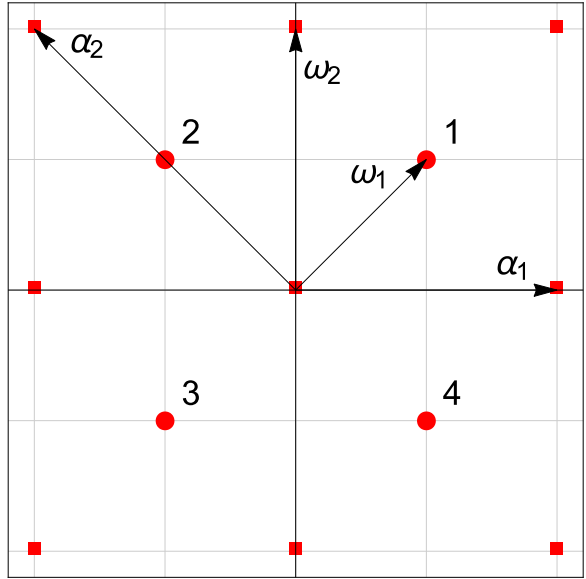}
\caption{\label{fig:USp4WeightLattice}Depicted is the $\mathfrak{usp}(4)\simeq\mathfrak{so}(5)$ weight lattice generated by the fundamental weights $\omega_1$ and $\omega_2$. The weights of the defining representation~$\mathbf{4}$ are pictured as red circles and  numerated for reference. The ten roots --- including two Cartan elements at the origin --- are represented by red squares, and $\alpha_1$ and $\alpha_2$ are the simple roots.}
\end{figure}

Let us recall the structure of the Lie algebra $\mathfrak{usp}(4)\simeq\mathfrak{so}(5)$, which is summarized in terms of the weight lattice in Figure~\ref{fig:USp4WeightLattice} and the Cartan matrix
\begin{equation}
  A_{\mathfrak{usp}(4)} \,=\, \begin{pmatrix} 2 & -1 \\ -2 & 2 \end{pmatrix} \ .
\end{equation}
We identify the Dynkin labels~$(\lambda_1,\lambda_2)$ --- specifying the weight $\lambda_1\omega_1+\lambda_2\omega_2$ --- of the positive roots $\Delta^+$ as
\begin{equation}
   \Delta^+ \,=\, \left\{\, (2,-1),\, (2,0),\, (0,1),\, (-2,2) \,\right\} \ ,
\end{equation}
and the Dynkin labels of the defining representation $\mathbf{4}$ as
\begin{equation}
  w(\mathbf{4})\,=\, \left\{\, (1,0),\, (-1,1),\, (-1,0),\, (1,-1)\,\right\} \ .
\end{equation}
The pairing of the weight lattice is given by the symmetric quadratic form matrix of $\mathfrak{usp}(4)$, which yields
\begin{equation}
    \langle \omega_1, \omega_1 \rangle \,=\,  \langle \omega_1, \omega_2 \rangle \,=\, \frac12 \ , \quad
    \langle \omega_2, \omega_2 \rangle \,=\, 1 \ .
\end{equation}

As before, for the two sphere partition function we need to determine the magnetic quantum numbers $\mathfrak{m}$ with respect to the gauge group $G$. We first observe that the representations of the gauge group~$G$ are induced from those representations of the double cover~$U(1)\times\operatorname{USp}(4)$ corresponding to highest weight vectors $\lambda\in\Lambda_w$ of $\mathfrak{u}(1)\times\mathfrak{usp}(4)$ with
\begin{equation} \label{eq:hw2}
   \lambda\,=\,\lambda_a q + \lambda_1 \omega_1 + \lambda_2 \omega_2 \quad
   \text{for} \quad \lambda_a + \lambda_1 \in 2\mathbb{Z} \ .
\end{equation}
The Abelian charge $q$ is canonically normalized as
\begin{equation}
  \langle q , q \rangle \,=\, 1 \ .
\end{equation}  
Then we determine again the magnetic charge quanta $\mathfrak{m}$ by evaluating the generalized Dirac quantization condition \cite{Goddard:1976qe}. These are given by all those $\mathfrak{m}$ with $\langle \lambda , \mathfrak{m} \rangle\in\mathbb{Z}$ for all weights of the representations of the gauge group~$G$. For the highest weights~\eqref{eq:hw2}, we arrive at
\begin{equation}
  \mathfrak{m} \,=\, \frac12 m_a q + m_1 \omega_1 + m_2 \omega_2  \quad\text{with}\quad m_1,m_a + m_2 \in 2\mathbb{Z} \ ,
\end{equation}  
and the magnetic charge lattice
\begin{equation}
   \Lambda_\mathfrak{m} \,=\, \left\{ \, (m_a,m_1,m_2)\in\mathbb{Z}^3 \,\middle|\, m_1, m_a+m_2 \in 2\mathbb{Z} \, \right\} \ .
\end{equation}
Analogously to the other \model, we set
\begin{equation}
   \boldsymbol{\sigma} \,=\, \frac12 \sigma_a q + \sigma_1 \omega_1+\sigma_2 \omega_2 \ , \qquad
   \boldsymbol{r} = 2 r q \ , \qquad 
   \boldsymbol{\theta}=2 \theta q \quad \text{with} \quad  \theta \sim \theta + 2 \pi \ .
\end{equation} 
Now we are ready to spell out the two sphere partition function
\begin{equation} \label{eq:ZModel2Raw}
    Z_{S^2}^{2,9,6}(r,\theta) \,=\, \frac{1}{64\pi^3}\!\!\sum_{(m_a,\vec m)\in\Lambda_\mathfrak{m}}\!\!
  \int_{\mathbb{R}^3}\!\! d^3\boldsymbol{\sigma}\,
    Z_G(\vec m,\vec\sigma)
    Z_\text{matter}(\mathfrak{m},\boldsymbol{\sigma})
    Z_{\text{cl}}(m_a,\sigma_a,r,\theta) \ ,
\end{equation}
where $\boldsymbol{\sigma}=(\sigma_a,\sigma_1,\sigma_2)=(\sigma_a,\vec\sigma)$, $\vec m=(m_1,m_2)$, $d^3\boldsymbol{\sigma} = \frac12 d\sigma_ad\sigma_1d\sigma_2$ and
\begin{equation}
\begin{aligned}
   &Z_G(\vec m,\vec\sigma)\,=\,(-1)^{m_2}\cdot\frac{m_1^2 +4\sigma_1^2}{16}\cdot\frac{\left(m_1+m_2\right)^2+4\left(\sigma_1+\sigma_2\right)^2}{4}\\
     &\qquad\qquad\qquad\cdot\frac{\left(\frac{m_1}{2}+m_2\right)^2+\left(\sigma_1+2\sigma_2\right)^2}{4}\cdot\frac{m_2^2 +4\sigma_2^2}{4} \ , \\[1ex]
   &Z_{\text{cl}}(m_a,\sigma_a,r,\theta) \,=\, e^{-4 \pi i  r \,\sigma_{a}-i\theta \, m_{a}} \ , \\[1ex]
   &Z_\text{matter}(\mathfrak{m},\boldsymbol{\sigma}) \,=\,
      Z_P(m_a,\sigma_a)^{15}\,Z_\phi(m_a,\sigma_a)^{9}\,Z_Q(\mathfrak{m},\boldsymbol{\sigma})\,
      Z_X(\mathfrak{m},\boldsymbol{\sigma})^6 \ .
\end{aligned}
\end{equation}
The individual contributions to $Z_\text{matter}$ are given by
\small
$$
\begin{aligned}
Z_P  &= \frac{\Gamma \left(\frac{m_{a}}{2}-q+i \sigma_{a}+1\right)}{\Gamma
   \left(\frac{m_{a}}{2}+q-i \sigma_{a}\right)} \ , \hspace{33mm} 
   Z_\phi = \frac{\Gamma \left(-\frac{m_{a}}{2}+q-i \sigma_{a}\right)}{\Gamma \left(-\frac{m_{a}}{2}-q+i \sigma_{a}+1\right)} \ , \\
Z_Q &=\underbrace{\frac{\Gamma\left(\frac{3m_a}{4}-\frac{m_1}{4}-\frac{m_2}{4}-\frac{3q}{2}+\frac{3i\sigma_a}{2}-\frac{i\sigma_1}{2}-\frac{i\sigma_2}{2}+1\right)}{\Gamma\left(\frac{3m_a}{4}-\frac{m_1}{4}-\frac{m_2}{4}+\frac{3q}{2}-\frac{3i\sigma_a}{2}+\frac{i\sigma_1}{2}+\frac{i\sigma_2}{2}\right)}}_{Z_{Q_1}} \underbrace{\frac{\Gamma\left(\frac{3m_a}{4}-\frac{m_2}{4}-\frac{3q}{2}+\frac{3i\sigma_a}{2}-\frac{i\sigma_2}{2}+1\right)}{\Gamma\left(\frac{3m_a}{4}-\frac{m_2}{4}+\frac{3q}{2}-\frac{3i\sigma_a}{2}+\frac{i\sigma_2}{2}\right)}}_{Z_{Q_2}}\\
&\cdot\underbrace{\frac{\Gamma\left(\frac{3m_a}{4}+\frac{m_1}{4}+\frac{m_2}{4}-\frac{3q}{2}+\frac{3i\sigma_a}{2}+\frac{i\sigma_1}{2}+\frac{i\sigma_2}{2}+1\right)}{\Gamma\left(\frac{3m_a}{4}+\frac{m_1}{4}+\frac{m_2}{4}+\frac{3q}{2}-\frac{3i\sigma_a}{2}-\frac{i\sigma_1}{2}-\frac{i\sigma_2}{2}\right)}}_{Z_{Q_3}} \underbrace{\frac{\Gamma\left(\frac{3m_a}{4}+\frac{m_2}{4}-\frac{3q}{2}+\frac{3i\sigma_a}{2}+\frac{i\sigma_2}{2}+1\right)}{\Gamma\left(\frac{3m_a}{4}+\frac{m_2}{4}+\frac{3q}{2}-\frac{3i\sigma_a}{2}-\frac{i\sigma_2}{2}\right)}}_{Z_{Q_4}}\ ,\\
Z_X &=\underbrace{\frac{\Gamma\left(-\frac{m_a}{4}-\frac{m_1}{4}-\frac{m_2}{4}+\frac{q}{2}-\frac{i\sigma_a}{2}-\frac{i\sigma_1}{2}-\frac{i\sigma_2}{2}+1\right)}{\Gamma\left(-\frac{m_a}{4}-\frac{m_1}{4}-\frac{m_2}{4}-\frac{q}{2}+\frac{i\sigma_a}{2}+\frac{i\sigma_1}{2}+\frac{i\sigma_2}{2}\right)}}_{Z_{X_1}} \underbrace{\frac{\Gamma\left(-\frac{m_a}{4}-\frac{m_2}{4}+\frac{q}{2}-\frac{i\sigma_a}{2}-\frac{i\sigma_2}{2}+1\right)}{\Gamma\left(-\frac{m_a}{4}-\frac{m_2}{4}-\frac{q}{2}+\frac{i\sigma_a}{2}+\frac{i\sigma_2}{2}\right)}}_{Z_{X_2}}\\
&\cdot \underbrace{\frac{\Gamma\left(-\frac{m_a}{4}+\frac{m_1}{4}+\frac{m_2}{4}+\frac{q}{2}-\frac{i\sigma_a}{2}+\frac{i\sigma_1}{2}+\frac{i\sigma_2}{2}+1\right)}{\Gamma\left(-\frac{m_a}{4}+\frac{m_1}{4}+\frac{m_2}{4}-\frac{q}{2}+\frac{i\sigma_a}{2}-\frac{i\sigma_1}{2}-\frac{i\sigma_2}{2}\right)}}_{Z_{X_3}}\underbrace{\frac{\Gamma\left(-\frac{m_a}{4}+\frac{m_2}{4}+\frac{q}{2}-\frac{i\sigma_a}{2}+\frac{i\sigma_2}{2}+1\right)}{\Gamma\left(-\frac{m_a}{4}+\frac{m_2}{4}-\frac{q}{2}+\frac{i\sigma_a}{2}-\frac{i\sigma_2}{2}\right)}}_{Z_{X_4}}\ .
\end{aligned}
$$
\normalsize
Here the indices of the individual factors $Z_{Q_i}$ and $Z_{X_i}$ of $Z_Q$ and $Z_X$ number the weights of the defining representation~$\mathbf{4}$ as in Figure~\ref{fig:USp4WeightLattice}. Using the reflection formula~\eqref{eq:GammaReflect} for Gamma functions, we confirm that the defined two sphere partition function~$Z_{S^2}^{2,9,6}$ is indeed real.

Let us also observe that the integrand of the two sphere partition function is invariant with respect to the Weyl symmetry $\mathcal{W}_{\mathfrak{usp}(4)} \simeq S_2 \ltimes (\mathbb{Z}_2\times\mathbb{Z}_2) \simeq D_4$ --- i.e., the order eight dihedral group $D_4$ --- generated by
\begin{equation}
\begin{aligned}
   &\mathfrak{s} :  (m_1,m_2,\sigma_1,\sigma_2) \longmapsto (-m_1,m_1+m_2,-\sigma_1+\sigma_1+\sigma_2) \ , \\
   &\mathfrak{r} :  (m_1,m_2,\sigma_1,\sigma_2) \longmapsto (-m_1-2m_2,m_1+m_2,-\sigma_1-2\sigma_2,\sigma_1+\sigma_2) \ . 
\end{aligned}   \label{eq:WeylM2A}
\end{equation}

To further evaluate the partition function~$Z_{S^2}^{2,9,6}$, we substitute the integration variables $\boldsymbol{\sigma}=(\sigma_a,\vec\sigma)$ by $\boldsymbol{x}=(x_1,x_2,x_3)$ according to
\begin{equation}  \label{eq:SubsModel2}
  \sigma_a = i \left(x_1+x_2+x_3-\mathfrak{q}\right) \ , \quad \sigma_1 = 2i \left(x_1-x_2\right) \ , 
  \quad \sigma_2 = -i \left(3x_1+x_2+x_3\right) \ .
\end{equation}
Then we arrive at
\begin{equation}
   Z_{S^2}^{2,9,6}(r,\theta) = i\, \frac{e^{-4\pi r \mathfrak{q}}}{32\pi^3}\!\!\!
 \sum_{(m_a,\vec m)\in\Lambda_\mathfrak{m}}\!\!\!
 \int_{\gamma+i\mathbb{R}^3}\!\!\! \omega(\mathfrak{m},\boldsymbol{x}) \,dx_1 \wedge dx_2 \wedge dx_3 \ , \quad
 \gamma = \left(-\tfrac{\mathfrak{q}}2,-\tfrac{\mathfrak{q}}2,2\mathfrak{q}\right) \ ,\label{eq:ZModel2Formal}
\end{equation}
with
\begin{equation} \label{eq:ZModel2}
   \omega(\mathfrak{m},\boldsymbol{x})\,=\,
   Z'_G(\vec m,\boldsymbol{x})Z'_P(m_a,\boldsymbol{x})^{15}Z'_\phi(m_a,\boldsymbol{x})^{9}
   Z'_Q(\mathfrak{m},\boldsymbol{x})Z'_X(\mathfrak{m},\boldsymbol{x})^6
   Z'_\text{cl}(r,\theta,m_a,\boldsymbol{x}) \ ,
\end{equation}
where $Z'_\bullet=\left.Z_\bullet\right|_{\boldsymbol{\sigma}=\boldsymbol{\sigma}(\boldsymbol{x})}$ for $\bullet=G,P,\phi,Q,X$, and
\begin{equation}
     Z'_{\text{cl}}(r,\theta,m_a,\boldsymbol{x}) \,=\, e^{4 \pi r \left(x_1+x_2+x_3\right) - i\,\theta \, m_a} \ .
\end{equation}
The integral in eq.~\eqref{eq:ZModel2Formal} is now of the type~\eqref{eq:MBInt} and we proceed with its evaluation as in Appendix~\ref{sec:MB}. We record that in terms of the new variables $\boldsymbol{x}$ the generators eq.~\eqref{eq:WeylM2A} of the Weyl group $\mathcal{W}_{\mathfrak{usp}(4)}$ take the form
\begin{equation}\label{eq:WeylM2B}
\begin{aligned}
   &\mathfrak{s} :  (m_1,m_2,x_1,x_2) \mapsto (-m_1,m_1+m_2,x_2,x_1) \, , \\
   &\mathfrak{r} : (m_1,m_2,x_1,x_2,x_3) \mapsto (-m_1-2m_2,m_1+m_2,x_2,-2x_1-x_2-x_3,3x_1+x_2+2x_3) \, . 
\end{aligned}  
\end{equation}
which induces the action on the signed volume form
\begin{equation} \label{eq:Wusp(4)vol}
\begin{aligned}
  \mathfrak{s}: dx_1 \wedge dx_2 \wedge dx_3\longmapsto -dx_1 \wedge dx_2 \wedge dx_3\ , \\
  \mathfrak{r}: dx_1 \wedge dx_2 \wedge dx_3\longmapsto +dx_1 \wedge dx_2 \wedge dx_3\ , 
  \end{aligned}
\end{equation}

Firstly, we again determine the divisors for the poles of the integrand $\omega(\mathfrak{m},\boldsymbol{x})$. Taking into account cancellation between the poles and zeros, we find the following divisors of poles in terms of the constraint integers $n_P, n_{Q_1},n_{Q_2}, n_{Q_3},n_{Q_4},n_{X_1},n_{X_2},n_{X_3},n_{X_4}$
\small
\begin{equation}\label{eq:DivisModel2}
\begin{aligned}
D^{n_P}_{P} &= x_1+x_2+x_3-n_P -\tfrac{m_a}{2}-1 &&\text{ for } n_P \geq \operatorname{Max}\left[0,-m_a\right] \ ,\\
D^{n_{Q_1}}_{Q_1} &= 2x_1+3x_2+2x_3-\tfrac{4+4n_{Q_1}+3m_a-m_1-m_2}{4} &&\text{ for }n_{Q_1} \geq \operatorname{Max}\left[0,\tfrac{-3m_a+m_1+m_2}{2}\right]\ ,\\
D^{n_{Q_2}}_{Q_2} &= 3x_1+2x_2+2x_3-\tfrac{4+4n_{Q_2}+3m_a-m_2}{4} &&\text{ for } n_{Q_2} \geq \operatorname{Max}\left[0,\tfrac{-3m_a+m_2}{2}\right]\ ,\\
D^{n_{Q_3}}_{Q_3} &= x_1+x_3-\tfrac{4+4n_{Q_3}+3m_a+m_1+m_2}{4} &&\text{ for } n_{Q_3} \geq \operatorname{Max}\left[0,\tfrac{-3m_a-m_1-m_2}{2}\right]\ ,\\
D^{n_{Q_4}}_{Q_4} &= x_2+x_3-\tfrac{4+4n_{Q_4}+3m_a+m_2}{4} &&\text{ for } n_{Q_4} \geq \operatorname{Max}\left[0,\tfrac{-3m_a-m_2}{2}\right]\ ,\\
D^{n_{X_1}}_{X_1} &= x_2-\tfrac{4n_{X_1}-m_a-m_1-m_2}{4} &&\text{ for } n_{X_1} \geq \operatorname{Max}\left[0,\tfrac{m_a+m_1+m_2}{2}\right]\ ,\\
D^{n_{X_2}}_{X_2} &= x_1-\tfrac{4n_{X_2}-m_a-m_2}{4} &&\text{ for } n_{X_2} \geq \operatorname{Max}\left[0,\tfrac{m_a+m_2}{2}\right]\ ,\\
D^{n_{X_3}}_{X_3} &= x_1+2x_2+x_3-\tfrac{-4n_{X_3}+m_a-m_1-m_2}{4} &&\text{ for } n_{X_3} \geq \operatorname{Max}\left[0,\tfrac{m_a-m_1-m_2}{2}\right]\ ,\\
D^{n_{X_4}}_{X_4} &= 2x_1+x_2+x_3-\tfrac{-4n_{X_4}+m_a-m_2}{4} &&\text{ for } n_{X_4} \geq \operatorname{Max}\left[0,\tfrac{m_a-m_2}{2}\right]\ .
\end{aligned}
\end{equation}
\normalsize
There is no contribution from $Z'_\phi$ because all its poles are canceled by the denominator of $Z'_P$.

Secondly, we identify $\boldsymbol{p}$ in eq.~\eqref{eq:MBw} as $\boldsymbol{p}= -4\pi r(1,1,1)$ from which we find the critical plane $\partial H$ introduced in eq.~\eqref{eq:MBHHigher} to be
\begin{equation}
\partial H = \left\{ x \in \mathbb{R}^3 \,|\, x_1 + x_2 + x_3 =\mathfrak{q} \right\}. \label{eq:UntiltedHM2}
\end{equation}
Since this critical line leads to degenerate simplices~\eqref{eq:MBSimplex}, it is necessary to apply the method of Appendix \ref{sec:MBParallel}. We thus introduce an additional factor such that
\begin{equation} \label{eq:OmegaPrimeM2}
  \omega(\mathfrak{m},\boldsymbol{x}) \, \longrightarrow \,
  \omega'(\mathfrak{m},\boldsymbol{x}) = \omega(\mathfrak{m},\boldsymbol{x}) \cdot e^{+4\pi r \varepsilon\,\boldsymbol{\delta}\cdot\boldsymbol{x}} \ ,
  \qquad \boldsymbol{\delta}=(\delta_1,\delta_2,\delta_3) \ ,
\end{equation}
where the limit $\varepsilon \to 0^+$ is taken outside the integral. This modifies the critical plane to
\begin{equation}
\partial H' = \left\{ x \in \mathbb{R}^3 \,\bigg|\, \sum_{i=1}^3\left(1+\varepsilon\,\delta_{i}\right)x_i=\mathfrak{q}- \frac{\mathfrak{q}}{2}\left(\delta_{1}+\delta_{2}-4\delta_3\right) \right\}\ ,\label{eq:TiltedHM2}
\end{equation}
which for $\delta_1 \neq \delta_3$ and $\delta_2 \neq \delta_3$ indeed results in having no degenerate simplices. With the two disjoint half-spaces bounded by the critical plane
\begin{equation}
\begin{aligned}
 H'_1 &= \left\{ x \in \mathbb{R}^3 \,\bigg|\, \sum_{i=1}^3\left(1+\varepsilon\,\delta_{i}\right)x_i < \mathfrak{q}- \frac{\mathfrak{q}}{2}\left(\delta_{1}+\delta_{2}-4\delta_3\right) \right\}\ ,\\
 H'_2 &= \left\{ x \in \mathbb{R}^3 \,\bigg|\, \sum_{i=1}^3\left(1+\varepsilon\,\delta_{i}\right)x_i > \mathfrak{q}- \frac{\mathfrak{q}}{2}\left(\delta_{1}+\delta_{2}-4\delta_3\right) \right\}\ ,
\end{aligned}\label{eq:HalfSpacesM2}
\end{equation}
the relevant half-spaces $H$ for the respective \model{} phases are
\begin{equation}
  H \,=\,\begin{cases} H'_1 & \text{for }r\gg0 \\ H'_2 & \text{for }r\ll0 \end{cases} \ .
\end{equation}  

Thirdly, we have to determine the set $\Pi$ of contributing poles that is introduced in eq.~\eqref{eq:MBpoles}. This is a straight forward but tedious calculation, for the sake of brevity we specialize to the case $\delta_3=0$ and $\delta_1, \delta_2 > 0$. Then we find
\begin{equation}
\Pi = \begin{cases} \left\{\, p^{\vec n_i}_{\vec\jmath_i}\, | \, i = 1,2,3,4 \,\right\} & \text{ for } r\gg 0 \ , \\[1.5ex]
 \left\{\, p^{\vec n_i}_{\vec\jmath_i}\, | \, i=5, \ldots, 20 \,\right\} & \text{ for } r\ll 0 \ , \\[1.5ex]
 \end{cases}\label{eq:M2Pi}
\end{equation}
in terms of the oriented divisor intersections with labels
\begin{equation}\label{eq:M2jLables}
\begin{aligned}
 &(\vec\jmath_1,\vec n_1)\,=\, (X_3,X_1,X_4; n_{X_3},n_{X_1},n_{X_4}) \ , &&(\vec\jmath_2,\vec n_2)\,=\, (X_3,X_1,Q_4; n_{X_3},n_{X_1},n_{Q_4}) \ , \\
 &(\vec\jmath_3,\vec n_3)\,=\, (X_3,X_2,X_4; n_{X_3},n_{X_2},n_{X_4}) \ , &&(\vec\jmath_4,\vec n_4)\,=\, (X_2,X_4,Q_3; n_{X_2},n_{X_4},n_{Q_3}) \ , \\
 &(\vec\jmath_5,\vec n_6)\,=\, (P,X_1,X_2; n_{P},n_{X_1},n_{X_2}) \ , &&(\vec\jmath_6,\vec n_6)\,=\, (P,X_1,Q_2; n_{P},n_{X_1},n_{Q_2}) \ , \\
 &(\vec\jmath_7,\vec n_7)\,=\, (P,Q_1,X_2; n_{P},n_{Q_1},n_{X_2}) \ , &&(\vec\jmath_8,\vec n_8)\,=\, (P,Q_1,Q_2; n_{P},n_{Q_1},n_{Q_2}) \ , \\
 &(\vec\jmath_9,\vec n_9)\,=\, (X_1,X_2,Q_3; n_{X_1},n_{X_2},n_{Q_3}) \ , &&(\vec\jmath_{10},\vec n_{10})\,=\, (X_1,X_2,Q_4; n_{X_1},n_{X_2},n_{Q_4}) \ , \\
 &(\vec\jmath_{11},\vec n_{11})\,=\, (X_1,Q_2,X_4; n_{X_1},n_{Q_2},n_{X_4}) \ , &&(\vec\jmath_{12},\vec n_{12})\,=\, (X_1,Q_2,Q_3; n_{X_1},n_{Q_2},n_{Q_3}) \ , \\
 &(\vec\jmath_{13},\vec n_{13})\,=\, (X_1,Q_2,Q_4; n_{X_1},n_{Q_2},n_{Q_4}) \ , &&(\vec\jmath_{14},\vec n_{14})\,=\, (X_3,Q_1,X_2; n_{X_3},n_{Q_1},n_{X_2}) \ , \\
 &(\vec\jmath_{15},\vec n_{15})\,=\, (X_3,Q_1,Q_2; n_{X_3},n_{Q_1},n_{Q_2}) \ , &&(\vec\jmath_{16},\vec n_{16})\,=\, (X_2,Q_3,Q_1; n_{X_2},n_{Q_3},n_{Q_1}) \ , \\
 &(\vec\jmath_{17},\vec n_{19})\,=\, (X_2,Q_4,Q_1; n_{X_2},n_{Q_4},n_{Q_1}) \ , &&(\vec\jmath_{18},\vec n_{18})\,=\, (X_4,Q_1,Q_2; n_{X_4},n_{Q_1},n_{Q_2}) \ , \\
 &(\vec\jmath_{19},\vec n_{19})\,=\, (Q_1,Q_2,Q_3; n_{Q_1},n_{Q_2},n_{Q_3}) \ , &&(\vec\jmath_{20},\vec n_{20})\,=\, (Q_1,Q_2,Q_4; n_{Q_1},n_{Q_2},n_{Q_4}) \ , \\
\end{aligned}    
\end{equation}
and the pole loci given by
\begin{equation}
 p^{\vec n_i}_{\vec\jmath_i} = \left(\,\bigcap_{k=1}^3 \,\{\,D^{(\vec n_i)_k}_{(\vec\jmath_i)_k} = 0 \, \} \right)\ .
\end{equation}
Moreover, the labels~\eqref{eq:M2jLables} have been chosen such that $\operatorname{sign}\Delta^{\vec n_i}_{\vec\jmath_i} = 1$ for all $i$. According to eq.~\eqref{eq:MBFormula} the partition function thus takes the form
\begin{equation}
   Z_{S^2}^{2,9,6}(r,\theta) \,=\,\frac{e^{-4\pi r \mathfrak{q}}}{4} \,\lim_{\varepsilon\to0^+}\,\sum_{(m_a,\vec{m})\in\Lambda_\mathfrak{m}}
   \, \sum_{\boldsymbol{x}\in\Pi}\operatorname{Res}_{\boldsymbol{x}} \omega'(\mathfrak{m},\boldsymbol{x}) \ ,\label{eq:ZM2}
\end{equation}
where the residue is with respect to the order of divisors specified in eq.~\eqref{eq:M2jLables}.

\subsubsection{The $SSSM_{2,9,6}$ phase $r\gg0$}
We now specialize to the phase $r \gg 0$, for which eq.~\eqref{eq:M2Pi} shows that there are four relevant pole loci
\footnotesize
$$
\begin{aligned}
p^{\vec n_1}_{\vec\jmath_1} &= \left( 
n_{X_1}+n_{X_3}-n_{X_4}-\tfrac{m_a+m_2}{4} ,  n_{X_1}-\tfrac{m_a+m_1+m_2}{4},  n_{X_4}-2n_{X_3}-3n_{X_1}+\tfrac{4m_a+m_1+2m_2}{4} \right) \ , \\ 
p^{\vec n_2}_{\vec\jmath_2} &= \left( 
-1-n_{X_1}+n_{X_3}-n_{Q_4}-\tfrac{m_a+m_2}{4},n_{X_1}-\tfrac{m_a+m_1+m_2}{4},1+ n_{Q_4}-n_{X_1}+\tfrac{4m_a+m_1+2m_2}{4}\right),\\
p^{\vec n_3}_{\vec\jmath_2} &= \left(
n_{X_2}-\tfrac{m_a+m_2}{4}, n_{X_2}+n_{X_4}-n_{X_3}-\tfrac{m_a+m_1+m_2}{4}, n_{X_3}-2n_{X_4}-3n_{X_2}+\tfrac{4m_a+m_1+2m_2}{4}
\right)\ ,\\ 
p^{\vec n_4}_{\vec\jmath_4} &= \left( 
n_{X_2}-\tfrac{m_a+m_2}{4},-1-n_{X_2}-n_{X_4}-n_{Q_3}-\tfrac{m_a+m_1+m_2}{4},1+ n_{Q_3}-n_{X_2}+\tfrac{4m_a+m_1+2m_2}{4} \right)\ .
\end{aligned}
$$
\normalsize
By calculations similar to those in eq.~\eqref{eq:inclusion} it can be shown that
\begin{equation}
  \Pi = \left\{\, p^{\vec n_i}_{\vec\jmath_i}\, | \, i = 1,2,3,4 \,\right\} = \Pi^1 \cup \Pi^2\, \quad \text{with}\quad \Pi^1 = \left\{\, p^{\vec n_1}_{\vec\jmath_1}\,\right\} \ ,\
   \Pi^2 = \left\{\, p^{\vec n_3}_{\vec\jmath_3}\,\right\} \ .
\end{equation}
The sets $\Pi^1$ and $\Pi^2$ are, however, not disjoint. Taking care to not count poles in their intersection twice, the partition function reads
\begin{equation}
\begin{aligned}
Z_{S^2,r\gg0}^{2,9,6}(r,\theta) \,=\,&\frac{e^{-4\pi r \mathfrak{q}}}{4} \,\lim_{\varepsilon\to0^+}\,\sum_{(m_a,\vec{m})\in\Lambda_\mathfrak{m}}
   \, \sum_{\boldsymbol{x}\in\Pi^1}\alpha(\boldsymbol{x})\operatorname{Res}_{\boldsymbol{x}} \omega'(\mathfrak{m},\boldsymbol{x})\\
+&\frac{e^{-4\pi r \mathfrak{q}}}{4} \,\lim_{\varepsilon\to0^+}\,\sum_{(m_a,\vec{m})\in\Lambda_\mathfrak{m}}
   \, \sum_{\boldsymbol{x}\in\Pi^2}\alpha(\boldsymbol{x})\operatorname{Res}_{\boldsymbol{x}} \omega'(\mathfrak{m},\boldsymbol{x})\\
\text{with } \alpha(\boldsymbol{x})&= \begin{cases}\frac{1}{2} \quad \text{for } \boldsymbol{x} \in \Pi^1 \cap \Pi^2 \\ 1 \quad \text{otherwise}\end{cases}.
\end{aligned}\label{eq:ZModel22Terms}
\end{equation}
Further, the two terms in eq.~\eqref{eq:ZModel22Terms} are equal. This is due to the symmetry of the integrand of the two sphere partition function under the generator $\mathfrak{s}$ of Weyl transformations given in eq.~\eqref{eq:WeylM2A}, which acts on the relevant divisor intersections as
\begin{equation}
(\vec\jmath_1,\vec n_1) \stackrel{\mathfrak{s}}{\longmapsto} (\vec\jmath_3,\vec n_3) \ , \quad(\vec\jmath_3,\vec n_3) \stackrel{\mathfrak{s}}{\longmapsto} (\vec\jmath_1,\vec n_1) \ .
\end{equation}
Here we have accounted for that the minus sign from the orientation-reversal of the ordered triplet of divisors precisely cancels the sign-reversal obtained by transforming the signed volume form in eq.~\eqref{eq:Wusp(4)vol}. With this we find
\begin{equation}\label{eq:ZModel21Term}
\begin{aligned}
Z_{S^2,r\gg0}^{2,9,6}(r,\theta) \,=\,&\frac{e^{-4\pi r \mathfrak{q}}}{2} \,\lim_{\varepsilon\to0^+}\,\sum_{(m_a,\vec{m})\in\Lambda_\mathfrak{m}}
  \, \sum_{\boldsymbol{x}\in\Pi^1}\alpha(\boldsymbol{x})\operatorname{Res}_{\boldsymbol{x}} \omega'(\mathfrak{m},\boldsymbol{x}) \ .
\end{aligned}
\end{equation}

Similarly to the previous calculations we introduce new variables
\begin{equation}
\begin{aligned}
	&a = n_{X_1}+n_{X_3}-m_a\ , && b= n_{X_3} -\frac{m_a-m_1-m_2}{2}\ , && c = n_{X_1}+n_{X_3}\ , \\
	&d = n_{X_3}  \ , && k  = n_{X_4}-\frac{m_a-m_2}{2} \ ,  && l = n_{X_4}  \ ,
\end{aligned}
\end{equation}
in order to simplify the summation in eq.~\eqref{eq:ZModel21Term}. In terms of $z$ and $\overline{z}$ introduced in eq.~\eqref{eq:zvar}, and with $\delta_1 = \delta_2 = \frac1{2\pi r}$ the partition function is found to be
\begin{align}
Z_{S^2,r\gg 0}^{2,9,6}(r,\theta) \,=\,&\frac{(z\overline{z})^{\mathfrak{q}}}{2} \,\lim_{\varepsilon\to0^+} \operatorname{Res}_{\boldsymbol{x}=0}
\Bigg( e^{2(x_1+x_2)\varepsilon} Z_{1} \left(z\overline{z}\right)^{-\sum_{i} x_i}\sum_{a=0}^\infty z^a e^{2a\varepsilon}\sum_{c=0}^\infty (\overline{z})^c e^{2c\varepsilon}\notag \\
&\sum_{b=0}^a\sum_{d=0}^c\sum_{k=0}^\infty\sum_{l=0}^\infty\alpha\, Z_{(a,b,k)} Z_{(c,d,l)}e^{-(k+l+b+d)\varepsilon} \Bigg)\ , \label{eq:ZModel2FinalPhase1}
\end{align}
with
\begin{align*}
Z_1 &= \,\pi^2\frac{\operatorname{sin}\left(\pi x_1\right)^6\operatorname{sin}\left[\pi\left( x_1+x_2+x_3\right)\right]^6\operatorname{sin}\left[\pi\left( x_1+x_3\right)\right]\operatorname{sin}\left[\pi\left(x_2+x_3\right)\right]}{ \operatorname{sin}\left(\pi x_2\right)^6\operatorname{sin}\left[\pi\left(2 x_1+x_2+x_3\right)\right]^6\operatorname{sin}\left[\pi\left(x_1+2x_2+x_3\right)\right]^6}\\
&\,  \hspace{2mm}\cdot\operatorname{sin}\left[\pi\left(3 x_1+2x_2+2x_3\right)\right]\operatorname{sin}\left[\pi\left(2x_1+3x_2+2x_3\right)\right]\ ,\\
Z_{(a,b,k)} &= \left(a-2k+3x_1+x_2+x_3\right)\left(b-k+x_1-x_2\right)\left(a-b-k+2x_1+2x_2+x_3\right) \\
&\, \hspace{2mm}\cdot \frac{\left(a-2b+x_1+3x_2+x_3\right)\Gamma\left(-a+k-x_1\right)^6\Gamma\left(1+a-x_1-x_2-x_3\right)^6}{\Gamma\left(1+a-b+x_2\right)^6\Gamma\left(1+b-x_1-2x_2-x_3\right)^6\Gamma\left(1+k-2x_1-x_2-x_3\right)^6}\\
&\, \hspace{2mm}\cdot \Gamma\left(1+a+b-2x_1-3x_2-2x_3\right)\Gamma\left(1+2a-b-x_1-x_3\right)\\
&\, \hspace{2mm}\cdot \Gamma\left(1+a+k-3x_1-2x_2-2x_3\right)\Gamma\left(1+2a-k-x_2-x_3\right)\ ,\\
\alpha &= \begin{cases}\frac{1}{2}\quad \text{for } k\leq a \text{ and } l\leq c\\1\quad \text{otherwise}\end{cases}\ .
\end{align*}
Note that the factor $\alpha$ is the same as the one defined in eq.~\eqref{eq:ZModel22Terms}, it prevents double counting of poles in $\Pi^1\cap \Pi^2$. Due to its appearance, eq.~\eqref{eq:ZModel2FinalPhase1} could not be written as compactly as the partition functions in eqs.~\eqref{eq:ZModel1Final} and \eqref{eq:ZM1P2Final}. However, eq.~\eqref{eq:ZModel2FinalPhase1} is still manifestly real, and its complex form is not a complication in the explicit evaluation. Moreover, in analogy to the factor $Z_2$ in eq.~\eqref{eq:ZM1P2FinalTerms}, the terms $Z_{(a,b,k)}$ and $Z_{(c,d,l)}$ are not regular at the origin for all values of the summation variables. Hence, some of the relevant poles are located on more than two divisors. Further, for a given order in $z$ and $\overline{z}$ there are two finite and two infinite sums. The infinite sums are automatically regularized by the exponential factors.

The explicit evaluation of eq.~\eqref{eq:ZModel2FinalPhase1} now proceeds in the same way as for the $r\ll0$ phase of $SSSM_{1,12,6}$. The infinite sums are here, however, observed to converge even for $\varepsilon = 0$. Moreover, contributions to the $\left(\operatorname{log}z\right)^3$ term are observed to only stem from poles in $\Pi^1\cap \Pi^2$. Since there are only finitely many such poles at a given order, we have been able to determine the exact fundamental period. Although its expression is too complicated to write down, its expansion has been checked to coincide with eq.~\eqref{eq:M1P2FundPeriod}. After fixing the overall normalization of the partition function by using $\chi\left(\mathcal{X}_{2,9,6}\right) = -102$,\footnote{The required K\"ahler transformation is divison by $-16\pi^3(z\overline{z})^{\mathfrak{q}} \left|\omega^{1,12,6}_{0,r\ll0}(z)\right|^2$ with $\omega^{1,12,6}_{0,r\ll0}(z)$ as in eq.~\eqref{eq:M1P2FundPeriod}.} we read off
\begin{equation} \label{eq:X296Inv}
   \operatorname{deg}(\mathcal{X}_{2,9,6}) \,=\, 21\ , \qquad  N_d(\mathcal{X}_{2,9,6}) \,=\,  387, \ 4\,671, \  \ldots \  ,
\end{equation}
in agreement with eq.~\eqref{eq:DataCY2}. The extracted geometric data justifies the assumption made at the end of Section~\eqref{sec:Xphase}, namely that the naive semi-classical derivation of the smooth target space variety $\mathcal{X}_{2,9,6}$ is correct even though it contains discrete points, where the non-Abelian gauge group $\operatorname{USp}(4)$ is not entirely broken.

Let us pause to stress the significance of this result. By calculating the two sphere partition functions of the two models $SSSM_{1,12,6}$ and $SSSM_{2,9,6}$, we find agreement among the invariants~\eqref{eq:X296Inv} and \eqref{eq:X1126Inv} in their respective phase $r\ll0$ and $r\gg0$.\footnote{In the same fashion we could check that the other two geometric phases of the $SSSM_{1,12,6}$ and $SSSM_{2,9,6}$ agree, that is to say $SSSM_{1,12,6}$ in the phase $r\gg0$ and $SSSM_{2,9,6}$ in the phase $r\ll 0$. However, by analytic continuation of the above result such a match is guaranteed  as well.} On the level of the partition functions we find the equality
\begin{equation}
   Z^{1,12,6}_{S^2}(r,\theta,\mathfrak{q}) \,=\, - \frac12 \,Z^{2,9,6}_{S^2}(-r,-\theta,1-\mathfrak{q}) \ ,
\end{equation}
which implies that the partition functions $Z^{1,12,6}_{S^2}$ and $Z^{2,9,6}_{S^2}$ differ only by a K\"ahler transformation resulting in the relative normalization factor $-\frac12$. This identity is highly non-trivial on the level of the two distinct integral representations~\eqref{eq:ZS2} of the two sphere partition function for the two \model s $SSSM_{1,12,6}$ and $SSSM_{2,9,6}$, and thus furnishes strong evidence in support of our duality proposal~\eqref{eq:dualpair} in Section~\ref{sec:duality}.

\subsection{Other conformal \model s}
Let us now briefly examine the remaining \model s in the Table~\ref{tb:models}. As has been argued in Section \ref{sec:duality}, the model $SSSM_{1,5,4}$ is self-dual with two equivalent geometric $T^2$ phases for $r\gg0$ and $r\ll0$ according to eq.~\eqref{eq:dualT2s}. We here want to provide further evidence for this duality in terms of the two-sphere partition function. While eq.~\eqref{eq:ZtoK} still holds, the sign-reversed exponentiated K\"ahler potential~\eqref{eq:expK} takes in flat coordinates the simple form
\begin{equation} \label{eq:ZTor}
  e^{-K(t)} \,=\, i (t -\bar t)\ .
\end{equation}
It gives rise to the Zamolodchikov metric $ds^2= \frac{i\,dt d\bar t}{t-\bar t} $ of the Teichm\"uller space of the torus. While the Zamolodchikov metric does not contain any additional geometric information of any geometric invariants, we nevertheless expect that the two sphere partition functions $Z_{S^2}^{1,5,4}$ takes with the canonical K\"ahler gauge in the (singular) large volume  phases the asymptotic form 
\begin{equation}
  Z^{1,5,4}_{S^2}(z) \,=\, \frac{c}{2\pi i} \left| \omega^{1,5,4}_0(z) \right|^2 (\log z - \log\bar z) + \ldots \quad \text{for} \quad |z|\text{ small} \ ,
\end{equation}
with the fundamental period $\omega^{1,5,4}_0(z)$ and a real constant $c$. Thus, our aim is to calculate $\omega^{1,5,4}_0(z)$ with the two sphere partition function correspondence \cite{Jockers:2012dk}, so as to show that they agree in the two geometric toroidal large volume phases.

For K3 surfaces the quantum K\"ahler metric does not receive non-perturbative worldsheet instanton corrections and the sign-reversed exponentiated K\"ahler potential~\eqref{eq:expK} exhibits in flat coordinates the characteristic form \cite{Halverson:2013qca}
\begin{equation} \label{eq:ZK3}
  e^{-K(t)} \,=\, \frac12 \kappa_{ab}(t^a -\bar t^a)(t^b-\bar t^b)\ ,
\end{equation}
where $t^a$ are the complexified K\"ahler moduli of the analyzed polarized K3~surface and $\kappa_{ab}$ are their intersection numbers. For the specific model $SSSM_{1,8,5}$ the large volume polarized K3~surfaces $\mathcal{X}_{\mathbb{S}_5}$ and $\mathcal{Y}_{\mathbb{S}^*_5}$ depend on a single K\"ahler modulus $t$ such that 
\begin{equation}
   e^{-K(t)} \,=\, \frac12 \deg(\mathcal{X}) (t -\bar t)^2 \ .
\end{equation}
However, the overall normalization can be changed by a K\"ahler transformation. Nevertheless, analogously to the toriodal model, the two sphere partition function in the canonical K\"ahler gauge takes the form
\begin{equation}
   Z^{1,8,5}_{S^2}(z) \,=\, -\frac{c}{4\pi^2} \left| \omega^{1,8,5}_0(z) \right|^2 (\log z - \log\bar z)^2 + \ldots \quad \text{for} \quad |z|\text{ small} \ ,
\end{equation}
which we want to compare in the two proposed dual phases. 

Let us now examine the remaining models in Table~\ref{tb:models}. The conformal \model s with central charge three, six and twelve differ from $SSSM_{1,12,6}$ only in the integers $m$ and $n$, which determine the multiplicities of the irreducible representations of the matter multiplets. In particular, the gauge group and the set of different irreducible representations are unchanged. Since $\binom{n}2>m$ holds for all these models, poles arising from $Z_\phi$ are always canceled by the denominator of $Z_P$ --- as noted below eq.~\eqref{eq:DivisModel1} for $SSSM_{1,12,6}$. Therefore the discussion in Section~\ref{sec:M1General} directly carries over, only the powers of the respective terms in $Z_{\text{matter}}$ have to be changed. For the phase $r\gg0$ we find
\begin{equation}
Z_{S^2,r\gg0}^{1,m,n}(r,\theta)=(-1)^{m+\binom{n}2}\frac{\left(z \overline{z}\right)^\mathfrak{q}}{2} \operatorname{Res}_{\boldsymbol{x}=0} \left(Z_{\text{sing}} \left|z^{x_1+x_2}\sum_{a=0}^\infty (-1)^{a(1+n+m)} z^a \sum_{b=0}^a Z_{\text{reg}} \right|^2\right), \label{eq:ZGen1}
\end{equation}
with
\begin{equation}
\begin{aligned}
Z_{\text{sing}} &=\pi^{2(n-1)-\binom{n}2+m} \, \frac{\text{sin}\left[\pi (x_1+x_2)\right]^{\binom{n}2-m}\text{sin}\left[\pi (2x_1+x_2)\right]\text{sin}\left[\pi (x_1+2x_2)\right]}{\text{sin}\left(\pi x_1\right)^n\text{sin}\left(\pi x_2\right)^n}\ ,\\
Z_{\text{reg}} &=(a-2b-x_1+x_2) \frac{\Gamma (1+a+x_1+x_2)^{\binom{n}2-m}}{\Gamma(1+b+x_1)^n \Gamma(1+a-b+x_2)^n}\\
&\, \qquad \quad\cdot \Gamma(1+a+b+2x_1+x_2)\Gamma(1+2a-b+x_1+2x_2)\ .
\end{aligned}
\end{equation}
The residue is being taken with respect to the oriented pair of divisors $(x_1,x_2)$. An explicit evaluation of eq.~\eqref{eq:ZGen1} then gives
\begin{equation}\label{eq:PeriodGen1}
\begin{aligned}
  \omega^{1,m,n}_{0,r\gg0}(z)  = \sum_{a=0}^\infty \sum_{b=0}^a (-1)^{a(1+m+n)}&\frac{a!^{\binom{n}2-m}(2a-b)!(a+b)!}{(a-b)!^n b!^n}\\
  &\qquad\quad\cdot\bigg[1 +(2b-a)\left(H_{a+b}-n H_{b}\right) \bigg] z^a \ .
\end{aligned}
\end{equation}
Similarly, in the phase $r\ll0$ the partition function reads
\begin{equation}
\begin{aligned}
Z_{S^2,r\ll0}^{1,m,n}(r,\theta) =&(-1)^{n+1} \frac{\left(w \overline{w}\right)^{1-\mathfrak{q}}}{2} \lim_{\varepsilon \to 0^+}\\
& \operatorname{Res}_{\boldsymbol{x}=0} \left( e^{2x_1 \varepsilon} Z_{\text{sing}} \left|w^{-(x_1+x_2)}\sum_{a=0}^\infty (-1)^{a(m+n)} w^a e^{\varepsilon a} \sum_{k=0}^\infty (-1)^k Z_{2} e^{-\varepsilon k}\right|^2\right)\text{,}
\end{aligned} \label{eq:ZGen2}
\end{equation}
where
\begin{equation}
\begin{aligned}
Z_{\text{sing}} &= \frac{\operatorname{sin}(\pi x_1)^n\operatorname{sin}(\pi x_2)^n\operatorname{sin}\left[\pi \left(x_1+2x_2\right)\right]}{\pi^{2n-\binom{n}2+m} \operatorname{sin}\left[\pi (x_1+x_2)\right]^{\binom{n}2-m}\operatorname{sin}\left[\pi(2x_1+x_2)\right]}\ , \\
Z_2 &= \left(1+3a-2k+x_1-x_2\right) \cdot \\
&\,\qquad \frac{\Gamma (-a+k-x_1)^n\Gamma (1+2a-k-x_2)^n\Gamma(-1-3a+k+x_1+2x_2)}{\Gamma(1+k-2x_1-x_2)\Gamma(1+a-x_1-x_2)^{\binom{n}2-m}}\ .
\end{aligned}
\end{equation}
Here, the residue is taken with respect to the ordering of divisors specified in eq.~\eqref{eq:M1jLables}. In this phase, the fundamental period reads
\begin{equation}
\begin{aligned}
\omega^{1,m,n}_{0,r\ll0}(w) &= \sum_{a=0}^\infty  \sum_{k=0}^a (-1)^{k\,n+a\left[n+\binom{n}2\right]}\frac{(1+3a-2k)(2a-k)!^n}{k! (1+3a-k)! a!^{\binom{n}2-m} (a-k)!^n} w^a \label{eq:PeriodGen2}\ .
\end{aligned}
\end{equation}

Let us now first specialize to the toroidal model $SSSM_{1,5,4}$, which we expect to be self-dual. Using the general expressions~\eqref{eq:PeriodGen1} and \eqref{eq:PeriodGen2} we in fact find
\begin{equation}
\begin{aligned}
\omega^{1,5,4}_{0,r\gg0}(z) &=  1-3z+19z^2-147z^3 +1\,251 z^4-11\,253z^5 + \ldots \ , \\
\omega^{1,5,4}_{0,r\ll0}(w) &= 1+3w+19w^2+147w^3 +1\,251 w^4+11\,253w^5  + \ldots \ ,
\end{aligned}
\end{equation}
i.e., in the expansion we observe the equality
\begin{equation}
\omega^{1,5,4}_{0,r\gg0}(z) = \omega^{1,5,4}_{0,r\ll0}(w) \quad \text{for } w = -z \ .
\end{equation}
From this we deduce that the two-sphere partition function satisfies the equality
\begin{equation}
Z^{1,5,4}_{S^2}(r,\theta,\mathfrak{q}) \,=\, Z^{1,5,4}_{S^2}(-r,\pi-\theta,1-\mathfrak{q}) \ ,
\end{equation}
which provides further evidence for self-duality of the model $SSSM_{1,5,4}$. On the level of the associated Picard--Fuchs operator
\begin{equation}
\begin{aligned}
\mathcal{L}^{1,5,4}_1&=\mathcal{L}^{1,5,4}(z) = \theta_z^2 + z\left(3+11\theta_z+11\theta_z^2\right) - z^2\left(1+\theta_z\right)^2, \quad \theta_z = z \frac{\partial}{\partial z}\ ,\\
\mathcal{L}_2& = w^2\mathcal{L}(w^{-1}),
\end{aligned}
\end{equation}
which annihilates the fundamental period, the self-duality becomes also apparent as the Picard--Fuchs operator exhibits the symmetry 
\begin{equation}
\mathcal{L}^{1,5,4}_1\, \omega^{1,5,4}_{0,r\gg0}(z) = 0 \ ,\qquad \mathcal{L}^{1,5,4}_2 \,\left( w \,\omega^{1,5,4}_{0,r\ll0}(w) \right) = 0 \ .
\end{equation}
This symmetry of the Picard--Fuchs operator becomes even more manifest after the shift
\begin{equation}
\widehat{\mathcal{L}}^{\,1,5,4}_1 =\widehat{\mathcal{L}}^{\,1,5,4}(z) = \left(\theta_z-\frac{1}{2}\right)^2+z\left(11\theta_z^2+\frac{1}{4}\right) -z^2\left(\theta_z+\frac{1}{2}\right)^2,
\end{equation}
which fulfills
\begin{equation}
 \widehat{\mathcal{L}}^{\,1,5,4}_1 \,\left( z^{\frac{1}{2}}\,\omega^{1,5,4}_{0,r\gg0}(z) \right) = 0\ , \quad 
 \widehat{\mathcal{L}}^{\,1,5,4}_2 \, \left(w^{\frac{1}{2}}\,\omega^{1,5,4}_{0,r\ll0}(-w) \right)= 0\ , 
\end{equation}
with $\widehat{\mathcal{L}}^{\,1,5,4}_2 = w^2\widehat{\mathcal{L}}^{\,1,5,4}(-w^{-1}) = - \widehat{\mathcal{L}}^{\,1,5,4}_1$.

Next, we look at the model $SSSM_{1,8,5}$ describing a K3 surface, which is also expected to be self-dual according to eq.~\eqref{eq:K3sd}. In expansion the fundamental periods read
\begin{equation}
\begin{aligned}
\omega^{1,5,4}_{0,r\gg0}(z) &=  1-5z+73z^2-1\,445z^3 +33\,001 z^4-819\,005z^5 + \ldots \ , \\
\omega^{1,5,4}_{0,r\ll0}(w) &=  1-5w+73w^2-1\,445w^3 +33\,001 w^4-819\,005w^5 + \ldots  + \ldots \ ,
\end{aligned}
\end{equation}
from which we deduce
\begin{equation}
\begin{aligned}
\omega^{1,8,5}_{0,r\gg0}(z) &= \omega^{1,8,5}_{0,r\ll0}(w) \quad \text{for } w = z \ ,\\
Z^{1,8,5}_{S^2}(r,\theta,\mathfrak{q}) \,&=\, Z^{1,8,5}_{S^2}(-r,-\theta,1-\mathfrak{q}) \ ,
\end{aligned}
\end{equation}
in support of the self-duality. Similar to the previous case, the duality is also manifest in terms of the associated (shifted) Picard--Fuchs operator 
\begin{equation}
  \widehat{\mathcal{L}}^{\,1,8,5}_1 =\widehat{\mathcal{L}}^{\,1,8,5}(z) = \left( \theta_z - \frac{1}{2}\right)^3 
  + \frac{1}{2}z\, \theta_z \left(68 \theta_z+3\right) + z^2\left(\theta_z + \frac{1}{2}\right)^3\ ,
\end{equation}
which fulfills
\begin{equation}
 \widehat{\mathcal{L}}^{\,1,8,5}_1 \,\left( z^{\frac{1}{2}}\,\omega^{1,8,5}_{0,r\gg0}(z) \right) = 0\ , \quad 
 \widehat{\mathcal{L}}^{\,1,8,5}_2 \, \left(w^{\frac{1}{2}}\,\omega^{1,8,5}_{0,r\ll0}(w) \right)= 0\ , 
\end{equation}
with $\widehat{\mathcal{L}}^{\,1,8,5}_2 = w^2\widehat{\mathcal{L}}^{\,1,8,5}(w^{-1}) = - \widehat{\mathcal{L}}^{\,1,8,5}_1$.

Finally, we turn to the model $SSSM_{1,17,7}$ with central charge twelve. Geometric invariants of the large volume phases can be extracted from the two sphere partition function according to ref.~\cite{Honma:2013hma}. Here we do not further examine the geometric properties of the phases of this model, but instead just determine the respective fundamental periods
\begin{equation}
\begin{aligned}
   \omega^{1,17,7}_{0,r\gg0}(z) &=  1+9z+469z^2+38\,601z^3 +4\,008\,501 z^4+ \ldots \ , \\
   \omega^{1,17,7}_{0,r\ll0}(w) &=  1+21w+2\,989w^2+714\,549w^3 +217\,515\,501w^4+ \ldots \ .
\end{aligned}
\end{equation}
As expected for this model we do not find an indication for a self-duality correspondence.

\section{A derived equivalence of Calabi--Yau manifolds} \label{sec:derivedequiv}
The proposed duality between the \model s $SSSM_{1,12,6}$ and $SSSM_{2,9,6}$ --- together with the resulting identification of geometric large volume phases according to eq.~\eqref{eq:XYdual} --- implies that the Calabi--Yau threefolds $\mathcal{X}_{1,12,6}\simeq\mathcal{Y}_{2,9,6}$ and $\mathcal{X}_{2,9,6}\simeq\mathcal{Y}_{1,12,6}$ occur both as geometric phases in the quantum K\"ahler moduli space of the infrared worldsheet theory of the associated type~II string compactification. In this section we discuss the implications of our findings for the associated brane spectra.

First, we briefly summarize our findings concerning the target spaces $\mathcal{X}_{k,m,n}$ and $\mathcal{Y}_{k,m,n}$. In particular, we recall the explicit correspondence between their complex structure moduli. As formulated in eqs.~\eqref{eq:Xvariety} and \eqref{eq:Yvariety}, their respective complex structures are encoded in terms of their intersecting projective subspaces $\mathbb{P}(L) \subset \mathbb{P}(V,\Lambda^2V^*)$ and $\mathbb{P}(L^\perp) \subset \mathbb{P}(V^*,\Lambda^2V)$, which are related by the orthogonality condition~\eqref{eq:Lperp}. 

The result is that for given a \model{} $SSSM_{k,m,n}$ --- with all the previously assumed genericness conditions and the constraints on the integers $(k,m,n)$ fulfilled --- we find for even $n$ that the geometric phases $\mathcal{X}_{k,m,n}$ and $\mathcal{Y}_{k,m,n}$ are realized in terms of the projective varieties
\begin{equation} 
\begin{aligned}
   \mathcal{X}_{k,m,n} \,&=\, \left\{ [\phi,\tilde P] \in \mathbb{P}( V, \Lambda^2 V^*) \, \middle| \, 
   \operatorname{rk} \tilde P \le 2k \ \text{and}\ \phi \in \ker \tilde P \right\}\cap \mathbb{P}(L) \ , \\
  \mathcal{Y}_{k,m,n}\,&=\, 
   \left\{ [\tilde\phi,P] \in \mathbb{P}( V^*, \Lambda^2 V) \, \middle| \, 
   \operatorname{rk} P \le 2\tilde k \ \text{and}\ \tilde\phi \in \ker P \right\} \cap \mathbb{P}(L^\perp) \ ,
\end{aligned}
\end{equation}
with
\begin{equation}
\begin{aligned}
    &n \text{ even } \ , \quad \tilde k= \frac{n}2 - k \ , \quad L(L^\perp) = \emptyset \ \text{  for  } \
    L\subset V\oplus \Lambda^2 V^*\ ,\  L^\perp \subset V^*\oplus \Lambda^2 V \ ,  \\ 
    &\dim_\mathbb{C} V=n \ , \quad \dim_\mathbb{C} L = m \ , \quad \dim_\mathbb{C} L^\perp = \binom{n+1}2-m \ .
\end{aligned}    
\end{equation}
These projective spaces are constructed analogously to refs.~\cite{Galkin:2014Talk,Galkin:2015InPrep}.

Our main interest in this section is the model $SSSM_{1,12,6}$ with the two Calabi--Yau threefolds phases $\mathcal{X}_{1,12,6}$ and $ \mathcal{Y}_{1,12,6}$. In ref.~\cite{Miura:2013arxiv} Miura studies the Calabi--Yau threefold variety $\mathcal{X}_{1,12,6}$ with a different construction and anticipates the emergence of the Calabi--Yau variety $\mathcal{Y}_{1,12,6}$ in the same quantum K\"ahler moduli space by means of mirror symmetry. The above construction of both Calabi--Yau varieties and the relationship to Miura's Calabi--Yau threefold has already appeared in ref.~\cite{Galkin:2014Talk,Galkin:2015InPrep}. In this work this picture is confirmed, since the Calabi--Yau threefold~$\mathcal{X}_{1,12,6}$ arises as the weakly-coupled phase $r\gg0$ and $\mathcal{Y}_{1,12,6}$ as the strong-coupled phase $r\ll0$ of the model~$SSSM_{1,12,6}$. According to the duality relation eqs.~\eqref{eq:XYdual} and \eqref{eq:dualpair}, the same large volume Calabi--Yau threefold phases occur in the dual model $SSSM_{2,9,6}$ just with the role of weakly- and strongly-coupled phases exchanged.

From the combined analysis of $SSSM_{1,12,6}$ and $SSSM_{2,9,6}$ we conclude that the two distinct geometric Calabi--Yau threefold phases $\mathcal{X}_{1,12,6}$ and $\mathcal{Y}_{1,12,6}$ are path connected in the quantum K\"ahler moduli space. As already argued at the end of Section~\ref{sec:CYs} these threefolds are not birational to each other. 

The appearance of two geometric phases in the same connected quantum K\"ahler moduli space has an immediate consequence for the spectra of branes in the two geometric regimes. In the topological B-twisted string theory associated to the \model s we capture the topological B-branes, which after imposing a suitable stability condition give rise to B-type BPS branes in the physical theory \cite{Douglas:2000gi,Douglas:2000ah,Aspinwall:2001dz}. While the notion of stability depends on the K\"ahler moduli of the theory, the spectrum of topological B-branes are insensitive to K\"ahler deformation.\footnote{The spectrum of topological B-branes, however, does depend on the complex structure moduli in a non-trivial way, which for instance becomes explicit in the study of open-closed deformation spaces, see e.g. refs.~\cite{Jockers:2008pe,Alim:2009rf,Alim:2009bx}.} Therefore, by continuity we can determine the category of topological B-branes at any point in the quantum K\"ahler moduli space. In particular, for any value of the (complexified) Fayet--Iliopoulos parameter the infrared worldsheet theory comes with the same spectrum of topological B-branes \cite{Herbst:2008jq}.  

The categories of topological B-branes in the geometric Calabi--Yau threefold phases $\mathcal{X}_{1,12,6}$ and $\mathcal{Y}_{1,12,6}$ are mathematically described in terms of their derived categories of bounded complexes of coherent sheaves $\mathcal{D}^b(\mathcal{X}_{1,12,6})$ and $\mathcal{D}^b(\mathcal{Y}_{1,12,6})$  \cite{Douglas:2000gi,Diaconescu:2001ze}. As a consequence of the universality of the category of topological branes in the model $SSSM_{1,12,6}$, we propose a derived equivalence between the Calabi--Yau threefolds $\mathcal{X}_{1,12,6}$ and $\mathcal{Y}_{1,12,6}$, i.e.,
\begin{equation}
    \mathcal{D}^b(\mathcal{X}_{1,12,6}) \, \simeq \, \mathcal{D}^b(\mathcal{Y}_{1,12,6}) \ , \qquad
     \mathcal{D}^b(\mathcal{X}_{2,9,6}) \, \simeq \, \mathcal{D}^b(\mathcal{Y}_{2,9,6}) \ .
\end{equation}       
Using mirror symmetry Miura also proposes the existence of such a derived equivalence \cite{Miura:2013arxiv}, which --- combined with the work by Galkin, Kuznetsov and Movshev \cite{Galkin:2014Talk,Galkin:2015InPrep} --- directly relates to our proposal.\footnote{Due to Orlov's theorem such a derived equivalence can be proven by finding an invertible Fourier--Mukai transformation between the derived categories of bounded complexes of coherent sheaves of the Calabi--Yau varieties \cite{MR1465519}.}

Finally, we briefly comment on the model $SSSM_{1,8,5}$ with the self-duality correspondence~\eqref{eq:K3sd} and with two geometric phases of degree twelve  polarized K3~surfaces $\mathcal{X}_{\mathbb{S}_5}$ and $\mathcal{Y}_{\mathbb{S}_5^*}$ in eqs.~\eqref{eq:K31} and \eqref{eq:K32}. By the same arguments, their appearance in a connected component of the quantum K\"ahler moduli space of the model $SSSM_{1,8,5}$ implies the derived equivalence
\begin{equation}
   \mathcal{D}^b(\mathcal{X}_{\mathbb{S}_5}) \, \simeq \, \mathcal{D}^b(\mathcal{Y}_{\mathbb{S}_5^*}) \ .
\end{equation}
This derived autoequivalence among polarized K3 surfaces of degree twelve is well-known in the literature~\cite{MR1714828,MR2047679,Hosono:2014ty}. Here, it supports the validity of the presented gauged linear sigma model arguments.

\section{Conclusions} \label{sec:con}
The aim of this work was to introduce and to study \model s, which furnished a specific class of two-dimensional $N=(2,2)$ gauged linear sigma models based upon a non-Abelian symplectic gauge group factor. While part of our discussion even applied for an anomalous axial $U(1)_A$ R-symmetry, we mainly focused on \model s with non-anomalous axial $U(1)_A$ R-symmetries. Using standard arguments~\cite{Witten:1993yc}, at low energies they described interesting $N=(2,2)$ superconformal field theories that were suitable for worldsheet theories of type~II string compactifications. In particular, we identified \model s with two distinct and non-birational large volume Calabi--Yau threefold target spaces. The emergent Calabi--Yau spaces were projective varieties of the non-complete intersection type also discussed in refs.~\cite{Miura:2013arxiv,Galkin:2014Talk,Galkin:2015InPrep}. The phase structure of the gauged linear sigma models led towards a non-trivial duality proposal within the class of \model s, for which strong coupling effects of the symplectic gauge group factor became important at low energies.

To support our claims we used the two sphere partition function correspondence so as to extract perturbative and non-perturbative geometric invariants --- namely the intersection numbers and genus zero Gromov--Witten invariants of the semi-classical target space phases. In this way, we were able to confirm the geometric properties of the studied Calabi--Yau threefold target space varieties. Furthermore, by finding agreement of the two sphere partition functions of \model s, we presented strong evidence in favor of our duality proposal. As a side let us remark that the demonstrated match of the analytic expressions for dual two sphere partition functions was a result of rather non-trivial identities among nested infinite sums. Analyzing the observed identities and tracing in detail their emergence through gauge dualities on the level of two sphere partition functions could be rewarding both from a physics and a mathematics perspective.

In order to compute the two sphere partition functions of \model s, we developed a systematic way to evaluate higher-dimensional Mellin--Barnes type integrals in terms of local Grothendieck residues. Our techniques were inspired by the interesting work of Zhdanov and Tsikh in ref.~\cite{MR1631772}, where a method to evaluate two-dimensional Mellin--Barnes type integrals is established. Our extension to higher dimensions admits a systematic series expansion of integrals arising from two sphere partition functions of two-dimensional $N=(2,2)$ gauge theories with higher rank gauge groups. However, our approach is also applicable in other contexts where Mellin--Barnes type integrals appear, such as the analytic continuation of periods arising from systems of Picard--Fuchs differential equations. 

Interpreting the \model s with an axial $U(1)_A$ anomaly-free R-symmetry as worldsheet theories for type~II strings, we arrived at an interesting equivalence between the derived category of the bounded complexes of coherent sheaves for a pair of Calabi--Yau threefolds. For the explicit pairs of derived equivalent Calabi--Yau threefolds their embedding in projective and dual projective spaces played a prominent role. This suggests that the proposed derived equivalence is actually a consequence of homological projective duality as proposed by Kuznetsov \cite{MR2354207} and as proven for the R\o{}dland example in ref.~\cite{Kuznetsov:2006arxiv}. As the two respective Calabi--Yau threefold phases were related by a non-trivial two-dimensional $N=(2,2)$ gauge duality, it would be interesting to see more generally if there is a non-trivial relationship between two-dimensional gauge theory dualities --- as established in refs.~\cite{Hori:2006dk,Hori:2011pd} --- and homological projective duality in algebraic geometry \cite{Kuznetsov:2006arxiv}. While the connection between geometric invariant theory --- realizing mathematically the target space geometries of gauged linear sigma models --- and homological projective duality is for instance studied in refs.~\cite{Ballard:2012rt,Ballard:2013fxa}, the phase structure arising from non-trivial two-dimensional gauge theory dualities seems to require a more general notion of geometric invariant theory quotients \cite{Addington:2012zv,Addington:2014sla,Segal:2014jua}.

\bigskip
\subsection*{Acknowledgments}
We would like to thank
Lev Borisov,
Stefano Cremonesi,
Sergey Galkin,
Shinobu Hosono,
Daniel Huybrechts,
Albrecht Klemm,
and
Eric Sharpe
for discussions and correspondences. In particular, we are indebted to Kentaro Hori, who resolved conceptual shortcomings in a preliminary draft of this work, and as a result the discussion in Section~\ref{sec:strongphase} crucially builds upon his input.
A.G.~is supported by the graduate school BCGS and the Studienstiftung des deutschen Volkes, and
H.J.~is supported by the DFG grant KL 2271/1-1.

\newpage
\appendix

\section{Multidimensional Mellin--Barnes type integrals}\label{sec:MB}
In this Appendix we present an algorithm to evaluate multidimensional Mellin--Barnes type integrals in terms of suitable sums over local Grothendieck residues. Such integrals arise for instance in the context of periods of Calabi--Yau manifolds or in computing the two sphere partition function of two-dimensional supersymmetric $N=(2,2)$ gauge theories \cite{Doroud:2012xw,Benini:2012ui}, which is also our motivation to study them here. While Zhdanov and Tsikh have presented a beautiful combinatorial construction to evaluate two-dimensional Mellin--Barnes type integrals \cite{MR1631772}, we extend their method to integrals in arbitrary dimension. This generalization allows us to evaluate --- at least in principle --- the two sphere partition functions for Abelian $U(1)^n$ gauge theories with $n$ arbitrary and for non-Abelian gauge theories with higher rank gauge groups. In the context of gauged linear sigma models as worldsheet theories for closed strings on Calabi--Yau target spaces \cite{Witten:1993yc}, the presented extension admits for instance the explicit evaluation of the two sphere partition function for complete intersection Calabi--Yau geometries in toric ambient spaces with several K\"ahler moduli or for non-complete intersection Calabi-Yau projective varieties based on constructions involving higher rank non-Abelian gauge groups.

The relevant $n$-dimensional Mellin--Barnes type integrals are of the form
\begin{equation} \label{eq:MBInt} 
  I_\gamma \,=\,\int_{\gamma + i \mathbb{R}^n} \omega
  \,=\, \int_{\gamma + i \mathbb{R}^n} \omega(\boldsymbol{x})\,dx_1\wedge \ldots \wedge dx_n \ , \qquad \gamma \in \mathbb{R}^n \ ,
\end{equation}
in terms of the coordinates of $\boldsymbol{x}=(x_1,\ldots,x_n)$ of $\mathbb{C}^n$ and with
\begin{equation} \label{eq:MBw}
\begin{aligned}
  \omega(\boldsymbol{x}) \,=\, h(\boldsymbol{x}) 
    \prod_{j=1}^{j_{\text{max}}}&\Gamma(g_j(\boldsymbol{x}))^{\alpha_j}\cdot e^{-\boldsymbol{p}\cdot \boldsymbol{x}}\ , \qquad
    g_j(\boldsymbol{x}) \,=\, \boldsymbol{a}_j \cdot \boldsymbol{x}+b_j \ , \\
    &\alpha_j \in \mathbb{N},\ b_j \in \mathbb{R},\ \boldsymbol{a}_j,\boldsymbol{p} \in \mathbb{R}^n  \ .
\end{aligned}  
\end{equation}
Here $h(\boldsymbol{x})$ is an entire function on $\mathbb{C}^n$. Without loss of generality we assume that $g_j(\boldsymbol{x}) \neq g_k(\boldsymbol{x})$ for all $j \neq k$. Furthermore, we assume that $\omega(\boldsymbol{x})$ is regular at $\gamma$ and that its asymptotics is dominated by the exponential factor $e^{-\boldsymbol{p}\cdot\boldsymbol{x}}$.

For integrals~$I_\gamma$ associated to two sphere partition functions of two-dimensional $N=(2,2)$ gauge theories the vector $\boldsymbol{p}$ depends linearly on the Fayet--Iliopoulos parameters $\boldsymbol{r}$. Furthermore, if the axial $U(1)_A$ R-symmetry is anomaly free, the asymptotics of the integrand $\omega(\boldsymbol{x})$ is typically always governed by the exponential factor $e^{-\boldsymbol{p}\cdot\boldsymbol{x}}$, even for different choices of signs in the entries of the vector $\boldsymbol{p}$. Such choices thus tune the integral~$I_\gamma$ into different phases, which in the string theory context model different regimes in the quantum K\"ahler moduli space of the described string compactifications. 

\subsection{The strategy to analyze Mellin--Barnes type integrals} \label{sec:strategy}
The basic idea is to calculate $I_\gamma$ with $n$ successive contour integrations in the respective complex planes of the coordinates~$\boldsymbol{x}$. For definiteness, we assume the first integration to be in $x_1$. It then depends on the vector $\boldsymbol{p}$ within the exponential in which half plane the contour has to be chosen, such that no additional contributions arise from closing the contour. As it is in the one-dimensional case, the integral will be given as the sum over all encircled residues. The position of a specific pole now, however, is a hyperplane in $\mathbb{R}^n$ --- spanned by the real part of $\boldsymbol{x}$ --- that is parametrized by the $x_i$ for $i=2,\ldots,n$. To these hyperplanes we will refer as \emph{divisors}. After the first contour integration the original integral thus localized on certain divisors. Now, the second contour integration is carried out in the same manner. It picks out those locations at which another Gamma function diverges and thus localizes on the intersection of another divisor with the divisors on which the previous integration has already localized upon. Therefore, the values of the remaining $x_i$ with $i=3,\ldots,n$ are now restricted to certain $(n-2)$-dimensional planes in the original space~$\mathbb{R}^n$. Repeating this argument $(n-2)$ further times, non-zero contributions to $I_\gamma$ can only stem from residues at the intersection of $n$ divisors. The challenging question to answer is, which of those residues have to be taken into account.

Unlike in the one-dimensional case, the residue contributions arising from the higher-dimensional contour cannot directly be read off from the vector $\boldsymbol{p}$ in the exponential term. After contour integration in $x_1,\ldots, x_m$, these variables are replaced by the values obtained from solving the respective $m$ hyperplane equations. Since this solution in general depends on the remaining $x_{m+1},\ldots, x_n$, the coefficients of these remaining $(n-m)$ variables within the exponent may change. The coefficients can even reverse their sign, such that a contour potentially has to be closed in different half spaces for different divisors localized upon in the previous integrations. Moreover, the position of a pole in the $x_m$ plane can depend on the values of $x_{m+1}, ..., x_n$. As these are only fixed by later integrations, it is not obvious which divisors will turn out to be encircled.

Another complication arises when there are $n$ divisors whose intersection is of codimension $m<n$. Those $n$ divisors alone cannot give a contribution to $I_\gamma$. Intuitively this is understood, since $(n-m)$ variables of integration are not fixed by restricting to a codimension $m$ subspace. There can, however, be $(n-m)$ further divisors, whose intersection with any $m$ of the original $n$ divisors is a point. Such poles, located on strictly more than $n$ divisors, are referred to as \emph{degenerate}. In order to avoid a singular result in this case, they have to be treated as described in Section~\ref{sec:MBDegenerate}.

After the described combinatorial challenge has been addressed --- namely which intersections of poles contribute to the multidimensional residue calculus --- the higher dimensional residue is evaluated with the help of the local Grothendieck residue \cite{MR1631772}. For completeness, we recall the definition of the local Grothendieck residue here but refer, e.g., to ref.~\cite{MR1288523} for more details. Given an isolated solution $\boldsymbol{x}_0\in U$ to the holomorphic equations $f_1(\boldsymbol{x}_0)=f_2(\boldsymbol{x}_0)=\ldots=f_n(\boldsymbol{x}_0)=0$ in the open set $U\subset\mathbb{C}^n$, a real $n$-cycle $\mathcal{C}$ is defined as
\begin{equation}
  \mathcal{C} \,=\, \left\{\, \strut\boldsymbol{x} \in U \,\middle|\,  \left| f_k(\boldsymbol{x}) \right|=\varepsilon_k \, \right\} \subset U\ ,
\end{equation}
with $\varepsilon_k>0$ sufficiently small and oriented such that $d(\arg f_1) \wedge \ldots d(\arg f_n)$ is positive. Then the integral $\int_\mathcal{C} \omega$ over the meromorphic $n$-form
\begin{equation}
  \omega \,=\, \frac{g(\boldsymbol{x})}{f_1(\boldsymbol{x}) \cdots f_n(\boldsymbol{x})} dx_1 \wedge \ldots \wedge dx_n \ ,
\end{equation}
with $g(\boldsymbol{x})$ holomorphic in $U$, is determined by the local Grothendieck residue as
\begin{equation} 
  \int_\mathcal{C} \omega \,=\, (2\pi i)^n \operatorname{Res}_{\boldsymbol{x}_0} \omega \ .
\end{equation}
The residue symbol is evaluated by
\begin{equation} \label{eq:GrRes}
   \operatorname{Res}_{\boldsymbol{x}_0}\omega \,=\, \frac{g(\boldsymbol{x}_0)}{\mathcal{J}_f(\boldsymbol{x}_0)} \ ,
   \quad \mathcal{J}_f(\boldsymbol{x})\,=\, \det \frac{\partial(f_1(\boldsymbol{x}),\ldots,f_n(\boldsymbol{x}))}{\partial(x_1,\ldots,x_2)} \ .
\end{equation}
Note that the (sign of) the residue depends on the order of the divisors $f_1$ through $f_n$, so as to conform with the chosen orientation of the cycle $\mathcal{C}$.  

\subsection{Two-dimensional Mellin--Barnes type integrals}\label{sec:2dMB}
We first review the construction of Zhdanov and Tsikh in ref.~\cite{MR1631772}, which expresses two-dimensional Mellin-Barnes integrals as sums of local Grothendieck residues.

Consider the points $p_{j_1 j_2}^{m_1m_2}\in\mathbb{R}^2\subset \mathbb{C}^2$, at which the poles of two specific Gamma functions in eq.~\eqref{eq:MBInt} intersect transversely, i.e., 
\begin{equation} \label{eq:p2}
  p^{m_1 m_2}_{j_1 j_2} = \left(\,g^{-1}_{j_1}(-m_1)\cap g^{-1}_{j_2}(-m_2)\right) \ ,\quad (j_1,j_2,m_1,m_2) \in \mathcal{I}  \ ,
\end{equation}
with
\begin{equation} \label{eq:I2}
  \mathcal{I} = \left\{\,(j_1,j_2,m_1,m_2) \in \mathbb{N}^4_0\,\middle|\,
  1\leq j_1<j_2\leq j_{\text{max}}  \text{ and } \det(\boldsymbol{a}_{j_1},\boldsymbol{a}_{j_2})\ne 0\, \right\} \ .
\end{equation}
Furthermore, we consider the half space~$H$ and its boundary $\partial H$ given by
\begin{equation} \label{eq:MBHn2}
  H = \left\{ \boldsymbol{x} \in \mathbb{R}^2 \,\middle|\, \boldsymbol{p} \cdot \boldsymbol{x} > \boldsymbol{p} \cdot \gamma \right\} \ , \qquad 
   \quad \partial H = \left\{ \boldsymbol{x} \in \mathbb{R}^2 \, \middle| \, \boldsymbol{p} \cdot \boldsymbol{x} = \boldsymbol{p} \cdot \gamma \right\} \ . 
\end{equation}
The point $\gamma \in \partial H$ divides the line $\partial H$ into two rays $\partial H_\pm$ such that the ordered pair of rays~$\left(\partial H_+, \overrightarrow{\gamma p}\right)$ has positive orientation. Consider further the intersection points
\begin{equation}\label{eq:MBq2D}
q^{m}_{j}\,=\, \partial H \cap g^{-1}_{j}(-m)\,\in\,\partial H \ , \quad m\in\mathbb{N}_0 \ ,
\end{equation}
which we here assume to exist for any $j=1,\ldots,j_\text{max}$. The required modification for the more general case, namely with divisors $g^{-1}_{j}(-m)$ parallel to the line $\partial H$, is treated in Section~\ref{sec:MBParallel}. Finally with the help of the subsets
\begin{equation} \label{eq:I2pm}
  \widehat{\mathcal{I}}_\pm \,=\, 
  \left\{(j_1,j_2,m_1,m_2) \, \in \mathcal{I}\, \middle|\, q^{m_1}_{j_1} \, \in \partial H_\pm \text{ and } q^{m_2}_{j_2} \, \in \partial H_\mp 
  \text{ and } p^{m_1 m_2}_{j_1 j_2}\in H \right\} \ ,
\end{equation}  
we define the contributing sets of poles
\begin{equation} \label{eq:P2pm}
  \Pi_\pm \,=\, \coprod_{(j_1,j_2,m_1,m_2)\, \in \,\widehat{\mathcal{I}}_\pm}  p^{m_1 m_2}_{j_1 j_2}\ .
\end{equation}  
Then Zhdanov and Tsikh show that the two-dimensional Mellin--Barnes integral $I_\gamma$ is given by the sum of local Grothendieck residues
\begin{equation} \label{eq:MB2D}
  I_\gamma \,=\,(2\pi i)^ 2\left(\sum_{\boldsymbol{x} \, \in \, \Pi_+} \operatorname{Res}_{\boldsymbol{x}}\omega 
    -  \sum_{\boldsymbol{x} \, \in \, \Pi_-} \operatorname{Res}_{\boldsymbol{x}} \omega\right)  \ ,
\end{equation}
where the residues are evaluated with eq.~\eqref{eq:GrRes} in terms of the two divisors associated to the poles $p^{m_1 m_2}_{j_1 j_2}$ ordered according to their indices $j_1, j_2$. Note that if some of the poles $p^{m_1 m_2}_{j_1 j_2}$ in the set $\Pi_\pm$ coincide the evaluation of some of the residues is a priori not defined. We relegate the analysis of such a situation to Section~\ref{sec:MBDegenerate}.

\subsection{Higher-dimensional Mellin--Barnes type integrals}
Now we proceed with Mellin--Barnes type integrals $I_\gamma$ of arbitrary dimensions. While most relevant formulas readily generalize, the construction of the two sets $\mathcal{I}_\pm$ is specific to two dimensions and requires a more general formulation.

In analogy to eqs.~\eqref{eq:p2} and \eqref{eq:I2} in the two dimensions, we define the points $p^{\vec m}_{\vec\jmath}$ with $\vec\jmath=(j_1,\ldots,j_n)$, $\vec m=(m_1,\ldots,m_n)$ of transversely intersecting poles by
\begin{equation}
    p^{\vec m}_{\vec\jmath} \,=\, \left(\,g^{-1}_{j_1}(-m_1)\cap \ldots \cap g^{-1}_{j_n}(-m_n)\right) \ ,\quad (\vec\jmath,\vec m) \in \mathcal{I} \ ,
\end{equation}
in terms of the index set
\begin{equation} 
  \mathcal{I} = \left\{\,(\vec\jmath,\vec m) \in \mathbb{N}^{2n}_0\,\middle|\,
  1\leq j_1<j_2<\ldots<j_n\leq j_{\text{max}}  \text{ and } \det(\boldsymbol{a}_{j_1},\ldots,\boldsymbol{a}_{j_n})\ne 0\, \right\} \ .
\end{equation}
The definition of the half space~\eqref{eq:MBHn2} directly generalizes to higher dimensions
\begin{equation}
 H = \left\{ \boldsymbol{x} \in \mathbb{R}^n \,\middle|\, \boldsymbol{p} \cdot \boldsymbol{x} > \boldsymbol{p} \cdot \gamma \right\} \ , \qquad 
   \quad \partial H = \left\{ \boldsymbol{x} \in \mathbb{R}^n \, \middle| \, \boldsymbol{p} \cdot \boldsymbol{x} = \boldsymbol{p} \cdot \gamma \right\} \ ,\label{eq:MBHHigher}
\end{equation}  
whereas the analogs of intersection points~\eqref{eq:MBq2D} become
\begin{equation} \label{eq:MBPoint}
  q^{\vec{\jmath},\vec{m}}_{k} = \partial H \cap 
  \left(\bigcap_{\substack{\ell=1\\ \ell\ne k}}^n g_{j_\ell}^{-1}(-m_\ell) \right) \ , \quad 1\le k \le n \ .
\end{equation}
\begin{figure}[th]
\offinterlineskip
\halign{#\hfil&~#\hfil&~#\hfil\cr
\footnotesize{\emph{(i)}}&\footnotesize{\emph{(ii)}}&\footnotesize{\emph{(iii)}}\cr
\mbox{\includegraphics[height=270px]{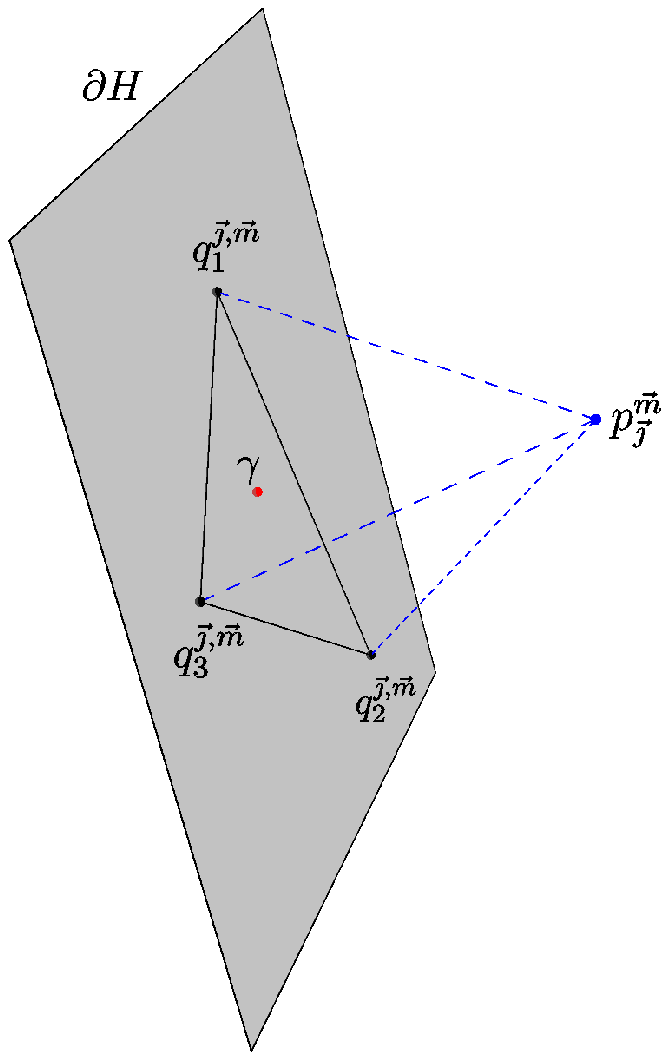}}&
\mbox{\includegraphics[height=270px]{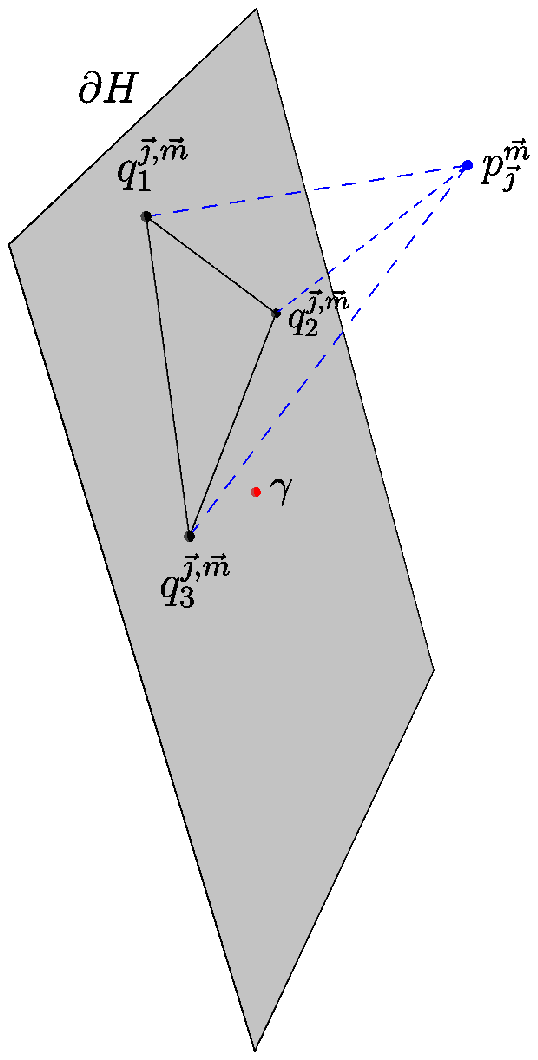}}&
\mbox{\includegraphics[height=270px]{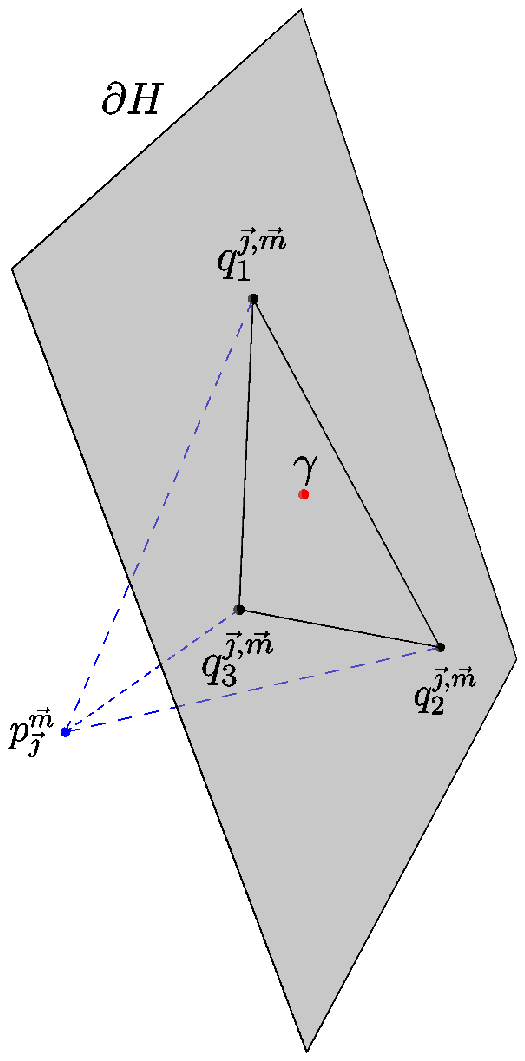}}\cr}
\caption{The images illustrate in a three-dimensional setting which poles $p^{\vec m}_{\vec\jmath}$ contribute to $\Pi$. Assuming that the half space $H$ is bounded by $\partial H$ to the left, Figure~(i) depicts a contributing pole as $\gamma\in\Delta^{\vec m}_{\vec\jmath}$ and $p^{\vec m}_{\vec\jmath}\in H$. The poles in the remaining pictures do not contribute because $\gamma\notin\Delta^{\vec m}_{\vec\jmath}$ in Figure~(ii) and $p^{\vec m}_{\vec\jmath}\notin H$ in Figure~(iii).}\label{fig:MBpoles}
\end{figure}
We assume in the following that the points~$q^{\vec{\jmath},\vec m}_{k}$, $k=1,\ldots,n$, for given $\vec{m}$ and $\vec{\jmath}$, are affinely independent. The necessary modifications for the non-generic cases, in which this assumption does not hold, is discussed in Section~\ref{sec:MBParallel}.\footnote{Such non-generic situations are actually a typical feature of Mellin--Barnes type integrals arising from the two sphere partition function of two-dimensional $N=(2,2)$ non-Abelian gauge linear sigma models.} To these $n$ affinely independent points $q^{\vec{\jmath},\vec{m}}_{k}$, we associate on $\partial H$ the $(n-1)$-dimensional simplexes
\begin{equation}
    \Delta^{\vec{m}}_{\vec\jmath}\,=\, \left\{\, \sum_{\ell=1}^n t_\ell\,q_\ell^{\vec\jmath,\vec m} \in \partial H \,\middle|\, 
    t_\ell\ge 0\text{ and }\sum_\ell t_\ell = 1\right\} \ .\label{eq:MBSimplex}
\end{equation}
This allows us to define with the index set
\begin{equation} \label{eq:MBIndexSet}
  \widehat{\mathcal{I}} \,=\, \left\{ (\vec\jmath,\vec m) \in \mathcal{I}\,\middle|\, \gamma\in \Delta^{\vec m}_{\vec\jmath}
  \text{ and } p^{\vec m}_{\vec\jmath} \in H \right\} \ ,
\end{equation}
the set $\Pi$ of contributing poles
\begin{equation} \label{eq:MBpoles}
  \Pi \,=\,  \coprod_{(\vec\jmath,\vec m)\, \in \,\widehat{\mathcal{I}}}  p^{\vec m}_{\vec\jmath}\ ,
\end{equation}  
as illustrated in Figure~\ref{fig:MBpoles}. With these definitions at hand, we can rewrite Mellin--Barnes type integral $I_\gamma$ in terms of a sum local Grothendieck residues as
\begin{equation} \label{eq:MBFormula}
   I_\gamma \,=\, (2\pi i)^n \sum_{\boldsymbol{x}\in\Pi}(\operatorname{sign}\Delta^{\vec m}_{\vec\jmath}) \operatorname{Res}_{\boldsymbol{x}} \omega \ .
 \end{equation}
Here, $\operatorname{sign}\Delta^{\vec m}_{\vec\jmath}$ denotes the orientation of the simplex $\Delta^{\vec m}_{\vec\jmath}$ --- as spanned by the ordered points $q_k^{\vec\jmath,\vec m}$, $k=1,\ldots,n$ --- in the hyperplane~$\partial H$, which itself is oriented such that the outer normal vector $-\boldsymbol{p}$ followed by a positive orientation in~$\partial H$ is the canonical positive orientation of $\mathbb{R}^n$. The residue is evaluated with eq.~\eqref{eq:GrRes} according to the given order of divisors $g^{-1}_{j_k}(-m_k)$, $k=1,\ldots,n$. 

The presented integration algorithm for Mellin--Barnes type integrals in arbitrary dimensions agrees in the two-dimensional case with the method by Zhdanov and Tsikh \cite{MR1631772}. As discussed in Section~\ref{sec:strategy} --- by performing one integration contour explicitly --- the presented generalization to arbitrary dimension can be shown via induction. 

In the degenerate case of a pole $p^{\vec m}_{\vec\jmath}\in\Pi$ residing on more than $n$ divisors, the evaluation of the integral~$I_\gamma$ needs to be further adjusted as discussed in Section~\ref{sec:MBDegenerate}.

\subsection{Non-generic integrals with degenerate simplexes}\label{sec:MBParallel}
In the above discussions, the points $q^{\vec{\jmath},\vec m}_{k}$ defined in eq.~\eqref{eq:MBPoint} were assumed to uniquely exist and to be affinely independent in $\partial H$ for given $\vec{\jmath}$ and $\vec{m}$, such that all the defined simplexes $\Delta^{\vec m}_{\vec\jmath}$ are non-degenerate.

While this assumption holds for the Mellin--Barnes type integrals of the two sphere partition function of Abelian gauged linear sigma models with generically chosen Fayet--Iliopoulos parameters $\boldsymbol{r}$, it often fails --- at least in some phases --- in the context of non-Abelian gauged linear sigma models, as observed in ref.~\cite{Jockers:2012dk} and as it is also the case for the \model s studied in this work. This can be traced back to the fact that the dimension of the integral of the two sphere partition functions for non-Abelian gauged linear sigma models is greater than the number of independent Fayet--Iliopoulos parameters. As a consequence the exponential term does not take its most generic form, which results in the described degeneracies of the simplexes $\Delta^{\vec m}_{\vec\jmath}$.

This problem is overcome by modifying the integrand in \eqref{eq:MBInt} by an additional exponential factor
\begin{equation} \label{eq:MBPar}
  \int_{\gamma + i\mathbb{R}^n} \omega \quad 
  \longrightarrow \quad \lim_{\varepsilon \to 0^+} \int_{\gamma +i \mathbb{R}^n} \omega 
  \cdot e^{-\varepsilon \,\tilde{\boldsymbol{p}} \cdot \boldsymbol{x}}\ ,
\end{equation}
with $\tilde{\boldsymbol{p}} \in \mathbb{R}^n$. This leads to the replacement $\boldsymbol{p} \rightarrow \boldsymbol{p}'= \boldsymbol{p}+ \varepsilon \,\tilde{\boldsymbol{p}}$, which results in a slightly tilted hyperplane $\partial H'$. For generically chosen $\tilde{\boldsymbol{p}}$ all simplexes $\Delta^{\vec{m}}_{\vec{j}}$ are rendered non-degenerate. If the limit on the right hand side of \eqref{eq:MBPar} exists, the specific choice of $\tilde{\boldsymbol{p}}$ is irrelevant for the end result~\eqref{eq:MBFormula}, even though the subset $\widehat{\mathcal{I}}$ defined in eq.~\eqref{eq:MBIndexSet} may depend on $\tilde{\boldsymbol{p}}$.

Let us close this section with a few general remarks concerning qualitative differences in introducing the shift $\varepsilon \tilde{\boldsymbol{p}}$. If the limit $\varepsilon \to 0^+$ described in eq.~\eqref{eq:MBPar} commutes with the integral, the limit can be taken on the level of individual terms in the sum of local Grothendieck residues. Then the original integral is well-defined and the introduced shift is really just a means to evaluate the integral by the described non-degenerate recipe of the previous section. If, however, the limit and the integrand do not commute, then the original integral $I_\gamma$ may not converge but instead only yield an asymptotic expansion. In the context of the two sphere partition function of two-dimensional $N=(2,2)$ gauged linear sigma models, the latter case indicates that in the analyzed phase the two sphere partition function is divergent and needs to be regulated. Then the shift in eq.~\eqref{eq:MBPar} furnishes a sensible regularization scheme.

\subsection{Non-generic integrals with degenerate poles}\label{sec:MBDegenerate}
If a pole in $\Pi$ as defined in eq.~\eqref{eq:MBpoles} is located on $m >n$ divisors, the residue is a priori not well-defined. It can be regularized by slightly shifting the divisors, i.e., by introducing in eq.~\eqref{eq:MBw} the shifts
\begin{equation}
   g_{j} \longrightarrow g_{j} + \delta \cdot \tilde{b}_j \ , \quad j=1,\ldots,j_\text{max} \ ,
\end{equation}
with $\tilde{b}_j \in \mathbb{R}$ and $\delta$ infinitesimal. For a generic choice of the parameters $\tilde{b}_j$, a degenerate pole arising from $m$ divisors is pulled apart into $\binom{m}{n}$ non-degenerate poles of $n$ intersecting divisors. After summing up the residues from all those poles in $\Pi$, the limit $\delta \to 0$ can be taken. We observe in examples that the existence of this limit crucially depends on the restriction to the set of poles $\Pi$, such that the final expression is unambiguous and well-defined.

\bigskip
\bibliographystyle{amsmod}
\bibliography{GJ}
\end{document}